\documentclass [10pt]{article}
\usepackage{latexsym}
\usepackage{amssymb}
\usepackage{amsfonts,lineno}
\usepackage{subfigure}
\usepackage{amsmath,mathrsfs}
\usepackage[dvips]{graphicx}
\usepackage[table, dvipsnames]{xcolor}
\usepackage{pstricks,geometry}
\usepackage[square, comma, sort&compress,numbers]{natbib}
\usepackage{float}
\usepackage{tikz}
\usetikzlibrary{shapes,arrows}
\usetikzlibrary{calc}
\usepackage{microtype}
\usepackage{color}
\usepackage{multirow}

\newif\ifnotesw \noteswtrue

\tikzstyle{decision} = [diamond, draw, text width=4.5em, text badly centered, node distance=3cm, inner sep=0pt]
\tikzstyle{block} = [rectangle, draw, text width=5em, text centered, rounded corners, minimum height=4em]

\title{Intermittent Preventive Treatment (IPT): Its role in averting disease-induced mortalities in children and in promoting the spread of antimalarial drug resistance}

\author{Carrie A. Manore$^{1}$, Miranda I. Teboh-Ewungkem$^{2}$, Olivia Prosper$^{3}$,  Angela L. Peace$^{4}$, \\ Katharine Gurski$^{5}$ ,  Zhilan Feng$^{6}$\thanks{Corresponding author: Phone: (1) ******; Fax: (1) *******; Email: *****@****.**** or ****@*****.****},
\\
$^{1}$Los Alamos National Laboratory and The New Mexico Consortium, Los Alamos, NM  \\
$^{2}$Department of Mathematics, Lehigh University, Bethlehem, PA 18015 USA,\\
$^{3}$Department of Mathematics, University of Kentucky,\\
$^{4}$Department of Mathematics and Statistics, Texas Tech University, Lubbock, TX 79409-1042. \\
$^{5}$Department of Mathematics, Howard University, Washington, DC 20059 USA\\
$^{6}$Department of Mathematics, Purdue University, West Lafayette, IN 47907-1395 USA}

\begin{document}

\date{\today}

\maketitle

\begin{abstract}
We develop a variable population age-structured ODE model to investigate the role of Intermittent Preventive Treatment (IPT) in averting malaria-induced mortalities in children, as well as its related cost in promoting the spread of anti-malarial drug resistance. IPT, a malaria control strategy in which a full curative dose of an antimalarial medication is administered to vulnerable asymptomatic individuals at specified intervals, has been shown to have a positive impact on reducing malaria transmission and deaths in children and pregnant women. However, it can also promote drug resistance spread. Our mathematical model is used to explore IPT effects on drug resistance in holoendemic malaria regions while quantifying the benefits in deaths averted. Our model includes both drug-sensitive and drug-resistant strains of the parasite as well as interactions between human hosts and mosquitoes. The basic reproduction numbers for both strains as well as the invasion reproduction numbers are derived and used to examine the role of IPT on drug resistance. Numerical simulations show the individual and combined effects of IPT and treatment of symptomatic infections on the prevalence levels of both parasite strains and on the number of lives saved. The results suggest that while IPT can indeed save lives, particularly in the high transmission region, certain combinations of drugs used for IPT and drugs used to treat symptomatic infection may result in more deaths when resistant parasite strains are circulating. Moreover, the half-lives of the treatment and IPT drugs used play an important role in the extent to which IPT may influence the rate of spread of the resistant strain. A sensitivity analysis indicates the model outcomes are most sensitive to the reduction factor of transmission for the resistant strain, rate of immunity loss, and the clearance rate of sensitive infections.
\end{abstract}


\section{Introduction}

Malaria continues to be a burden in many parts of the world, especially in the African continent. An estimated 214 million new malaria cases (range 149-303 million) were reported worldwide in 2015, with Africa contributing the most, about 88\%, followed by South-East Asia and the Eastern Mediterranean region, each contributing 10\% and 2\%, respectively \cite{WHO2015}. The estimated 2015 worldwide number of deaths was $438,000$, a decline from the 2012 estimates. Of these deaths, 90\% came from the African region, 7\% from South-East Asia and 2\%  from the Eastern Mediterranean region \cite{WHO2015, WHOFACTS2014, WHO2014}. Although malaria mortality rates are dropping (down by 60\% worldwide between 2000 and 2015), many people still suffer the burdens of illness, infection and death, with children under five more susceptible to these burdens. In fact, the 2015 globally estimated under five deaths was $306,000$  \cite{WHO2015}. Thus, strategies for reducing infection and disease burden in infants and children, groups bearing the highest burden of the disease, are increasingly urgent. Intermittent Preventive Treatment (IPT) is one such strategy employed.

IPT is a preventative malaria control strategy used as a tool to reduce disease burden and death among infants, children and pregnant women \cite{Goslingetal2010}. During IPT, these vulnerable humans are given a full curative antimalarial medication dose regardless of their infection status. IPT has been shown to be efficacious in reducing malaria incidence and burden in pregnant women, infants and children \cite{deloron2010sulfadoxine, ter2007effect, Matangila2015, konate2011intermittent}. In particular, its use in pregnant women (via IPTp) with the drug Sulfadoxine-pyrimethamine (SP) was shown to be efficacious \cite{deloron2010sulfadoxine, ter2007effect, Matangila2015}. In infants (via IPTi) and children (via IPTc), with the combination drug Sulfadoxine-pyrimethamine plus amodiaquine (SP+AQ), it was shown to be efficacious in reducing malaria incidence and burden \cite{Matangila2015, konate2011intermittent}, with significant protection for children sleeping under insecticide-treated bednets (ITNs) \cite{Matangila2015, konate2011intermittent}.

Although IPT (IPTp, IPTi, IPTc) as a malaria control strategy has been shown to have positive impact in averting disease deaths in IPT treated individuals, it faces challenges due to the emergence of resistance to the drugs used for IPT treatment \cite{deloron2010sulfadoxine, Goslingetal2010}. Thus understanding the interacting relationship between IPT use as a control strategy and the emergence and rate of spread of drug resistance is important. Previous modeling studies have shown that IPTi/IPTc is likely to accelerate drug resistance spread \cite{Omearaetal2006, teboh2014IPTa, teboh2015intermittent}.  Teboh-Ewungkem et al. in \cite{teboh2015intermittent} found that while treatment of symptomatic infections is the main driver for drug resistance, IPT can increase drug-resistant malaria, particularly when a long half-life drug such as SP is used. The IPT treatment schedule can also affect the intensity of acceleration, with a critical threshold above which drug resistant invasion is certain. 

The models used to examine the role of IPT in drug resistance did not consider the direct benefits of IPT in deaths (and/or cases) averted \cite{teboh2015intermittent, teboh2014IPTa, Omearaetal2006}. In order to better understand the trade off between deaths averted and increasing drug resistance, we adapted the Teboh-Ewungkem et al. 2015 \cite{teboh2015intermittent} model to include age structure, death due to disease, and high or low transmission regions with year-round transmission. This allowed us to quantify the relative impact of IPT and inform strategies for using IPT that will maximize number of deaths averted while minimizing resistance. In particular, we considered the following quantities of interest: number of deaths averted by IPT, ratio of sensitive to resistant strains in the population across time, total number of malaria deaths, basic reproduction number, and invasion reproduction number.
Our goals were to (1) determine the critical level of IPT treatment that would minimize the spread of drug resistance and maximize the positive impact in lives saved; (2) determine the role of IPT in saving lives and potentially facilitating drug resistance for low and high transmission regions; and (3) understand the relative roles of symptomatic treatment and IPT in the establishment of drug resistant strains of malaria while also considering partial resistance. In order to explicitly consider the sustainability of particular approaches, we modeled our time-varying quantities of interest for 1, 5, and 10 years.
Our model differs from that of O'Meara et al. \cite{Omearaetal2006} and Teboh-Ewungkem et al. \cite{teboh2014IPTa} in that the transmission dynamics of the vector population are explicitly modeled as well as age structure for the human hosts. The model explicitly accounted for humans with different levels of immunity as well as incorporated the dynamics of the resistant malaria strain.

The paper is divided as follows: Section \ref{S:Model1} describes the model, giving the associated variables and parameters, while Section \ref{S:analysis} gives a detailed analysis of the disease-free, non-trivial boundary, and endemic equilibria of the model. In Section \ref{S:results}, we present the model results and associated figures, with a parameter sensitivity analyses carried out in Section \ref{S:sensitivity}. Section \ref{S:discussion} then gives a discussion and conclusion. We found that although IPT treatment can increase the levels and timing of resistant strain invasion, treatment of symptomatic individuals plays a much larger role in promoting resistance under our assumptions and parameter values. We also found that the resistant strain is highly sensitive to the half-life of the drug being administered. Successful establishment of the resistant strain is more likely when the drug being used for IPT and treatment has a long half-life. Finally, in the the scenario where the symptomatic treatment drug has a short half-life and low or little resistance to the treatment drug is present in the circulating malaria strains, then using SP as an IPT drug in high transmission regions will result in many lives saved without significantly increasing resistance levels. It should be noted, however, that if strains with high resistance to the symptomatic treatment drug and the IPT drug emerge, then IPT could drive higher resistance proportions and result in an increase in number of deaths. Therefore, close monitoring of resistant strains is suggested by our model when IPT is in use.



\section{Description of the Model}\label{S:Model1}

\begin{figure}[!ht]
 \begin{center}
 \tikzstyle{block} = [rectangle, draw, fill=gray!40,
    text width=2em, text centered, rounded corners, minimum height=3em, minimum width=3em]
  \begin{tikzpicture}[node distance=2.0cm, auto, >=stealth]
   \coordinate(S) at (0,0);
   	\node[block] (S)             {\Large$S$};
        \node [] (Inode) [right of=S, node distance=5cm]{};
	\node[block](Is)[above of = Inode, node distance=1cm] 	{\Large$I_s$};
        \node[block](Ia)[above of=Is, node distance=2cm] 	{\Large$I_a$};
        \node[block](Ja)[below of = Inode, node distance=1cm] 	{\Large$J_a$};
        \node[block](Js)[below of=Ja, node distance=2cm] 	{\Large$J_s$};
        \node[block](T)[left of = S, above of = S, node distance=1.75cm] 	{\Large$T$};
        \node[block](Ta)[right of = Ja, node distance=5.25cm] 	{\Large$T_a$};
        \node[block](Ts)[below of = Ta, node distance=2cm] 	{\Large$T_s$};
        \node[block](R)[right of = Inode, node distance=7.25cm] 	{\Large$R$};

	\coordinate (Imiddle) at ($(Ia)!0.5!(Is)$);
	\coordinate (Jmiddle) at ($(Ja)!0.5!(Js)$);
        \node [] (splitI) [left of=Imiddle, node distance=2cm]{};
        \node [] (splitJ) [left of=Jmiddle, node distance=2cm]{};
	
	 \draw[<-](S)to[out=160,in=-90]node[pos=.5,left]{$r$}(T.-60);
        \draw[<-](T.0)to[out=0,in=90]node[pos=.6,left]{$c$}(S.130);
        \draw[dashed](S.45)--node[pos=.5,left,sloped,anchor=center,above]{$\beta_hM_s/N_h$}(splitI);
        \draw[dashed](S.-45)--node[pos=.5,left,sloped, anchor=center,below]{$\beta_h\kappa_hM_r/N_h$}(splitJ);
        \draw[dashed,->](splitI)--node[pos=.5,left,sloped,anchor=center,above]{$1-\lambda$}(Ia);
	\draw[dashed,->](splitI)--node[pos=.5,left,sloped,anchor=center,below]{$\lambda$}(Is);
        \draw[dashed,->](splitJ)--node[pos=.5,left,sloped,anchor=center,above]{$1-\lambda$}(Ja);
	\draw[dashed,->](splitJ)--node[pos=.5,left,sloped,anchor=center,below]{$\lambda$}(Js);
	\draw[dashed,<-](Ja.10)--node[]{$(1-\lambda)\kappa_h\beta_hM_r/N_h$}(Ta.170);
	\draw[dashed,<-](Js.-10)--node[]{$\lambda\kappa_h\beta_hM_r/N_h$}(Ts.190);
	\draw[dashed,<-](Ja.-10)--node[pos=.09,sloped]{$(1-\lambda)\kappa_h\beta_hM_r/N_h$}(Ts.140);
	\draw[dashed,<-](Js.20)--node[sloped]{$\lambda\kappa_h\beta_hM_r/N_h$}(Ta.200);
	\draw[->](Ja.70) |- node[pos=0.7, above]{$\xi\sigma_a$}(R.180);
	\draw[->](Ia.35) -| node[pos=0.2, above]{$\xi\sigma_a$}(R.120);
	\draw[->](Ia.-25) -| node[pos=0.2, above]{$c$}(Ta.110);
	\draw[dashed,->](T.-120) |- ++(0,-5.75)-- ++(1,0) -| node[pos=0.2, below]{$\lambda\kappa_h\beta_hM_r/N_h$}(Js.-120);
	\draw[dashed,->](T.-90) |- ++(0,-3.5)-- ++(1,0) -| node[pos=0.2, below]{$(1-\lambda)\kappa_h\beta_hM_r/N_h$}(Ja.-120);
	\draw[<-](S.60) |- node[pos=0.7, above]{$(1-\xi)\sigma_a$}(Ia.180);
	\draw[<-](S.0) -| node[pos=0.25, above]{$(1-\xi)\sigma_a$}(Ja.90);
	\draw[<-](S.85) |- ++(0,3.75)-- ++(1,0) -| node[pos=0.2, above]{$(1-b)r$}(Ta.60);
	\draw[<-](S.115) |- ++(0,4.75)-- ++(1,0) -| node[pos=0.2, above]{$\omega$}(R.70);
	\draw[->](Ja.-65) -- node[left]{$\nu$}(Js.65);
	\draw[->](Js.-60) -| ++(0,-0.5)-- ++(1,0) -| node[pos=0.2, below]{$pa$}(Ts.-90);
	\draw[->](Ia.-90) -- node[right]{$\nu$}(Is.90);
	\draw[<-](Ts.30)-|++(0.6,0)-- ++(0,3.7)--node[pos=0.75, above]{a}(Is.0);
	\draw[->](Ts.-30) -| node[pos=0.3, below]{$r_s$}(R.-100);
	\draw[->](Ta.0) -| node[pos=0.1, above]{$br$}(R.-120);
  \end{tikzpicture}
 \end{center}
 \caption{Transfer diagram for human infection within the naive-immune population. Dashed lines represent parasite transmission via infected mosquitoes. $I$ infections are with sensitive strains and $J$ with resistant strains of malaria with subscripts $a$ and $s$ representing asymptomatic and symptomatic cases. $T$ and $T_a$ are susceptible and asymptomatic individuals, respectively, that received IPT, while $T_s$ is individuals receiving treatment for a symptomatic case. $S$ is fully susceptible and $R$ is temporarily immune.}\label{fig:model1naive}
\end{figure}
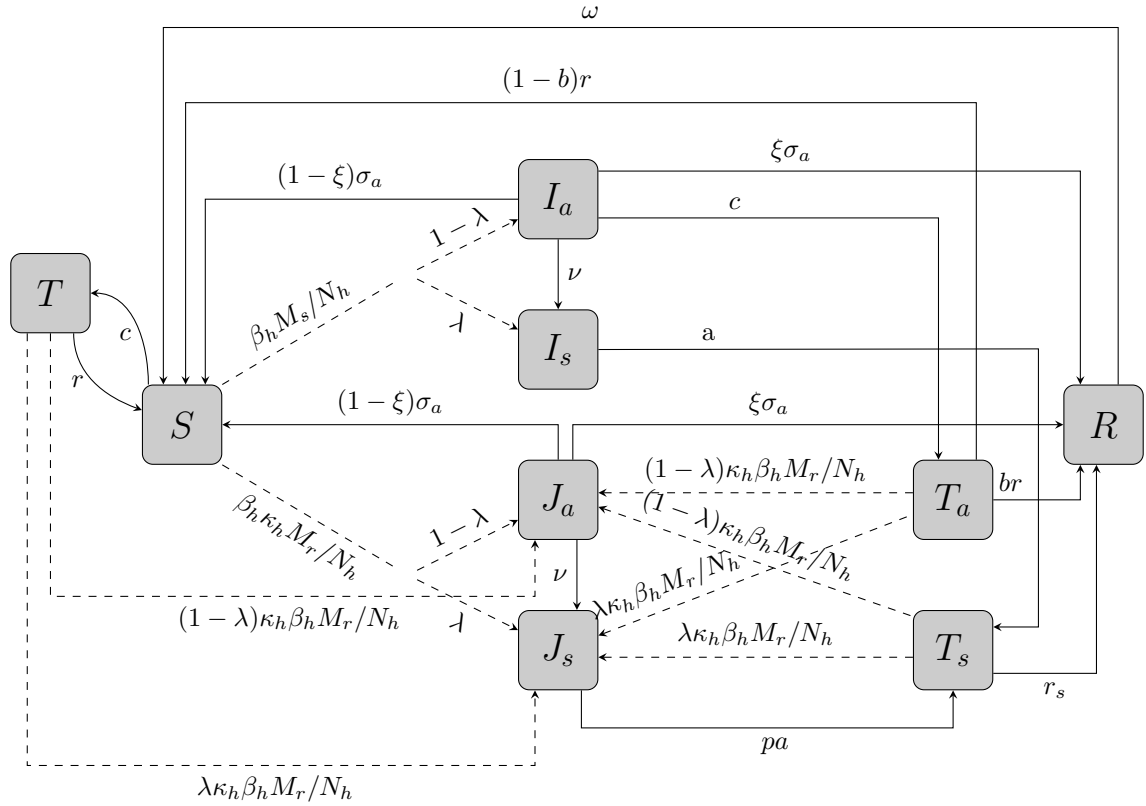

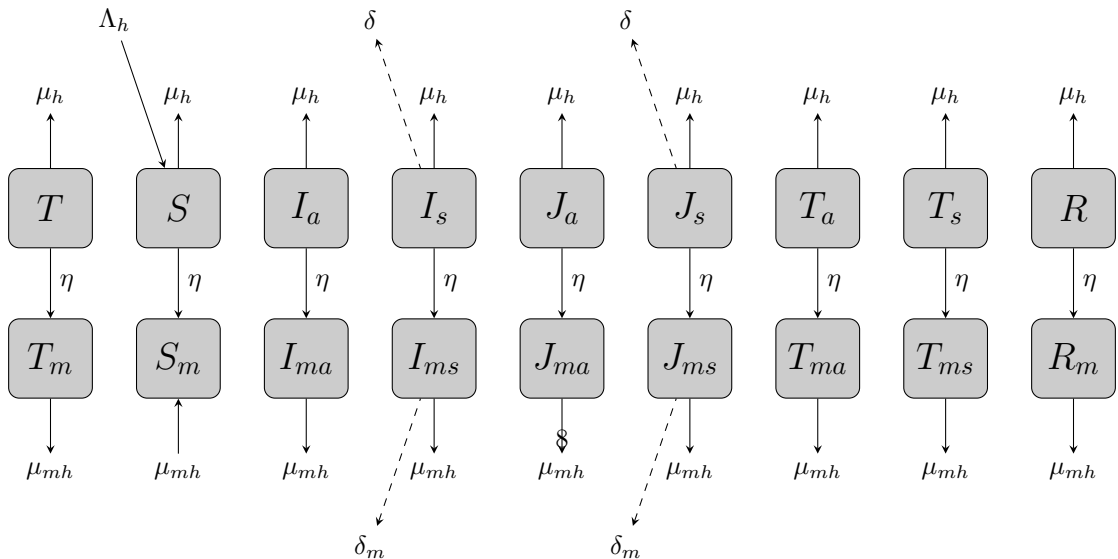
\begin{figure}[H]
 \begin{center}
 \tikzstyle{block} = [rectangle, draw, fill=gray!40,
    text width=2.5em, text centered, rounded corners, minimum height=3em, minimum width=3em]
  \begin{tikzpicture}[node distance=2.0cm, auto, >=stealth]
   \coordinate(S) at (0,0);
   	\node[block] (S)             {\Large$S$};
        \node[block](T)[left of=S, node distance=1.7cm] 	{\Large$T$};
 	\node[block](Tm)[below of=T, node distance=2cm] 	{\Large$T_{m}$};
	\node[block](Sm)[below of=S, node distance=2cm] 	{\Large$S_{m}$};
        \node[block](Ia)[right of=S, node distance=1.7cm] 	{\Large$I_{a}$};
 	\node[block](Ima)[below of=Ia, node distance=2cm] 	{\Large$I_{ma}$};
        \node[block](Is)[right of=Ia, node distance=1.7cm] 	{\Large$I_{s}$};
 	\node[block](Ims)[below of=Is, node distance=2cm] 	{\Large$I_{ms}$};
        \node[block](Ja)[right of=Is, node distance=1.7cm] 	{\Large$J_{a}$};
 	\node[block](Jma)[below of=Ja, node distance=2cm] 	{\Large$J_{ma}$};
        \node[block](Js)[right of=Ja, node distance=1.7cm] 	{\Large$J_{s}$};
 	\node[block](Jms)[below of=Js, node distance=2cm] 	{\Large$J_{ms}$};
        \node[block](Ta)[right of=Js, node distance=1.7cm] 	{\Large$T_{a}$};
 	\node[block](Tma)[below of=Ta, node distance=2cm] 	{\Large$T_{ma}$};
        \node[block](Ts)[right of=Ta, node distance=1.7cm] 	{\Large$T_{s}$};
 	\node[block](Tms)[below of=Ts, node distance=2cm] 	{\Large$T_{ms}$};
        \node[block](R)[right of=Ts, node distance=1.7cm] 	{\Large$R$};
 	\node[block](Rm)[below of=R, node distance=2cm] 	{\Large$R_{m}$};

	\node[](Tdeath)[above of = T, node distance=1.5cm]{$\mu_h$};
	\node[](Sdeath)[above of = S, node distance=1.5cm]{$\mu_h$};
	\node[](Iadeath)[above of = Ia, node distance=1.5cm]{$\mu_h$};
	\node[](Isdeath)[above of = Is, node distance=1.5cm]{$\mu_h$};
	\node[](Jadeath)[above of = Ja, node distance=1.5cm]{$\mu_h$};
	\node[](Jsdeath)[above of = Js, node distance=1.5cm]{$\mu_h$};
	\node[](Tadeath)[above of = Ta, node distance=1.5cm]{$\mu_h$};
	\node[](Tsdeath)[above of = Ts, node distance=1.5cm]{$\mu_h$};
	\node[](Rdeath)[above of = R, node distance=1.5cm]{$\mu_h$};
	\node[](Tmdeath)[below of = Tm, node distance=1.5cm]{$\mu_{mh}$};
	\node[](Smdeath)[below of = Sm, node distance=1.5cm]{$\mu_{mh}$};
	\node[](Imadeath)[below of = Ima, node distance=1.5cm]{$\mu_{mh}$};
	\node[](Imsdeath)[below of = Ims, node distance=1.5cm]{$\mu_{mh}$};
	\node[](Jmadeath)[below of = Jma, node distance=1.5cm]{$\mu_{mh}$};
	\node[](Jmsdeath)[below of = Jms, node distance=1.5cm]{$\mu_{mh}$};
	\node[](Tmadeath)[below of = Tma, node distance=1.5cm]{$\mu_{mh}$};
	\node[](Tmsdeath)[below of = Tms, node distance=1.5cm]{$\mu_{mh}$};
	\node[](Rmdeath)[below of = Rm, node distance=1.5cm]{$\mu_{mh}$};	
	
	\coordinate (TSmiddle) at ($(T)!0.5!(S)$);
	\node[](birth)[above of = TSmiddle, node distance=2.5cm]{$\Lambda_h$};
	\coordinate (Imiddle) at ($(Ia)!0.5!(Is)$);
	\node[](deltaI)[above of = Imiddle, node distance=2.5cm]{$\delta$};
	\node[](deltaIm)[below of = Imiddle, node distance=4.5cm]{$\delta_m$};
	\coordinate (Jmiddle) at ($(Ja)!0.5!(Js)$);
	\node[](deltaJ)[above of = Jmiddle, node distance=2.5cm]{$\delta$};
	\node[](deltaJm)[below of = Jmiddle, node distance=4.5cm]{$\delta_m$};
		
	\draw[->](T.-90) -- node[right]{$\eta$}(Tm.90);
	\draw[->](S.-90) -- node[right]{$\eta$}(Sm.90);
	\draw[->](Ia.-90) -- node[right]{$\eta$}(Ima.90);
	\draw[->](Is.-90) -- node[right]{$\eta$}(Ims.90);
	\draw[->](Ja.-90) -- node[right]{$\eta$}(Jma.90);
	\draw[->](Js.-90) -- node[right]{$\eta$}(Jms.90);
	\draw[->](Ta.-90) -- node[right]{$\eta$}(Tma.90);
	\draw[->](Ts.-90) -- node[right]{$\eta$}(Tms.90);
	\draw[->](R.-90) -- node[right]{$\eta$}(Rm.90);

	\draw[<-](Tdeath.-90) -- node[right]{}(T.90);
	\draw[<-](Sdeath.-90) -- node[right]{}(S.90);
	\draw[<-](Iadeath.-90) -- node[right]{}(Ia.90);
	\draw[<-](Isdeath.-90) -- node[left]{}(Is.90);
	\draw[<-](Jadeath.-90) -- node[right]{}(Ja.90);
	\draw[<-](Jsdeath.-90) -- node[left]{}(Js.90);
	\draw[<-](Tadeath.-90) -- node[right]{}(Ta.90);
	\draw[<-](Tsdeath.-90) -- node[right]{}(Ts.90);
	\draw[<-](Rdeath.-90) -- node[right]{}(R.90);
	
	\draw[<-](Tmdeath.90) -- node[right]{}(Tm.-90);
	\draw[->](Smdeath.90) -- node[left]{}(Sm.-90);
	\draw[<-](Imadeath.90) -- node[right]{}(Ima.-90);
	\draw[<-](Imsdeath.90) -- node[left]{}(Ims.-90);
	\draw[<-](Jmadeath.90) -- node[right]{}(Jma.-90);
	\draw[<-](Jmsdeath.90) -- node[left]{}(Jms.-90);
	\draw[<-](Tmadeath.90) -- node[right]{}(Tma.-90);
	\draw[<-](Tmsdeath.90) -- node[right]{}(Tms.-90);
	\draw[<-](Rmdeath.90) -- node[right]{}(Rm.-90);
	
	\draw[->](birth)-- node[]{}(S);
	\draw[dashed, <-](deltaI)-- node[]{}(Is);
	\draw[dashed,<-](deltaJ)-- node[]{}(Js);
	\draw[dashed, <-](deltaIm)-- node[]{}(Ims);
	\draw[dashed,<-](deltaJm)-- node[]{}(Jms);
		
  \end{tikzpicture}
 \end{center}
 \caption{Transfer diagram between the naive-immune juvenile human population and the mature human population. Dashed lines represent disease-induced mortality.  An average time of $1/\eta$ is spent in the naive-immune class.   }\label{fig:model1population}
\end{figure}

Our mathematical model of the transmission dynamics of the malaria parasite takes into account the following three interacting components of the parasite's life cycle: (1) the parasite that causes the disease, (2) the human hosts who can be infected by the parasite, and (3) the vector hosts (mosquitoes) that transmits the disease from one human to another. The model incorporates the use of IPT and symptomatic treatment employed in the human population. Humans infected with the transmissible forms of the parasites could carry parasites that are either sensitive or refractory (resistant) to drugs used to treat the malaria infection, or drugs used as chemoprophylaxis via IPT.  The model developed expands the model in Teboh-Ewungkem et al. 2015 \cite{teboh2015intermittent} to include explicit age structure and  disease-induced mortalities in the human populations.  We consider scenarios with non-seasonal high transmission as well as low transmissions. Model flow diagrams are shown in Figures \ref{fig:model1naive}, \ref{fig:model1population} and \ref{fig:model1mature}, while the definitions of the variables and parameters are given in Tables \ref{T:vars}, \ref{T:parmsmerge}, and \ref{T:parmsmergeb}.

The model utilized is a nonlinear deterministic age-structured variable-population model described by a system of ordinary differential equations with IPT usage incorporated. In the model, the human population is split into two main groups: (1) juveniles with naive or no clinical immunity, and (2) mature humans who have a higher level of clinical immunity to malaria, due to frequent exposure to the parasites \cite{klein2008clinically, teboh2014IPTa}. By “clinical immunity”, we mean the gradual acquisition of parasite-exposed-primed immune response enabling an individual to be symptom-free even though they might have the transmissible forms of the parasites in their blood stream \cite{cohen1961gamma}.  Thus mature humans, those considered to have higher immune levels, usually do not feel sick from the malaria parasite infection \cite{klein2008clinically, teboh2014IPTa}, which can be associated with less severe malaria symptoms. Thus the rates of anti-malarial drug use among mature individuals will be considered to be lower \cite{klein2008clinically, teboh2014IPTa}.

Thus juveniles, the infants and children, are those receiving IPTi or IPTc, respectively, while mature individuals do not receive any form of IPT. Typically, the population of juveniles will consist of the $0-5$ years old age group. However, this age group can be extended or made shorter depending on the transmission intensity of the region (low or high) and/or whether the region has stable or unstable transmission with transmission either occurring all year round (holoendemicity) or intermittently with periods of intense transmission (hyperendemicity) \cite{hay2008measuring}.  For example, for a region with high malaria transmission intensity, we consider mature humans to be those who have been repeatedly re-exposed to the malaria parasite and thus have developed a more superior immunity \cite{klein2008clinically, teboh2014IPTa}. We consider that age group to be the $>5$ years old group.
We note, however that, even within the same endemic country, there might be regions of high transmission intensity or low transmission intensity depending on whether the region is a highland or lowland region or a rural or urban region. Foe example, the Kenyan highland has low transmission and the Kenyan lowland has high transmission. In addition, the urban city of Nairobi in Kenya is considered to be a low transmission region while Lake Victoria, a rural area, is considered be a high transmission region \cite{teboh2014IPTa}.

In our model, both the juvenile and mature human populations are subdivided into mutually exclusive compartments categorized by their malaria strain-type disease infection or treatment status. In our presentation below, we will refer to IPTi and IPTc as just IPT. Then compartments for the juveniles at any time $t$ are: susceptible juveniles (denoted by $S$), symptomatic juveniles infected with the sensitive strain ($I_{s}$) or the resistant strain ($J_{s}$), asymptomatic juveniles infected with the sensitive strain ($I_{a}$) or the resistant strain ($J_{a}$), susceptible juveniles who've received IPT ($T$), asymptomatic infected juveniles who received IPT ($T_{a}$), treated symptomatic infected juveniles ($T_{s}$) and the temporarily immune juveniles ($R$), see Figure \ref{fig:model1naive}. As juveniles age, they join a corresponding mature human population class (see Figure \ref{fig:model1population}). Denoting the corresponding mature human classes by the subscript $m$, the compartments for the mature human population at time $t$ are: susceptible individuals ($S_m$), symptomatic infected with the sensitive strain ($I_{ms}$) or the resistant strain ($J_{ms}$), asymptomatic individuals infected with the sensitive strain ($I_{ma}$) or the resistant strain ($J_{ma}$), uninfected juveniles who received IPT and aged, aging into the mature class ($T_m$), infected asymptomatic juveniles who received IPT and aged, aging into the mature class ($T_{ma}$), treated symptomatic infected humans ($T_{ms}$) and temporarily immune humans ($R_m$), see Figure \ref{fig:model1mature}. Additionally, at any time $t$, there are a number $S_v$ (susceptible mosquitoes) and $M$ (infectious mosquitoes) that define the mosquito classes. The $M$ mosquitoes are further sub-divided into subclasses $M_r$ and $M_s$ which determines the type of parasite they are infected with, sensitive or resistant. Thus the total mosquito population at time $t$, denoted by $N_v$ is $N_v=S_v + M_r + M_s$. A detailed description of all the variable classes are given in Table \ref{T:vars}.

When a susceptible human comes in contact with an infectious mosquito, the human may become infected at a certain rate, $\beta_h$, following a standard incidence infection term. Some of those infected humans may show symptoms while others may not. Hence the split of the infected human class (considered here to be those with the parasite in their blood stream with the potential to infect a mosquito) into two subgroups: the asymptomatic subgroup (identified by the subscript $a$ and considered to be those who do not show clinical symptoms), and the symptomatic subgroup (identified by the subscript $s$). We consider that a proportion of the susceptible individuals ($\lambda$ for the juveniles and $\lambda^{\prime}$ for the mature individuals) may show symptoms upon infection, while the remaining proportion ($1-\lambda$ for the for the juveniles and $1-\lambda^{\prime}$ for the mature individuals) are assumed to be asymptomatic. We note that the population of mature asymptomatic individuals is typically much larger than that for the juveniles in high transmission areas because of higher levels of clinical immunity to malaria for these mature individuals due to their frequent exposure to the parasites \cite{klein2008clinically, teboh2014IPTa} which enables them to be symptom-free even when they have parasites in their blood stream \cite{cohen1961gamma, klein2008clinically, teboh2014IPTa}. Thus we will expect $\lambda > \lambda^{\prime}$ in a high transmission region, but to be of similar size in a low transmission region.

Additionally, contact between an infected mosquito and a susceptible human may lead to the human being infected with the sensitive parasite strain, identified by the variable $I$, if their bite came from an $M_s$-type mosquito, or a resistant parasite strain, identified by the variable $J$, if their bite came from an $M_r$-type mosquito. It is possible for the strains to differ in fitness, noted by $\kappa_h$, the fitness difference for the resistant strain. The factor $\kappa_h$ multiplies the transmission terms for individuals (whether mosquito or human) infected with the resistant strain. We assume $0\leq \kappa_h \leq 1$. In summary, an infectious human, naive or mature -immune, may be symptomatic and infected with the sensitive parasite strain (classes $I_s$ and $I_{ms}$), or the resistant parasite strain (classes $J_s$ and $J_{ms}$), or asymptomatic and infected with the sensitive parasite strain (classes $I_a$ and $I_{ma}$), or the resistant parasite strain (classes $J_a$ and $J_{ma}$).  We note that we do not consider co-infection in our model. Thus any individual co-infected with the sensitive or resistant parasite strain is considered a resistant infectious human.

In our model, we assume that only the symptomatic humans (juveniles or mature) will seek treatment. In particular, we assume that symptomatic naive-immune individuals clear their symptomatic parasite infections only via treatment else they will die from the infection (thus all symptomatic children who do not die of the disease receive treatment). This assumption is related to the less developed immune system for these individuals.  On the other hand, in addition to treatment, symptomatic mature-immune individuals can also clear their parasite naturally, because of their developed immune response. Symptomatic individuals who do not clear their infections via treatment or naturally (for the case of mature-immune humans) can die due to the disease. This death rate differs between naive-immune, $\delta$, and mature-immune, $\delta_m$. Typically, the disease-induced death rate for the naive-immune individuals is much higher than for the mature individuals \cite{desai2014age}, up to $10$ folds higher. Thus, we will assume that  $\delta > \delta_m$.

The baseline drugs considered for treatment of symptomatic malaria infections are WHO recommended combination therapy drugs such as Artemether-lumefantrine (also called Coartem, to be referred henceforth as the AL drug) or other approved Artemisinin-based combination therapy drugs (ACT drugs) \cite{WHO2015, WHOTreatmentGuidelines2015}. However, we will investigate the impact of a long half-life drug such as sulphadoxine-pyrimethamine (SP) as a treatment drug for symptoms. If a symptomatic individual infected with the sensitive parasite strain receives treatment, they move to the treatment class $T_s$ for the naive-immune individual or $T_{ms}$ for the mature-immune individual. This occurs at rate $a$, where $1/a$ is the average time from the beginning of the treatment to the clearance of the sensitive parasite. If the human (naive or mature -immune) is infected with the resistant parasite strain, we assume that the drug is ineffective against the resistant parasite. Thus such infectious humans, type $J_s$ and $J_{ms}$ individuals, move to their corresponding treatment classes, class $T_s$, respectively $T_{ms}$, at rate $pa$, where $p$ is measures the efficacy of the drug against a resistant infection. We note that $p$ can account for full resistance (in which case $p=0$) or partial resistance (in which case $p>0$). In addition, mature-immune symptomatic humans can also clear their infection naturally at rate $\sigma_{ms}$, with a proportion $\xi_m$ developing temporal immunity to join the temporal immune class $R$, and the remainder $1-\xi_{ms}$ instead joining the susceptible mature human class.

Asymptomatic infectious individuals (naive or mature - immune) do not seek treatment because they do not show symptoms even though considered to be clinically sick and infectious. However, these naive-immune and mature-immune individuals can clear their parasitic infections naturally at rate $\sigma_{a}$ and $\sigma_{ma}$, respectively, with a proportion $\xi$ and $\xi_m$, respectively, developing temporal immunity to join the temporal immune classes $R$ and $R_m$. The remainder, $1-\xi$ and $1-\xi_m$, instead join the susceptible naive immune ($S$) and mature human ($S_m$) classes. We also assumed that asymptomatic infectious humans (naive-immune and mature-immune) can develop symptoms at rates $\nu$ and $\nu^\prime$, respectively.

As a preventative measure, both susceptible and asymptomatic naive-immune individuals receive intermittent preventive treatment (IPT), as was the case in \cite{Omearaetal2006, teboh2014IPTa, teboh2015intermittent}. IPT is administered at a constant per-capita rate $c$ where $1/c$ is the average time between IPT treatments. We will use the WHO recommended drug for IPT treatment, sulphadoxine-pyrimethamine (SP), a long-half life drug \cite{teboh2014IPTa, teboh2015intermittent, WHO2015, WHOTreatmentGuidelines2015} as the baseline IPT treatment drug. Naive-immune juveniles who receive IPT will move to the IPT treated class $T$, for the case where the IPT was administered to a susceptible juvenile, and to the IPT treated class $T_a$, for the case where the IPT was administered to an asymptomatic infectious juvenile.

All individuals, mature or naive-immune, who've received treatment and are in the treated classes are assumed to have drugs at therapeutic levels in their system that can clear sensitive parasites. This is regardless of whether the treatment was due to a symptomatic infection (classes $T_s$ and $T_{ms}$ individuals), or due to IPT (for the case of naive-immune individuals) classes $T$ and $T_a$. As the drug concentration in these treated individuals declines,  the treated individuals may either join the temporarily immune class or the susceptible class. In particular, as the drug concentration in treated individuals who were treated as a result of a symptomatic infection declines at rate $r_s$, these treated individuals are assumed to join the temporary immune class ($R$ or $R_m$) with class $T_s$ moving to class $R$ and class $T_{ms}$ moving to class $R_{m}$. The rate $r_s$ is dependent on the half-life of the drug used for treatment, with $1/r_s$ the time in days the treatment drug reaches levels that do not have therapeutic effects on a sensitive parasite infection. We've assumed here that an immune response is triggered as a result of malaria symptoms, hence the development of temporary immunity. For individuals who receive IPT, the rate of decline of the drug in their system is $r$. If the IPT was administered to a susceptible naive-immune, generating a type $T$ naive-immune juvenile, the individual will move to the susceptible class, $S$, as their drug concentration declines at that rate $r$. However, if the IPT was administered to an asymptomatic infectious naive-immune juvenile generating a type $T_a$ naive-immune juvenile, a proportion $b$ of these treated juveniles will move to the temporary immune class $R$, while the remaining proportion $1-b$ join the susceptible class, $S$, both at rate $r$. The separation is justified in that an asymptomatic infection is as a result of some naive level of temporal immunity bolstered by the IPT drug. Here $1/r$ is the time in days the IPT drug is at levels that do not have therapeutic effects on a sensitive parasite. Temporarily immune individuals (in classes $R$ and $R_m$) lose their temporary immune status to join the susceptible class at a rate $\omega$ for naive-immune and $\omega^{\prime}$ for mature-immune individuals.

We further assume in our model that after age $5$, which could be shorter depending on whether the region is a stable high transmission region, a naive-immune juvenile matures to join an equivalent corresponding mature class. This maturation happens at a constant per-capita rate of $\eta$ where $1/\eta$ is the age considered for the naive-immune individual to have developed a reasonable immune response due to repeated re-exposure to the malaria parasite. For naive-immune treated individuals who received IPT, we assume if they mature while receiving IPT, they move into a temporary IPT treatment compartment in the mature group represented by classes $T_m$ and $T_{ma}$. When the drug concentration of the individuals in these classes decline at rates $r$, where $r$ is as earlier defined, they either join the susceptible mature human class $S_m$, or the temporary immune mature human class $R_m$. If the individuals are coming from class $T_m$ then they will move to class $S_m$. On the other hand, If the individuals are coming from class $T_{ma}$ then a proportion $b_m$ will move to class $R_m$ while the remaining proportion $1-b_m$ will move to class $S_m$. None of the mature humans receive IPT, and thus there is no movement of mature-immune individuals into class $T_m$ or $T_{ma}$.

Additionally, we assume that all recruitment via births occur at a constant rate $\Lambda_h$ into the susceptible naive-immune class and that natural death can occur from all compartments at a constant per-capita death rate of $\mu_{h}$ for the naive-immune compartments, and a constant per-capita death rate of $\mu_{mh}$ for the mature immune individuals. Figure \ref{fig:model1population} shows the the movement due to maturation from every naive-immune compartment into the parallel compartment in the mature-immune classes, indicating where there is disease-induced deaths, natural death and recruitment. The equations governing the human disease dynamics are given in equations \eqref{juvenilehuman_odesfirst}-\eqref{juvenilehuman_odeslast} and \eqref{Maturehuman_odesfirst}-\eqref{Maturehuman_odeslast}, where equations \eqref{juvenilehuman_odesfirst}-\eqref{juvenilehuman_odeslast} model the dynamics of the naive-immune human population, and equations \eqref{Maturehuman_odesfirst}-\eqref{Maturehuman_odeslast} model that of the mature-immune human population. The total human population as well as the sub total naive and mature -immune human populations are modeled by equations \eqref{Totalchildren_odes}-\eqref{Totalhuman_odes}.

When a susceptible mosquito feeds, successfully taking blood from an infectious human, the mosquito may acquire the malaria parasite from the human at rate $\beta_{v}$, moving to either the $M_s$ or $ M_r$ class. If the blood meal was from an infectious human infected with the sensitive parasite strain, then the mosquito, upon infection, will become a type $M_s$ mosquito, infected with the sensitive parasite strain. If on the other hand, the blood meal was from an infectious human infected with the resistant parasite strain, then the mosquito, upon infection, will become a type $M_r$ mosquito, infected with the resistant parasite strain. Here, we also assume that the transmission success to mosquitoes by humans infected with the resistant parasite is less than that from humans infected with the sensitive parasite. Thus, the transmission rate of resistant parasites to susceptible mosquitoes is $\kappa_{v}\beta_{v}$, where  $0 < \kappa_{v} < 1$ is the transmission reduction factor. We further assume that a mosquito cannot be co-infected, that is, if a mosquito is infected with a particular strain of malaria, the mosquito will not acquire nor successfully transmit a second distinct strain of malaria. Thus there is no movement between the $M_s$ and $M_r$ compartments; once a mosquito is infected, it remains so until it dies; and natural death occurs from each mosquito compartment at rate $\mu_{v}$. The equations governing the mosquito dynamics are given in equations \eqref{susmosquito_odes}-\eqref{resinfmosquito_odes}, with the total mosquito population modeled by equation \eqref{Totalmosquito_ode}.



\begin{figure}[H]
 \begin{center}
 \tikzstyle{block} = [rectangle, draw, fill=gray!40,
    text width=2.5em, text centered, rounded corners, minimum height=3em, minimum width=3em]
  \begin{tikzpicture}[node distance=2.0cm, auto, >=stealth]
   \coordinate(S) at (0,0);
   	\node[block] (S)             {\Large$S_m$};
        \node [] (Inode) [right of=S, node distance=5cm]{};
	\node[block](Is)[above of = Inode, node distance=1cm] 	{\Large$I_{ms}$};
        \node[block](Ia)[above of=Is, node distance=2cm] 	{\Large$I_{ma}$};
        \node[block](Ja)[below of = Inode, node distance=1cm] 	{\Large$J_{ma}$};
        \node[block](Js)[below of=Ja, node distance=2cm] 	{\Large$J_{ms}$};
        \node[block](T)[left of = S, above of = S, node distance=1.5cm] 	{\Large$T_m$};
        \node[block](Ta)[right of = Ja, node distance=5.5cm] 	{\Large$T_{ma}$};
        \node[block](Ts)[below of = Ta, node distance=2cm] 	{\Large$T_{ms}$};
        \node[block](R)[right of = Inode, node distance=7.5cm] 	{\Large$R_m$};

	\coordinate (Imiddle) at ($(Ia)!0.5!(Is)$);
	\coordinate (Jmiddle) at ($(Ja)!0.5!(Js)$);
        \node [] (splitI) [left of=Imiddle, node distance=2cm]{};
        \node [] (splitJ) [left of=Jmiddle, node distance=2cm]{};
	
	 \draw[<-](S)to[out=160,in=-90]node[pos=.5,left]{$r$}(T.-60);
        \draw[dashed](S.45)--node[pos=.5,left,sloped,anchor=center,above]{$\beta_hM_s/N_h$}(splitI);
        \draw[dashed](S.-45)--node[pos=.5,left,sloped, anchor=center,below]{$\beta_h\kappa_hM_r/N_h$}(splitJ);
        \draw[dashed,->](splitI)--node[pos=.5,left,sloped,anchor=center,above]{$1-\lambda'$}(Ia);
	\draw[dashed,->](splitI)--node[pos=.5,left,sloped,anchor=center,below]{$\lambda'$}(Is);
        \draw[dashed,->](splitJ)--node[pos=.5,left,sloped,anchor=center,above]{$1-\lambda'$}(Ja);
	\draw[dashed,->](splitJ)--node[pos=.5,left,sloped,anchor=center,below]{$\lambda'$}(Js);
	\draw[dashed,<-](Ja.10)--node[]{$(1-\lambda')\kappa_h\beta_hM_r/N_h$}(Ta.170);
	\draw[dashed,<-](Js.-10)--node[]{$\lambda'\kappa_h\beta_hM_r/N_h$}(Ts.190);
	\draw[dashed,<-](Ja.-10)--node[pos=.09,sloped]{$(1-\lambda')\kappa_h\beta_hM_r/N_h$}(Ts.140);
	\draw[dashed,<-](Js.20)--node[sloped]{$\lambda'\kappa_h\beta_hM_r/N_h$}(Ta.200);
	\draw[->](Ja.70) |- node[pos=0.7, above]{$\xi_m\sigma_{ma}$}(R.180);
	\draw[->](Ia.0) -| node[pos=0.2, above]{$\xi_m\sigma_{ma}$}(R.100);
	\draw[->](Is.35) -| node[pos=0.15, above]{$\xi_m\sigma_{ms}$}(R.120);
	\draw[dashed,->](T.-120) |- ++(0,-5.75)-- ++(1,0) -| node[pos=0.2, below]{$\lambda'\kappa_h\beta_hM_r/N_h$}(Js.-120);
	\draw[<-](S.-120) |- node[pos=0.8, below]{$(1-\xi_m)\sigma_{ms}$}(Js.220);
	\draw[dashed,->](T.-90) |- ++(0,-3.25)-- ++(1,0) -| node[pos=0.2, below]{$(1-\lambda')\kappa_h\beta_hM_r/N_h$}(Ja.-120);
	\draw[<-](S.60) |- node[pos=0.7, above]{$(1-\xi_m)\sigma_{ma}$}(Ia.180);
	\draw[<-](S.-10) -| node[pos=0.25, below]{$(1-\xi_m)\sigma_{ma}$}(Ja.90);
	\draw[<-](S.10) -| node[pos=0.25, above]{$(1-\xi_m)\sigma_{ms}$}(Is.-90);	
	\draw[<-](S.85) |- ++(0,3.75)-- ++(1,0) -| node[pos=0.2, above]{$(1-b_{m})r$}(Ta.60);
	\draw[<-](S.115) |- ++(0,4.75)-- ++(1,0) -| node[pos=0.2, above]{$\omega'$}(R.70);
	\draw[->](Ja.-65) -- node[left]{$\nu'$}(Js.65);
	\draw[->](Js.-60) -| ++(0,-0.5)-- ++(1,0) -| node[pos=0.2, below]{$pa$}(Ts.-90);
	\draw[->](Js.-80) -| ++(0,-1.5)-- ++(1,0) -| node[pos=0.2, below]{$\xi_m\sigma_{ms}$}(R.-60);
	\draw[->](Ia.-90) -- node[right]{$\nu'$}(Is.90);
	\draw[<-](Ts.30)-|++(0.7,0)-- ++(0,3.4)--node[pos=0.75, above]{a}(Is.-30);
	\draw[->](Ts.-30) -| node[pos=0.3, below]{$r_s$}(R.-100);
	\draw[->](Ta.0) -| node[pos=0.15, above]{$b_{m}r$}(R.-120);
  \end{tikzpicture}
 \end{center}
 \caption{Transfer diagram for human infection within the mature population. Dashed lines represent parasite transmission via infected mosquitoes. $T_{ma}$ and $T_m$ are holding compartments for individuals that mature while in an IPT treatment class (so drug is still circulating in their system). The subscript $m$ indicates immune-mature individuals, but all other notation is the same as in Figure \ref{fig:model1naive}.}\label{fig:model1mature}
\end{figure}

\begin{table}[htbp]
  \caption{State variables and their descriptions}
  \label{T:vars}
  \rowcolors{2}{gray!25}{white}
  \begin{tabular}{l p{13cm}} \hline
Variable & Description of Variable    \\ \hline
$S_{v}$ & Number of susceptible mosquitoes. \\
$M_{s}$ & Number of mosquitoes infected with the sensitive strain.  \\
$M_{r}$ & Number of mosquitoes infected with the resistant strain .  \\
$S$ & Number of susceptible juveniles.\\ 
$I_{s}$ & Number of symptomatic infected juveniles infected with the sensitive parasite strain.\\ 
$I_{a}$ & Number of asymptomatic infected juveniles infected with the sensitive parasite strain.\\ 
$J_{s}$ & Number of symptomatic infected juveniles infected with the resistant parasite strain.\\ 
$J_{a}$ & Number of asymptomatic infected juveniles infected with the resistant parasite strain.\\ 
$T_{s}$ & Number of symptomatic infected juveniles who are treated due to their symptoms.\\ 
$T$ & Number of susceptible juveniles who've received IPT treatment.\\ 
$T_{a}$ & Number of asymptomatic infected juveniles who've received IPT treatment.\\ 
$R$ & Number of infected juveniles who clear their parasite either naturally or via treatment and develop temporary immunity.\\
$S_m$ & Number of susceptible mature humans. \\ 
$I_{ms}$ & Number of symptomatic infectious mature humans infected with the
sensitive strain.   \\ 
$I_{ma}$ & Number of asymptomatic infected mature humans infected with
sensitive strain.   \\ 
$J_{ms}$ & Number of symptomatic infected mature humans infected with the
resistant strain.   \\ 
$J_{ma}$ & Number of asymptomatic infected mature humans infected with the
resistant strain.   \\ 
$T_{m}$ & Number of susceptible juveniles who had received IPT and aged prior to their  drug levels declining to the levels that rendered them susceptible.   \\ 
$T_{ma}$ & Number of asymptomatic juveniles who had received IPT and aged prior to their drug levels declining to the levels that rendered them temporary immune or susceptible.  \\ 
$T_{ms}$ & Number of mature humans who receive treatment due to their symptomatic infection. \\ 
$R_m$ & Number of  infected mature humans who clear their parasite either naturally or via treatment and develop  temporary immunity.\\
$N_{c}$ & Total Number of juvenile population.  \\ 
$N_{m}$ & Total Number of mature human population. \\ 
$N_h$ & Total human population.
 \\ \hline
  \end{tabular}
\end{table}

\clearpage

\begin{subequations}
\begin{align}
\frac{dS}{dt} &= \Lambda_h - \mu_h S- \beta_h(M_s +\kappa_h M_r)S/N_h - c S + (1-\xi)\sigma_a (I_a + J_a) \\
& + (1-b) r T_a+ r T + \omega R-\eta S, \label{juvenilehuman_odesfirst} \\
\frac{dI_s}{dt} &= \lambda \beta_h M_s S/N_h + \nu I_a - (a + \mu_h + \eta + \delta) I_s,\\
\frac{dI_a}{dt} &= (1-\lambda)\beta_h M_s S/N_h - (c + \nu + \sigma_a + \mu_h + \eta) I_a, \\
\frac{dJ_s}{dt} &= \lambda \kappa_h \beta_h M_r[S + T_s + T + T_a]/N_h + \nu J_a - (pa + \mu_h + \eta + \delta )J_s, \\
\frac{dJ_a}{dt} &= (1-\lambda) \kappa_h \beta_h M_r[S + T_s + T + T_a]/N_h - (\sigma_a + \nu + \mu_h + \eta) J_a , \\
\frac{dT_s}{dt} &= a I_s + pa J_s- r_s T_s - \kappa_h \beta_h M_r T_s/N_h - (\mu_h + \eta) T_s,\\
\frac{dT}{dt} &= c S - r T - \kappa_h \beta_h T M_r/N_h - (\mu_h + \eta) T, \\
\frac{dT_a}{dt} &= c I_a - r T_a - \kappa_h \beta_h T_a M_r/N_h - \mu_h T_a - \eta T_a,\\
\frac{dR}{dt} &= r_s T_s + b r T_a + \xi \sigma_a (I_a + J_a) - (\omega + \mu_h + \eta) R,\label{juvenilehuman_odeslast},
\end{align}
\end{subequations}

\begin{subequations}
\begin{align}
\frac{dS_m}{dt} &=  \eta S - \mu_{mh} S_m - \beta_h(M_s + \kappa_h M_r)S_m/N_h  + (1-\xi_{m})\sigma_{ma} (I_{ma} + J_{ma}) \\
& +(1-\xi_{m})\sigma_{ms} (I_{ms} + J_{ms})+ \omega^{\prime} R_m +  r T_m +(1-b_{m}) r T_{ma},\label{Maturehuman_odesfirst} \\
\frac{dI_{ms}}{dt} &= \eta I_s + \lambda^{\prime} \beta_h M_s S_m/N_h + \nu^{\prime} I_{ma} - (a + \mu_{mh}  + \delta_m + \sigma_{ms}) I_{ms}, \\
\frac{dI_{ma}}{dt} &= \eta I_a  + (1-\lambda^{\prime})\beta_h M_s S_m/N_h - (\sigma_{ma} + \nu^{\prime} +\mu_{mh}) I_{ma} ,\\
\frac{dJ_{ms}}{dt} &= \eta J_{s} + \lambda^{\prime} \kappa_h \beta_h M_r[S_m+ T_{ms} + T_m + T_{ma}]/N_h + \nu' J_{ma} - (pa +\sigma_{ms}  + \mu_{mh} + \delta_m)J_{ms}, \\
\frac{dJ_{ma}}{dt} &= \eta J_a+ (1-\lambda^{\prime})\kappa_h \beta_h M_r[S_m + T_{ms} +  T_m + T_{ma}]/N_h - (\sigma_{ma} + \nu^{\prime} + \mu_{mh})J_{ma}, \\
\frac{dT_{ms}}{dt} &= \eta T_s + a I_{ms} + pa J_{ms}- \kappa_h \beta_h M_r T_{ms}/N_h - (\mu_{mh} + r_s) T_{ms},\\
\frac{dT_m}{dt} &= \eta T - \kappa_{h} \beta_h T_m M_r/N_h - (\mu_{mh} + r) T_m, \\
\frac{dT_{ma}}{dt} &= \eta T_a - \kappa_h \beta_h T_{ma} M_r/N_h - (\mu_{mh} + r) T_{ma},\\
\frac{dR_m}{dt} &= \eta R + r_s T_{ms} + b_{m} r T_{ma} + \xi_{m} \sigma_{ma} (I_{ma}+ J_{ma} ) + \xi_{m} \sigma_{ms}(I_{ms}+J_{ms})- \omega^{\prime} R_m - \mu_{mh} R_m, \label{Maturehuman_odeslast}
\end{align}
\end{subequations}

In our model, the total juvenile population is $N_{c}=S + I_{s}+I_{a} + J_{s}+ J_{a} + T + T_{s} + T_{a} + R$, the total mature population is $N_{m}=S_m + I_{ms}+I_{ma} + J_{ms}+ J_{ma} + T_m + T_{ms} + T_{ma} + R_m$, so that the total human population $N_h = N_c + N_m$. The equations that model the $N_{c}$, $N_{m}$ and $N_h$ populations are:

\begin{subequations}
\begin{align}
\frac{dN_c}{dt} &= \Lambda_h - \eta N_c - \mu_h N_c - \delta (I_s +J_s),\label{Totalchildren_odes} \\
\frac{dN_m}{dt} &= \eta N_c - \mu_{mh} N_m - \delta_m (I_{ms} +J_{ms}), \label{Totalmature_odes}\\
\frac{dN_h}{dt} &= \Lambda_h - \mu_h N_c - \mu_{mh} N_m - \delta (I_s +J_s) - \delta_m (I_{ms} +J_{ms}).\label{Totalhuman_odes}
\end{align}
\end{subequations}
The total human population has a disease-free carrying capacity of $N_h^*=\Lambda_h /(\psi\mu_h+(1-\psi)\mu_{mh})$, where $\psi N_h = N_c$ is the total naive-immune human population, and $(1-\psi) N_h^* = N_{m}^*$ is the total mature-immune human population and $N_c^* = \Lambda_h/(\nu + \mu_h)$ and $N_m^* = \eta N_c/\mu_{mh}$ are the equilibria of the juvenile and mature populations without death from malaria. Thus, $\psi$ gives the ratios of naive - immune to the total human populations so that $N_c^* + N_m^* = N_h^*$, the total human population.

The equations that govern the mosquito dynamics  are
\begin{subequations}
\begin{align}
\frac{dS_v}{dt} &= \Lambda_v - \beta_v \left[I_a+I_s+ I_{ma} + I_{ms} + \kappa_{v} (J_a + J_s+ J_{ma} + J_{ms}) \right]S_v/N_h -\mu_v S_v, \label{susmosquito_odes}\\
\frac{dM_s}{dt} &= \beta_v (I_a + I_s + I_{ma} + I_{ms})S_v/N_h - \mu_v M_s, \label{susinfmosquito_odes}\\
\frac{dM_r}{dt} &=\kappa_{v}  \beta_v (J_a + J_s+ J_{ma} + J_{ms}) S_v/N_h - \mu_v M_r,\label{resinfmosquito_odes}
\end{align}
\end{subequations}
where the total mosquito population is $N_{v}=S_v + M_{s} + M_{r}$ and is modeled by the equation

\begin{subequations}
\begin{align}
\frac{dN_v}{dt} &= \Lambda_v - \mu_v N_v.\label{Totalmosquito_ode}
\end{align}
\end{subequations}
The total mosquito population is also non-constant, with a disease free carrying capacity of $\Lambda_v /\mu_v$.

We remark that in our model discussions, we consider the number of bites per day a human gets to be limited by mosquito density, not human density, i.e. every mosquito gets to bite as often as they desire. Therefore the total number of bites per day is defined as (the number of bites desired per day by a mosquito) * (total number of mosquitoes)  = $\alpha N_v$, where $N_v$ is the total number of mosquitoes and $\alpha$ is the number of bites per mosquito per day. Thus the number of bites per person per day is $\alpha N_v / N_h$, where $N_h$ is the total number of humans. See \cite{CHITNIS2006} for a discussion of alternative biting rates as the vector-to-host ratio becomes either very low or very high.  Thus, $\beta_h$ is then the product of the mosquito biting rate ($\alpha$, or number of bites on humans per mosquito per day) times the probability that transmission occurs if the bite is from an infectious mosquito (represented by $\beta_{hv}$). On the other hand, $\beta_v$ is the product of the mosquito biting rate ($\alpha$, or number of bites on humans per mosquito per day) times the probability that transmission occurs if the bite is on an infectious individual (represented by $\beta_{vh}$).

Table \ref{T:vars} summarizes the state variable descriptions.  All parameters, as defined in Tables \ref{T:parmsmerge} and \ref{T:parmsmergeb}, are non-negative. Details about their interpretation and values will be presented in Section \ref{S:parameters}. With non-negative initial conditions, it can be verified that the solutions to the model equations remain non-negative.

\begin{table}[htbp]
\caption{Descriptions and dimensions for parameters related to the natural transmission cycle}
\label{T:parmsmerge}
\rowcolors{2}{gray!25}{white}
 \begin{tabular}{lp{10cm}p{3cm}} \hline
Parameter & Description & Dimension \\ \hline
$\Lambda_{h}$ & Total human birth rate & humans  $T^{-1}$ \\
$\Lambda_{v}$ & Total  mosquito birth rate & mosquitoes  $T^{-1}$ \\
$\mu_{mh}$ & Per Capita death rate of mature humans & $T^{-1}$ \\
$\mu_{h}$ & Per Capita death rate of juveniles & $T^{-1}$ \\
$\delta_{m}$ & Malaria disease-induced mortality rate for mature humans & $T^{-1}$ \\
$\delta$ & Malaria disease-induced mortality rate for juveniles & $T^{-1}$ \\
$\mu_{v}$ & Natural mosquito death rate & $T^{-1}$ \\
$\eta$ & Rate of aging, i.e. rate at which juveniles become mature humans and no longer receive IPT & $T^{-1}$ \\
$\beta_{h}$ & Transmission rate of sensitive parasites from mosquitoes to humans ($\alpha \beta_{hv}$)  & mosquito$^{-1}$ $T^{-1}$ \\
$\beta_{v}$ & Transmission rate of sensitive parasites from humans to mosquitoes ($\alpha \beta_{vh}$)  &  mosquito$^{-1}$$T^{-1}$ \\
$\kappa_{h}$ & Reduction factor of human transmission rate by the resistant parasite strain & 1 \\
$\kappa_{v}$ & Reduction factor of mosquito transmission rate by the resistant parasite strain & 1 \\
$\lambda$ & Fraction of juveniles who become symptomatic upon infection & 1 \\
$\lambda^\prime$ & Fraction of matures who become symptomatic upon infection & 1 \\
$\omega$ & Rate of loss of temporary immunity in juveniles& $T^{-1}$ \\
$\omega^\prime$ & Rate of loss of temporary immunity in mature adults& $T^{-1}$ \\
$\lambda$ & Fraction of juveniles who become symptomatic upon infection & 1 \\
$\lambda^\prime$ & Fraction of matures who become symptomatic upon infection & 1 \\
$\nu$ & Rate at which juveniles progress from asymptomatic to symptomatic infections & $T^{-1}$ \\
$\nu^\prime$ & Rate at which mature humans progress from asymptomatic to symptomatic infections & $T^{-1}$ \\
$\sigma_{s}$ & Rate of naturally clearing a symptomatic infection for juveniles & $T^{-1}$ \\
$\sigma_{a}$ & Rate of naturally clearing an asymptomatic infection for juveniles & $T^{-1}$ \\
$\sigma_{ms}$ & Rate of naturally clearing a symptomatic infection for matures & $T^{-1}$ \\
$\sigma_{ma}$ & Rate of naturally clearing an asymptomatic infection for matures & $T^{-1}$ \\
$\xi$ & Proportion of asymptomatic juveniles who naturally clear their infection and develop temporary immunity & 1 \\
$\xi_m$ & Proportion of mature humans who naturally clear their infection and develop temporary immunity & 1 \\
$\delta$ &  Disease-induced death rate for juveniles & $T^{-1}$ \\
$\delta_m$ & Disease-induced death rate for mature humans & $T^{-1}$ \\
\hline
\end{tabular}%
\end{table}

\begin{table}[htbp]
\caption{Descriptions and dimensions for parameters related to symptomatic treatment and IPT}
\label{T:parmsmergeb}
\rowcolors{2}{gray!25}{white}
 \begin{tabular}{lp{10cm}p{3cm}} \hline
Parameter & Description & Dimension \\ \hline
$1/a$ & Days to clear a sensitive infection after treatment& $T$ \\
$c$ & Per Capita rate of IPT treatment administration & $T^{-1}$ \\
$1/r$ & Time chemoprophylaxis lasts in IPT treated humans & $T$ \\
$1/r_{s}$ & Time chemoprophylaxis lasts in symptomatic treated humans & $T$ \\
$b$ & Fraction of asymptomatic infected treated juveniles who become temporarily immune protected & 1 \\
$b_m$ & Fraction of asymptomatic infected treated mature humans who become temporarily immune protected & 1 \\
$p$ &  Efficacy of drugs used to clear resistant infections & $1$ \\
\hline
\end{tabular}%
\end{table}

\subsection{Parameters}\label{S:parameters}

In this section, we present a discussion of the parameters used in the model.
The chemoprophylaxis IPT drug considered here is sulphadoxine-pyrimethamine (SP), a drug with a long half-life (148-256 hours).
Drugs with long half-lives are slowly eliminated from the body compared to those with shorter half-lives, and are therefore expected to impose greater selective pressure for drug resistance than those with shorter half-lives \cite{Babikeretal2009resistance}. The expectation is that drugs that persist longer in the body at sub-therapeutic levels will provide more opportunities for non-resistant (susceptible) parasites to acquire resistant traits, and for partially resistant parasites to become fully resistant. Resistance to SP, a long half-life drug, is common, while resistance to Artemether-lumefantrine (AL ) or other approved Artemisinin-based combination therapy drugs (ACT), short-half life drugs, has not been reported in most African countries. Typically, SP, the long half-life drug, is used for IPT, while the short-half-life drugs ACT or AL are used to treat infections. ACT and AL currently work against both sensitive and resistant parasites in most regions, so are associated with values of $p$ closer to $1$. If resistance develops to these, then the value of $p$ for treatment drugs will be closer to $0$. On the other hand, SP clears sensitive parasites but not resistant parasites. Note that since short half-life drugs such as ACT and AL at therapeutic levels are effective against resistant parasites, if we consider their use as IPT drugs, then we may need to add an additional link from $J_a$ to $T_a$ but with much lower effectiveness. The lower effectiveness against clearance of resistant parasites comes as a result of the way IPT is administered, with long intervals between administration, allowing for opportunities for the drug to dip below therapeutic levels between treatments \cite{greenwood2010anti}. In this manuscript, we assume that asymptomatic infection by resistant parasites are untreated, since these individuals do not seek treatment and for those receiving IPT we assume a negligible impact on clearance. On the other hand, symptomatic infections by resistant parasites have higher clearance success rates if treated with an AL or ACT drug, or are partially treatable if treated with SP (this as a result of symptoms making it possible for the drug to bolster the symptom-initiated body's natural and adaptive immune response aiding in parasite clearance \footnote{This assumption comes from evidence in \cite{cravo2001antimalarial} suggesting higher success in parasite clearance under some background immunity. We note, however, that the original study was performed on the rodent malaria \emph{Plasmodium chabaudi}, where it was shown that drug-resistant parasites could be cleared in partially immune individuals.}.

The parameters $1/r_s$ and $1/r$, give the respective average time chemoprophylaxis lasts in symptomatic treated and  IPT-treated humans, respectively. These values were estimated based on reported half-lives values for antimalarial drugs. Omeara et al. in \cite{Omearaetal2006} reported that for a drug with a long half-life such as sulfadoxine-pyrimethamine (SP), it takes about 52 days for the drug concentration to drop below a threshold value that it cannot clear malaria parasites, while for a drug with a short half-life, such as AL or ACT, this time period is about 6 days \cite{makanga2009clinical}. These are the same values used in \cite{teboh2015intermittent}. For the number of IPT treatments given per person per day, $c$, we use the value $0.016$ day$^{-1}$ as in \cite{Omearaetal2006, teboh2015intermittent}. This value corresponds to IPT being given once every $60$ days, or $1/c$. Since a goal of this manuscript is to see the impact of IPT in averting disease induced deaths, we will vary $c$ to see the role frequency of IPT administration might have on the number of child disease-induced mortality and the rate of resistance spread.

The average number of days needed to clear an infection with appropriate treatment is $1/a$.  Assuming that treatment is pursued immediately, and a WHO recommended dosage is taken within the required dosage time frame, then $1/a$ is about $5$  days \cite{Omearaetal2006}. If the strain of malaria is not fully responsive to the drug, then $pa$ measures the rate of clearing an infection via treatment where $0\leq p <1$. If $p=0$ then the malaria strain is fully resistant to the drug and treatment is ineffective. For values of $0<p\leq 1$, the resistant strain of malaria partially responds to treatment.
We also assumed that asymptomatic and symptomatic infections of mature individuals are naturally cleared at the same rate ($\sigma_{ma}=\sigma_{ms}$), as in \cite{Omearaetal2006}, where a value of $1/33$ days$^{-1}$  was used. Mean rates of immune-response related clearance of $1/180$ days$^{-1}$ have also been cited in \cite{filipe2007determination}. Here, we chose a baseline value based on a weighted average.

Our focus was on regions were malaria is holoendemic. These regions could either have low or high malaria transmission intensity. Low transmission intensity areas are typically upland sites (see, e.g. \cite{craig1999climate}) and tend to exhibit conditions that make them less conducive for the malaria transmitting mosquito to reproduce \cite{teboh2014IPTa}. Such conditions may include lower rainfall accumulations and cooler temperatures due to the altitude. Thus, with fewer mosquitoes, there are less contacts, on average, between humans and infectious female mosquitoes \cite{Omearaetal2006, teboh2014IPTa}. On the other hand, high transmission regions, typically at lower elevations \cite{craig1999climate}, have conditions that enhance the breeding and hence growth and reproduction of the female mosquito population. Thus, in high transmission regions, there is a higher on average contact between humans and infectious female mosquitoes \cite{Omearaetal2006, teboh2014IPTa}. We used estimates from Chitnis et al. \cite{Chitnis2008} to inform our high and low mosquito biting, vector-to-host ratio, and transmission parameters.

Malaria mortality rates have been monitored since 2001 by Kenya Medical Research Institute (KEMRI) and the U.S. Centers for Disease Control and Prevention (CDC) as part of the KEMRI/CDC Health and Demographic Surveillance System (HDSS) in rural
western Kenya \cite{desai2014age}. The results published in \cite{desai2014age} show a declining malaria disease-induced mortality rate in all age groups, with the 2010 data reported as  $3.7$ deaths per 1000 person-years for children under five, with a $95\%$ confidence interval reported to be between $3.0$ and $4.5$ per 1000 person-years. For individuals five and above, the malaria mortalities were estimated for 2010 as $0.4$ deaths per 1000 person-years, with  a $95\%$ confidence interval reported to be between $0.3$ and $0.6$ per 1000 person-years. The study appears to have accumulated the deaths yearly during the time frame used. 
The area of the study, around where KEMRI/CDC HDSS is located, is in the lake region of western Kenya, a malaria endemic region considered to be of high transmission intensity \cite{desai2014age}. For disease mortality in regions of low transmission intensity, we assume a 3.5 times reduction in the under five malaria-related mortality. This assumption comes from the findings in \cite{Snow2006} that reported an approximately 3.5 times overall malaria-specific mortality in children in areas of higher stable transmission than in areas of low malaria transmission intensity in Sub-Saharan Africa, excluding southern Africa.


To initialize our simulations, we used a human density (in a 500 km$^2$ region of the KEMRI/CDC HDSS area the population density is 135,000 per km$^2$) and estimated mosquito density to be 3 times the human density for high transmission regions and 1 time the human density for low transmission regions \cite{amek2012spatial}. We assumed that both human and mosquito populations are constant in the absence of the disease, which implies equal birth and death rates for each species. Using the data in  Table \ref{T:humans}, we computed the human birth rate to be $\Lambda_h = \frac{(\# \text{births per 1000 people per year})}{\text{1000 people}}\times\frac{1 \text{ year}}{365 \text{ days}}\times N_h^*$ where $N_h^*$ is the total human population. To keep the total population constant (apart from malaria deaths), the juvenile natural death rate was computed to be  $\mu_h = \frac{\Lambda_h}{N_c^*} - \eta$ where $N_c^*$ is the total number of juveniles. Then, the mature death rate is $\mu_{mh} = \frac{\psi\eta}{1 - \psi}$  where $\psi=N_c^*/N_h^*$ is the fraction of the population in the juvenile class.

The natural mosquito death rate, $\mu_v$, is assumed to be the reciprocal of the average lifetime of a mosquito. In the wild, mosquitoes
are thought to live for about two weeks, though other modeling efforts have used values ranging up to 28 days \cite{Teboh-ewungkem2010a, teboh-ewungkem2010b, ngonghala2012periodic}. We set the mosquito emergence rate to be $\Lambda_m = \mu_v Q N_h$, where $Q$ is the number of mosquitoes per human. We assume the mosquito biting rate range to be $\alpha\in(0.2,0.5)$ per day \cite{mandal2011mathematical}.
%
%
%
%
%
%
%

\section{Model Analysis}\label{S:analysis}

In this section, we derived the stability conditions of the disease-free equilibrium. We computed the basic reproduction number for the resistant and sensitive strains and present biological interpretations of the expressions. We also derived the invasion reproduction numbers and present invasion maps for the resistant and sensitive strains of malaria.

\subsection{The disease-free equilibrium (DFE)}
Let $\mathcal{X}=(I_s, I_a, J_s, J_a, I_{ms}, I_{ma}, J_{ms}, J_{ma}, M_s, M_r, S, T_s, T, T_a, R, S_m, T_{ms}, T_m, T_{ma}, R_m, S_v)$ denote an equilibrium of
the system described by equations \eqref{juvenilehuman_odesfirst}-\eqref{juvenilehuman_odeslast}, \eqref{Maturehuman_odesfirst}-\eqref{Maturehuman_odeslast} and \eqref{susmosquito_odes}-\eqref{resinfmosquito_odes}. The system has the DFE $\mathcal{E}_0=(0,0,0,0,0,0,0,0,0,0, S_0, 0, T_0, 0, S_{m0}, 0, T_{m0}, 0, S_{v0} )$, where
\begin{subequations}
\begin{align}
S_0=\frac{\Lambda_h\left( r+\mu_h+\eta \right) }{\left(\mu_h+c+\eta \right) \left(r+\mu_h+\eta \right)-rc}, ~~ T_0=\frac{c}{r+\mu_h+\eta}S_0~~~~~~~~~~~~~~~~~~~~~~ \\ \nonumber
 S_{m0}=\frac{\eta}{\mu_{mh}} \left(1+\frac{rc}{\left(\mu_{mh}+r \right) \left(r+\mu_h+\eta \right) } \right) S_0, ~~~
T_{m0} = \frac{\eta c}{\left( \mu_{mh}+r\right) \left(r+\mu_h+\eta \right)}S_0, ~~~S_{v0} =\frac{\Lambda_v}{\mu_v}.
\end{align}
\end{subequations}

\begin{table}[htbp]
\caption{Data from \cite{CIA2013} on the three African countries, Kenya Ghana and Tanzania, used to determine current natural death rates and to infer death rates for malaria in our model.}
\label{T:humans}
\rowcolors{2}{gray!25}{white}
\begin{tabular}{c c c c}
\hline
{\bf Data Information} &{\bf Kenya} & {\bf Ghana} & {\bf Tanzania} \\ \hline
{\bf Total Population} & 45,925,301 & 26,327,649 & 51,045,882 \\
{\bf $< 5$ years old in millions} & $\approx 3.3$ & $\approx 1.9$ & $\approx 4.1$\\
{\bf Infant mortality: deaths/1,000 live births)} & 39.38 & 37.37 & 42.43\\
{\bf Births/1,000 population} & 26.4  & 31.09 & 36.39\\
{\bf Deaths/1,000 population} & 6.89 & 7.22 & 8\\
{\bf Life expectancy at birth in years} & 63.77 & 66.18 & 61.71\\
{\bf Calculated proportion under 5} & 0.0719 & 0.0722 & 0.0804\\
\hline
\end{tabular}
\end{table}

\begin{table}[htbp]
\caption{Parameter values, ranges, and references that are unchanged across high/low transmission scenarios.}
\label{T:parmvalues}%
\begin{center}
\rowcolors{2}{gray!25}{white}
\begin{tabular}{llll}
\hline
Parameter & Value Range & Baseline Value & Reference \\\hline
$\Lambda_h$ &$(2.24\times 10^3,5.08\times 10^3)$ & $3.55 \times 10^3$ & CIA data \\
$\mu_{h}$ & $(4.583\times10^{-4},6.922\times10^{-4})$ & $5.94\times 10^{-4}$ & CIA data \\
$\mu_{mh}$ &$(4.25\times10^{-5},4.791\times10^{-5})$ & $4.43\times10^{-5}$ & CIA data \\
$\mu_{v}$ & $\left( 1/7,1/21\right) $ day$^{-1}$ & $1/14$ day$^{-1}$ &  \cite{teboh-ewungkem2010b} \\
$\delta_{m}$ & $\left( \frac{0.3}{1000*365},\frac{0.6}{1000*365}\right) $ day$^{-1}$ & $\frac{0.4}{1000*365}$ day$^{-1}$ &  \cite{desai2014age} \\
$\delta$ & $\left( \frac{3.0}{1000*365},\frac{4.5}{1000*365}\right) $ day$^{-1}$ & $\frac{3.7}{1000*365}$ day$^{-1}$ &  \cite{desai2014age} \\
$1/\omega$ & (28) & 28 day & \cite{Omearaetal2006} \\
$1/\omega'$ & (370) & 370 day & \cite{Omearaetal2006} \\
$\nu$ & $\left( 0.001,0.05\right) $ & 0.01 & \cite{Omearaetal2006} \\
$\nu'$ &$\left( 0.001,0.05\right) $ & 0.05 & \cite{Omearaetal2006} \\
$\sigma_{ms}$ & (1/28-1/365) & 1/33 day$^{-1}$ & \cite{filipe2007determination,Omearaetal2006} \\
$\sigma_{ma}$ &  (1/28-1/365) & 0.03 day$^{-1}$ & \cite{filipe2007determination,Omearaetal2006} \\
$1/a$ & (3,10) & 5 days &\cite{Omearaetal2006}  \\
$c$ & (0.005,0.03) & 0.016 day$^{-1}$ & \cite{Omearaetal2006} \\
$1/r,1/r_{s}$ & constant & $1/6$, $1/52$ day$^{-1}$ & \cite{Omearaetal2006} \\
 \hline
\end{tabular}%
\end{center}
\end{table}


\begin{table}[htbp]
\caption{Parameter values, ranges, and references that change across high/low transmission scenarios.}
\label{T:parmvalues2}%
\rowcolors{2}{gray!25}{white}
\begin{center}
\begin{tabular}{lllll}
\hline
Parameter & Value Range & High Baseline Value & Low Baseline Value & Reference \\\hline
$\Lambda_v$ &$(1-10)*N_h/\mu_v$ &$3*N_h/\mu_v$ &$1*N_h/\mu_v$& \cite{Chitnis2008,amek2012spatial} \\
$\beta_{v}$ & (0.03,0.2) & 0.0927 & 0.0313 &   \cite{Chitnis2008}\\
$\beta_{h}$ & (0.18,0.9)& 0.5561 & 0.1251 & \cite{Chitnis2008}\\
$\kappa_{v}$ & (0,1) & 0.6 & 0.6 & assumed  \\
$\kappa_{h}$ & (0,1) & 0.6 & 0.6 & assumed \\
$\sigma_{a}$ & (1/365-1/20) & 1/33 day$^{-1}$ & 1/180 day$^{-1}$ & \cite{filipe2007determination,Omearaetal2006} \\
$\sigma_{s}$ & (0.02-0.05) & 0.03 day$^{-1}$ &1/365$^{-1}$ & \cite{filipe2007determination,Omearaetal2006} \\
$p$ & (0,1) & 0.3 & 0.1 & assumed \\
$\lambda$ & (0.25,0.75)  & 0.5 & 0.7 & \cite{Omearaetal2006} \\
$\lambda'$ &(0.15,0.35) & 0.2 & 0.7 & \cite{Omearaetal2006,teboh2014IPTa,baliraine2009high}\\
$\xi_{m}$ & (0.8,1) & 0.9 &0.5 &  \cite{Omearaetal2006,teboh2014IPTa,baliraine2009high}\\
$\xi$ &(0.1,0.5) & 0.4 & 0.2 & \cite{Omearaetal2006,teboh2014IPTa,baliraine2009high}\\
$b$ & (0.25,0.50) & 0.5 & 0.25 & \cite{Omearaetal2006} \\
$b_m$ & (0.25,0.50)& 0.5 & 0.25 & \cite{Omearaetal2006,teboh2014IPTa,baliraine2009high} \\
$\delta$ & & 1.0137e-05 & 2.8963e-06 & \cite{desai2014age} \\
$1/\eta$ & & 5 yrs & 8 yrs & \cite{baliraine2009high}
\\ \hline
\end{tabular}%
\end{center}
\end{table}

\clearpage

\subsection{Basic reproductive numbers }\label{S:R0}
The basic reproduction numbers for the sensitive parasite strain $\mathcal{R}_s$ and the resistant parasite strain $\mathcal{R}_r$ were computed using the next generation matrix, as well as derived from biological interpretation of the model. Details of both approaches are listed in Appendix \ref{A:R0}.
The reproduction number for the sensitive strain of infection takes the following form:
\begin{equation}
\begin{aligned}
\mathcal{R}^2_s &=\frac{\beta_v\beta_hS_0S_{v0}}{\mu_vN_0^2}\left[\frac{1-\lambda}{A_a}+ \frac{\nu(1-\lambda)}{A_aA_s} + \frac{\eta\nu(1-\lambda)}{A_aA_{ms}A_s} + \frac{\eta(1-\lambda)}{A_a A_{ma}}
+\frac{\eta\nu'(1-\lambda)}{A_aA_{ma}A_{ms}} +\frac{\lambda}{A_s}+\frac{\eta\lambda}{A_sA_{ms}}   \right]
\\&+\frac{\beta_v\beta_hS_{m0}S_{v0}}{\mu_vN_0^2}\left[\frac{1-\lambda'}{A_{ma}}+ \frac{\nu'(1-\lambda')}{A_{ma}A_{ms}} +\frac{\nu'}{A_{ms}}  \right].
\end{aligned}
\label{Eq:Rs}
\end{equation}

The reproduction number for the resistant strain of infection takes the following form:
\begin{equation}
\begin{aligned}
\mathcal{R}^2_r &=\frac{\kappa_v\beta_v\kappa_h\beta_h(S_0+T_0)S_{v0}}{\mu_vN_0^2}\left[\frac{1-\lambda}{B_a}+ \frac{\nu(1-\lambda)}{B_aB_s} + \frac{\eta\nu(1-\lambda)}{B_aB_{ms}B_s} + \frac{\eta(1-\lambda)}{B_aA_{ma}}
+\frac{\eta\nu'(1-\lambda)}{B_aA_{ma}B_{ms}} +\frac{\lambda}{B_s}+\frac{\eta\lambda}{B_sB_{ms}}   \right]
\\&+\frac{\kappa_v\beta_v\kappa_h\beta_h(S_{m0}+T_{m0})S_{v0}}{\mu_vN_0^2}\left[\frac{1-\lambda'}{A_{ma}}+ \frac{\nu'(1-\lambda')}{A_{ma}B_{ms}} +\frac{\nu'}{B_{ms}}  \right].
\end{aligned}
\label{Eq:Rr}
\end{equation}

Where, the following parameters represent the durations of infections:
\begin{equation}
\begin{aligned}
A_s =a+\mu_h+\eta+\delta \qquad &\Rightarrow \qquad \frac{1}{A_s} = \text{duration of sensitive sym. naive infection}\\
A_a =c+\nu+\sigma_a+\mu_h+\eta \qquad &\Rightarrow \qquad \frac{1}{A_a} =  \text{duration of sensitive asym. naive infection}\\
A_{ms} =a+\mu_{mh}+\delta_m+\sigma_{ms} \qquad &\Rightarrow \qquad \frac{1}{A_{ms}} =  \text{duration of sensitive sym. mature infection}\\
A_{ma} =\nu'+\sigma_{ma}+\mu_{mh} \qquad &\Rightarrow \qquad \frac{1}{A_{ma}} =  \text{duration of sensitive asym. mature infection}\\
B_s =pa+\mu_h+\eta+\delta \qquad &\Rightarrow \qquad \frac{1}{B_s} =  \text{duration of resistant sym. naive infection}\\
B_a =\nu+\sigma_a+\mu_h+\eta \qquad &\Rightarrow \qquad \frac{1}{B_a} =  \text{duration of resistant asym. naive infection}\\
B_{ms} =pa+\mu_{mh}+\delta_m+\sigma_{ms} \qquad &\Rightarrow \qquad \frac{1}{B_{ms}} =  \text{duration of resistant sym. mature infection}\\
\end{aligned}
\label{Eq:durations}
\end{equation}
Note that for a mature individual, the duration of a resistant asymptomatic infection is equivalent to the duration of  a resistant symptomatic infection ($1/A_{ma}$).

\begin{figure}[h]
\centering
      \subfigure[]{\includegraphics[scale=0.35]{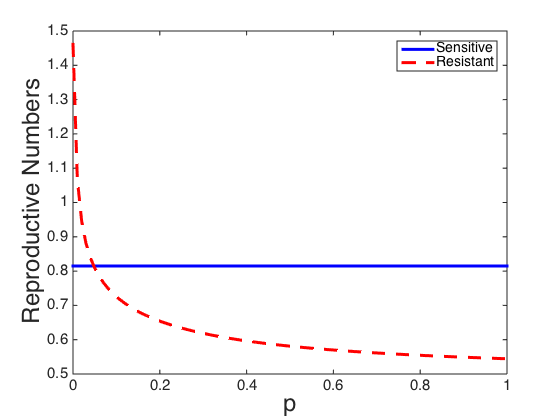}}
      \subfigure[]{\includegraphics[scale=0.35]{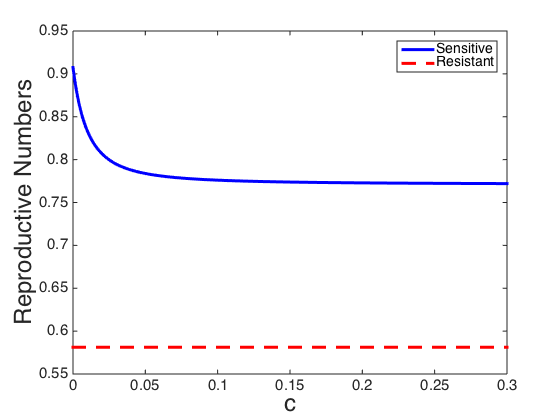}}
 \subfigure[]{\includegraphics[scale=0.35]{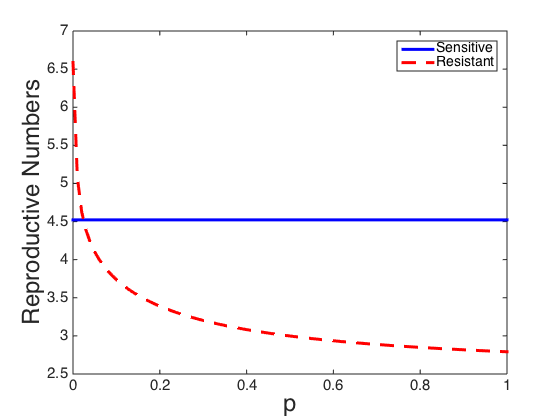}}
 \subfigure[]{\includegraphics[scale=0.35]{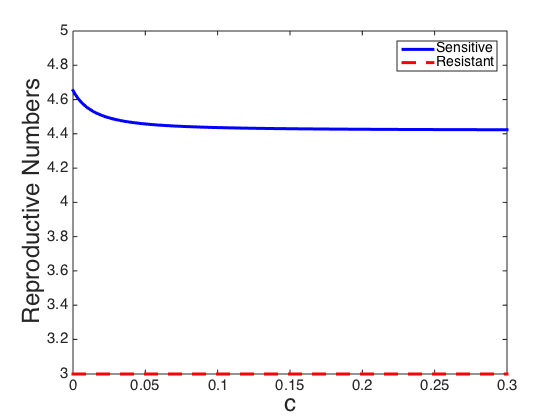}}
  \caption{Reproduction numbers for the low transmission scenario (top graphs (a) and (b)) and high transmission scenario (bottom graphs (c) and (d)) for varying values of $p$ and $c$. All other parameter values are given in Tables \ref{T:parmvalues} and \ref{T:parmvalues2}. Notice the different scales on the y-axes. In the low transmission region, $R_0$ is rarely above 1.}
  \label{F:RoPlots}
\end{figure}

The reproductive numbers depend on the IPT treatment regime and drug efficacy (Figure \ref{F:RoPlots}). The rate of IPT administration to individuals per day ($c$) has a small influence on $\mathcal{R}_s$ (Figure \ref{F:RoPlots} b, d). The drug efficacy ($p$) influences $\mathcal{R}_r$ (Figure \ref{F:RoPlots} a, c). For both low and high transmission scenarios, $\mathcal{R}_r$ decreases for increasing levels of $p$. While increasing $p$ decreases $\mathcal{R}_r$, it is unable to bring $\mathcal{R}_r<1$ in the high transmission scenario (Figure \ref{F:RoPlots} c).

\clearpage

\begin{table}[H]
\caption{Reproduction and invasion numbers for the low and high transmission scenarios using baseline parameter values from Tables  \ref{T:parmvalues} and \ref{T:parmvalues2}. Since the low transmission basic reproduction numbers are less than one (so no sensitive- or resistant-only equilibria exist), we do not compute the invasion reproductive numbers.}
\label{T:RoValues}
\begin{center}
\begin{tabular}{cc|c|c|c|c|c|}
\cline{2-7}
& \multicolumn{2}{ |c| }{Low Transmission } & \multicolumn{4}{ |c| }{High Transmission }  \\
\cline{2-7}
 & \multicolumn{1}{|c|}{$\mathcal{R}_s$}& $\mathcal{R}_r$ & $\mathcal{R}_s$ & $\mathcal{R}_r$ & $\mathcal{R}^r_s$ & $\mathcal{R}^s_r$ \\ \hline
\multicolumn{1}{|c|}{$r_s=1/6$}& 0.8148 & 0.5811 &4.5217  & 2.9984 &1.329 & 4.533 \\ \hline
\multicolumn{1}{|c|}{$r_s=1/52$}& 0.8148 & 0.5811 &4.5217  & 2.9984  & 1.0821 & 6.7323 \\ \hline
\end{tabular}
\end{center}
\end{table}

Table \ref{T:RoValues}, presents the reproductive numbers for the sensitive strain, $\mathcal{R}_s$, and resistant strain, $\mathcal{R}_r$, using baseline parameter values for the low and high transmission scenarios in equations \eqref{Eq:Rs} and  \eqref{Eq:Rr}. In the low transmission scenario both $\mathcal{R}_s$ and $\mathcal{R}_r$ are less than unity and malaria only persists in low transmission regions with regular introductions from outside. In the high transmission scenario both $\mathcal{R}_s$ and $\mathcal{R}_r$ are greater than unity and malaria persists.

\subsection{Invasion Reproduction Numbers}

The basic reproduction number is not sufficient to determine the competitive outcome of the resistant and sensitive strains.  In addition to $R_s$ and $R_r$, we must derive the invasion reproduction numbers $R_r^s$ and $R_s^r$, which are threshold quantities determining whether the resistant strain is able to invade the sensitive-strain boundary equilibrium, and vice versa.  The derivation follows the Next Generation Approach, but with the disease-free equilibrium replaced with either the sensitive-only boundary equilibrium, or the resistant-only boundary equilibrium.

The square of the thresholds determining whether the resistant strain can invade the sensitive-only boundary equilibrium, and whether the sensitive strain can invade the resistant-only boundary equilibrium is given by:
\begin{equation}
\begin{aligned}
(R_r^s)^2 = &\frac{\beta_v k_v S_v^*}{\mu_v N_h^*}\cdot \frac{\beta_h k_h}{N_h^*}\left\{(S_m^*+T_m^*+T_{ma}^*+T_{ms}^*)\left[\frac{(1 - \lambda^\prime)}{A_{ma}} + \frac{\lambda^\prime}{B_{ms}} + \frac{(1 - \lambda^\prime)\nu^\prime}{A_{ma} B_{ms}}\right] \right.\\
& \left. + (S^* + T_a^* + T_s^* + T^*)\left[\frac{1 - \lambda}{B_a} + \frac{\eta(1 - \lambda)}{A_{ma} B_a} + \frac{\lambda}{B_s} + \frac{\eta \lambda}{B_{ms} B_s} +  \frac{(1-\lambda) \nu}{B_a B_s} + \frac{\eta(1 - \lambda) (A_{ma}\nu + B_s \nu^\prime)}{A_{ma} B_a B_{ms} B_s}\right]\right\}\\
(R_s^r)^2= &\frac{\beta_h\beta_v S_v^*}{\mu_v (N_h^*)^2}\left\{S_m^*\left[\frac{\lambda^\prime}{A_{ms}} + (1-\lambda^\prime)\left(\frac{1}{A_{ma}} + \frac{\nu^\prime}{A_{ms}A_{ma}}\right)\right] \right.\\
& \left. + S^* \left[\lambda\left(\frac{1}{A_{ms}} + \frac{\eta}{A_sA_{ms}}\right) + (1-\lambda)\left(\frac{1}{A_a} + \frac{\eta}{A_aA_{ma}} + \frac{\nu}{A_sA_a} + \frac{\eta(A_{ma}\nu + A_s\nu^\prime)}{A_sA_aA_{ms}A_{ma}}\right)\right] \right\},
\label{Eq:Rsr}
\end{aligned}
\end{equation}
where the equilibrium values correspond to the sensitive-only, and resistant-only boundary equilibria, respectively. Table \ref{T:RoValues} presents the invasion reproductive numbers ($\mathcal{R}^r_s$, $\mathcal{R}^s_r$) using baseline parameter values for the low and high transmission scenarios in equation \eqref{Eq:Rsr}.  Here the variables notes with $*$ are at equilibrium for their respective strain-only equilibria.

%


\section{Numerical Results}\label{S:results}

In this section, we present results from numerical simulations for the high and low transmission regions. Our quantities of interest (QOI), or outputs, were number of children who died of malaria, number of adults who died of malaria, and the proportion of deaths that resulted from infection with the resistant strain. For both regions we consider two IPT/treatment regimes: (1) SP/SP where SP, a long half-life drug (and could be replaced with another similar long half-life drug) is used for both IPT and treatment, and (2) SP/ACT where SP (the long half-life drug) is used for IPT and ACT, a short half-life drug (and could also be replaced by another similar short half-life drug such as AL), is used for treatment of symptomatic infection. We denote these scenarios as long/long and long/short.  We also compute PRCC sensitivity indices for our outcomes to the parameters used. For simplification, and in an abundance of caution, we assume that the IPT drug and dose given is completely ineffective against the \textit{resistant} pathogen when given to asymptomatic juveniles. The drug and dosages used for symptomatic treatment of the resistant pathogen, however, may be partially effective depending on the value chosen for $p$.

\subsection{Numerical Results: High Transmission Region}

For the following figures we assume a high transmission region with an initial population of $N=35,000,000$ humans and a constant population of $105,000,000$ mosquitoes.  Initial conditions:  $N_{child}= 7.5\% N$, $S(0) = N_{child}$, $I_a(0) = I_s(0) = J_a(0)=J_s(0)=T(0) = T_a(0) = T_s(0) = R(0)=0$. For the adults, $N_{adult}=92.5\% N$, $S_m(0)=53\% N_{adult}$, $I_{ma}(0) = 10\% N_{adult}$, $I_{ms}(0)=5\% N_{adult}$, $J_{ma}(0) = J_{ms}(0)=1\% N_{adult}$, $R_m(0)=3 0\% N_{adult}$, with all other classes equal to zero. For the mosquitoes, we assume $S_v(0) = 90\% N_{mosquito}$, $M_r(0)=M_s(0)= 5\% N_{mosquito}$.

\begin{figure}[h]
\centering
\subfigure[SP Treatment]{\includegraphics[scale=0.15]{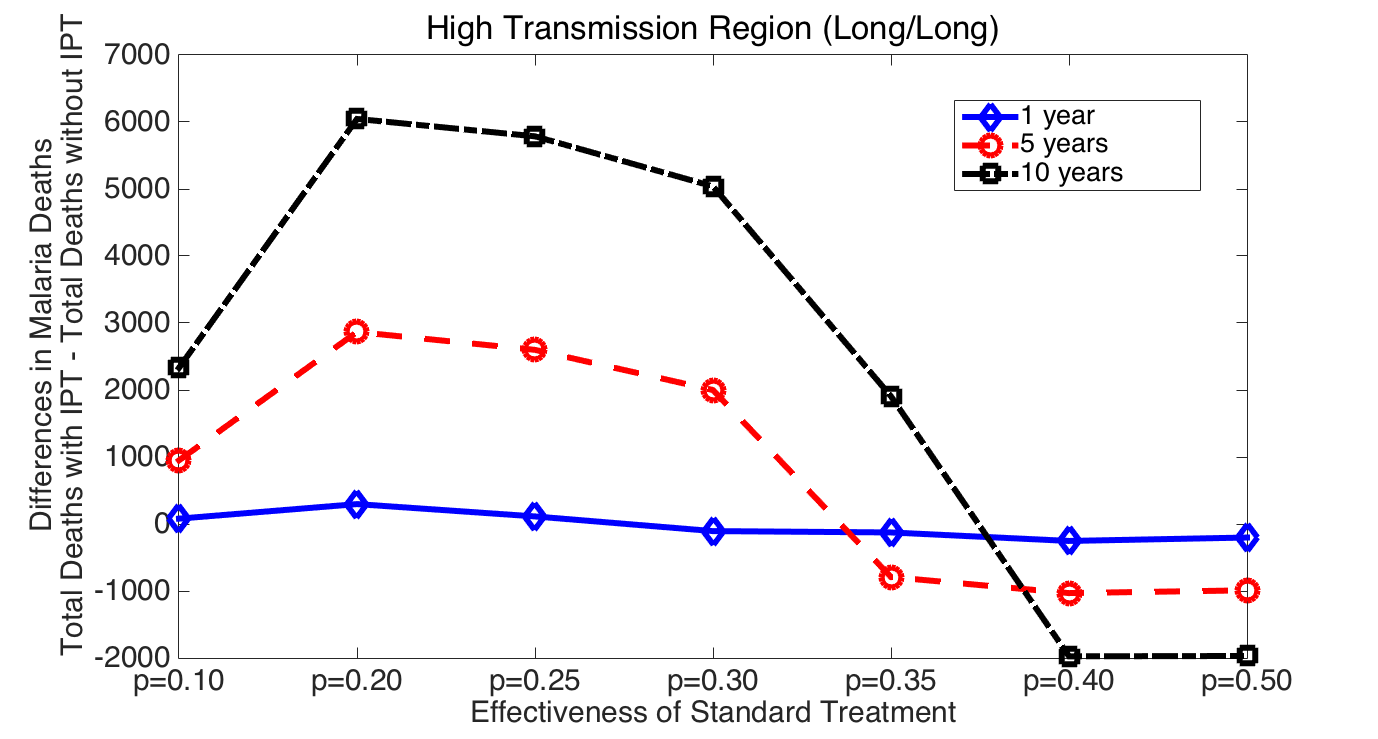}}
\subfigure[AL Treatment]{\includegraphics[scale=0.15]{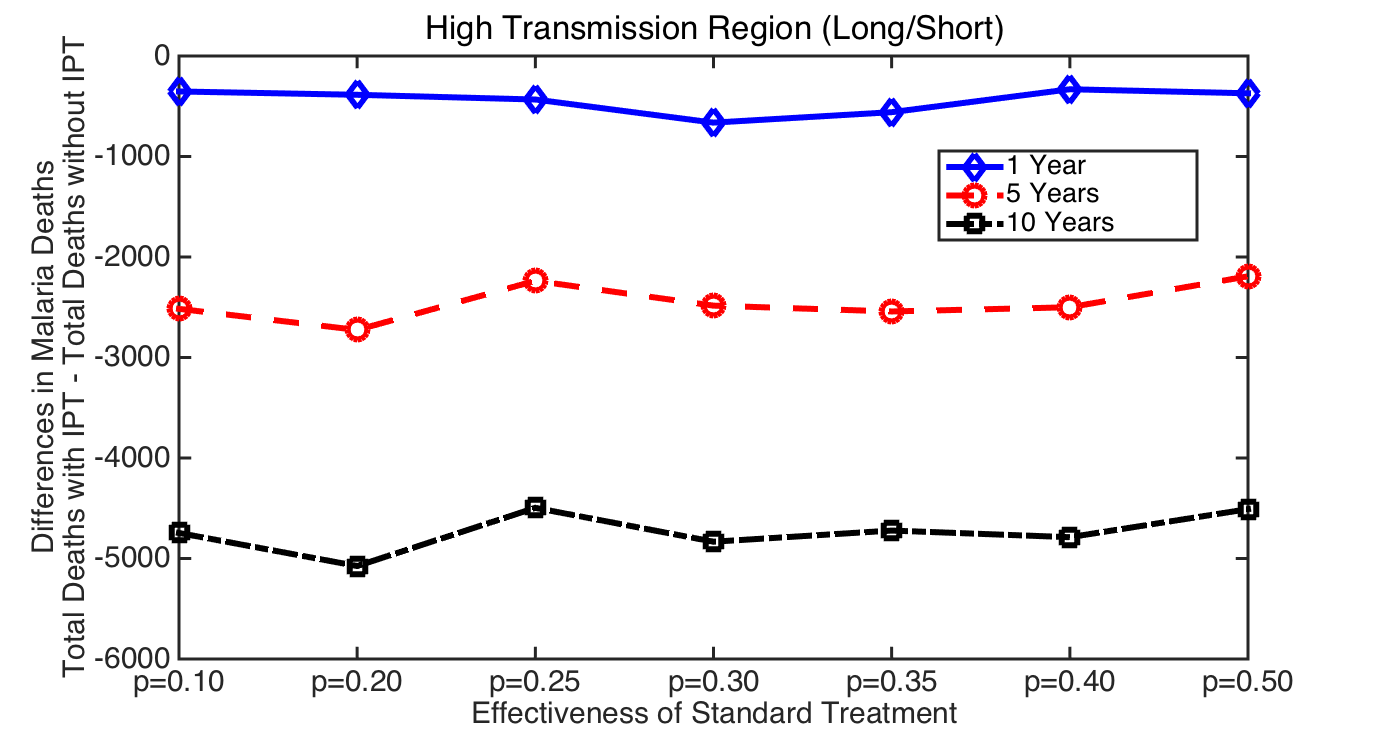}}
\caption{High transmission region: Net increase in deaths due to IPT usage, or \textit{(Total child deaths due to sensitive and resistant strains of malaria with IPT) - (total child deaths without IPT)} for 1 year, 5 years, and 10 years of IPT use for different levels of standard treatment effectiveness against the resistant strain, $p$. (a) is for long half-life SP used as the treatment drug and (b) is for short half-life AL used for symptomatic treatment. Negative numbers indicate lives saved due to IPT while positive numbers indicate more deaths from using IPT.}
\label{F:hightotal}
\end{figure}

Figure \ref{F:hightotal}(a) shows the total changes in number of child deaths due to IPT for various values of $p$ in the high transmission region using drugs with long half-life for both IPT and treatment (long/long). For $p=0.3$, some lives are saved over the course of 1 year, but by 5 or 10 years, IPT increases resistance enough that there is a net increase in number of deaths. In order to see a net number of lives saved over 10 years, $p$ must be greater than or equal to $0.4$. Figure \ref{F:hightotal}(b) shows the total changes in number of child deaths due to IPT for various values of $p$ in the high transmission region using a drug with long half-life for IPT and a drug with short half-life for treatment (long/short). Here we always see a net increase of number of lives save regardless of the value of $p$.

\begin{figure}[htpb]
\centering
\subfigure[Sensitive Infections]{\includegraphics[scale=.15]{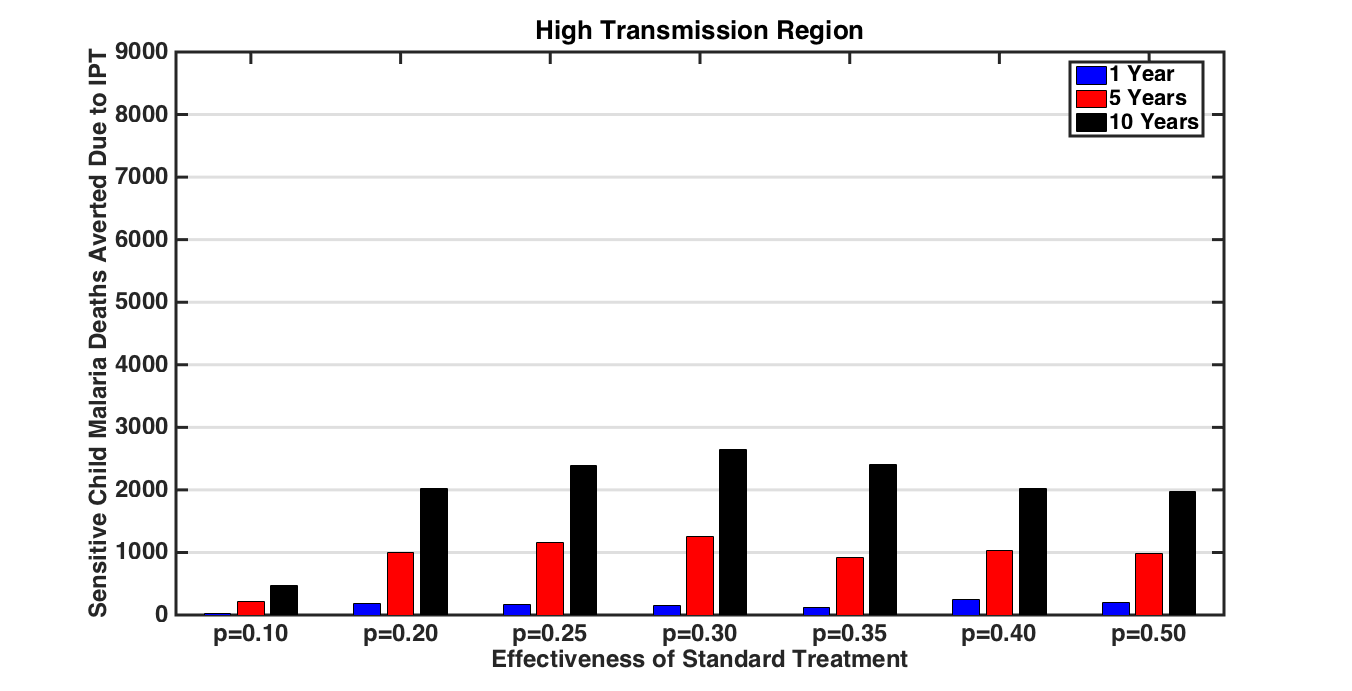}}
\subfigure[Resistant Infections]{\includegraphics[scale=.15]{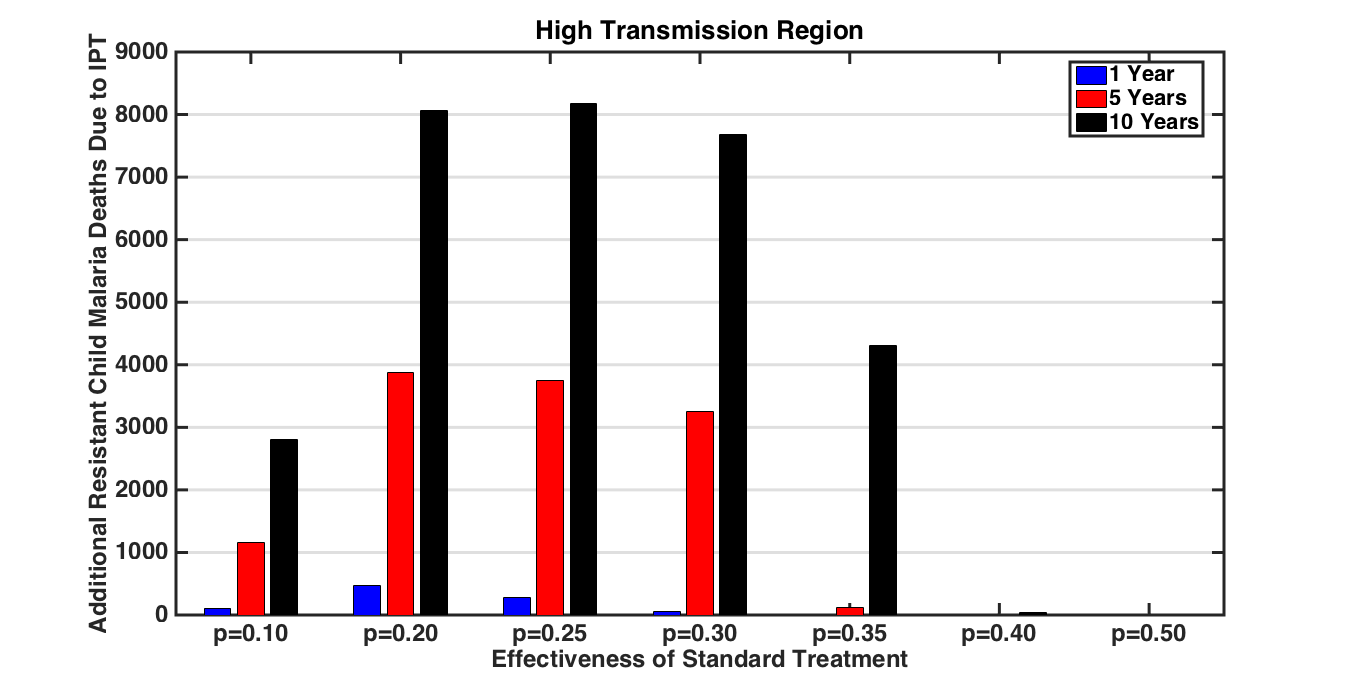}}
\caption{ High transmission region, SP for treatment: Change in child malaria deaths  by use of IPT separated by infection type, treatment level for the resistant infection, and number of years since beginning of IPT treatment. Subfigure (a) is child sensitive infection deaths averted (sensitive deaths in a scenario without IPT minus sensitive deaths in the same scenario with IPT). Subfigure (b) is the additional number of child deaths due to resistant infection with IPT usage. IPT treatment can reduce the number of child deaths due to the sensitive infection, but  increase the number of child deaths due to the resistant strain for some scenarios. Notice the different scales on subfigures (a) and (b).}
\label{F:highmortality}
\end{figure}

Figure \ref{F:highmortality} shows the effects of IPT on the number of sensitive strain malaria deaths averted under the long/long scenario.  It illustrates how the use of IPT reduces number of sensitive deaths but can in fact increase number of deaths from resistant infections for mid-range values of $p$.  For very low values of $p$ in the high transmission region, the resistant strain becomes dominant quickly, so IPT has very little impact on anything. Since we assume that the resistant strain has some kind natural competitive disadvantage compared to the sensitive strain, symptomatic treatment is driving this take over. The number of deaths from resistant infections is significantly reduced for $p\ge 0.4$.  High values of $p$ decrease the total number of malaria deaths by diminishing the duration of the resistant symptomatic infection. This is seen in the expressions for $B_s$ and $B_{ms}$ in equation \eqref{Eq:durations}.

%

%
%

\begin{figure}[h]
\centering
       \subfigure[No IPT]{
    \includegraphics[scale=0.18]{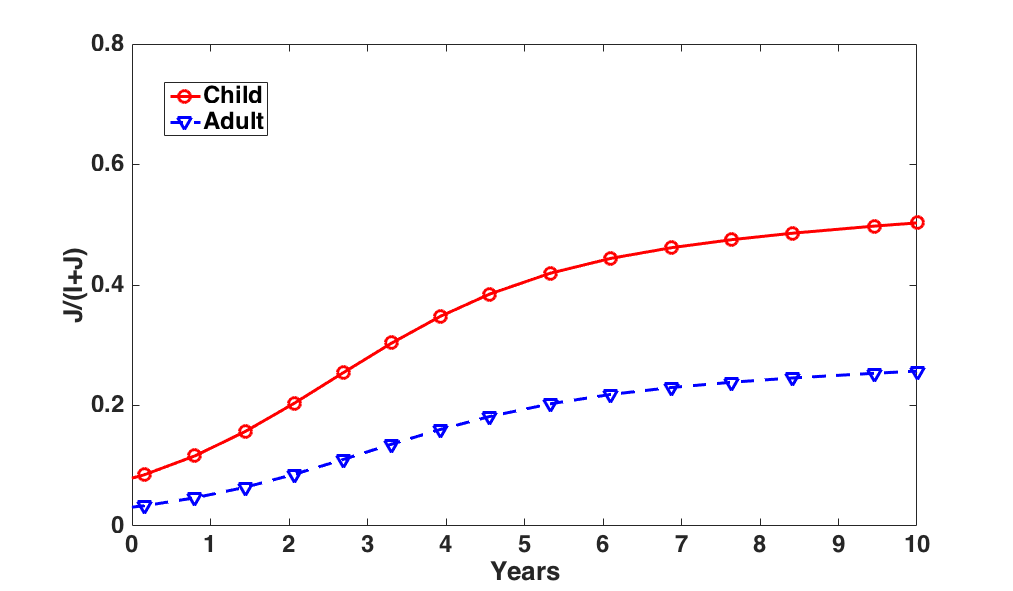} }
   \subfigure[With IPT]{
    \includegraphics[scale=0.18]{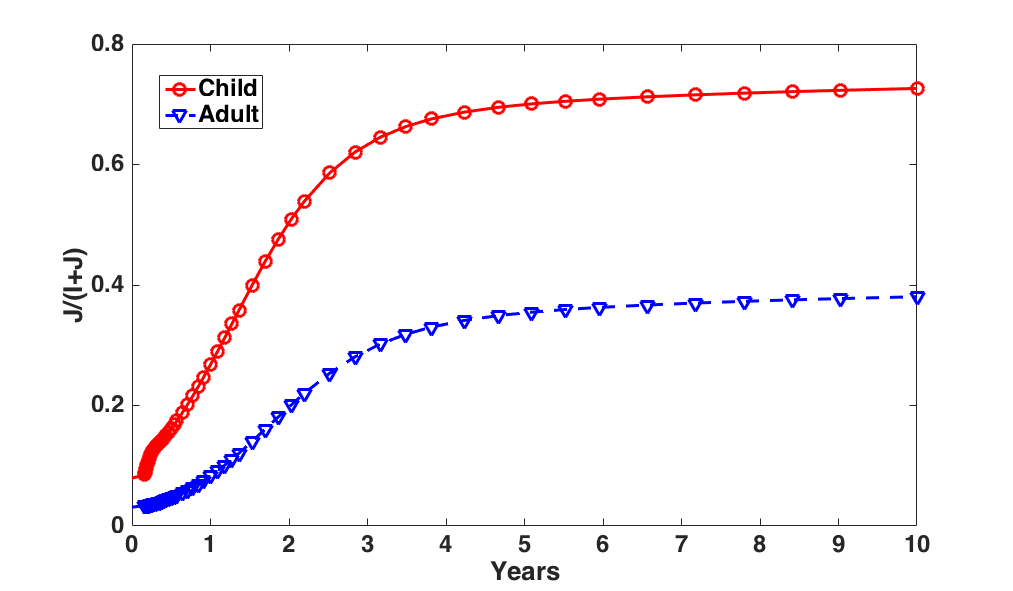} }
\caption{High transmission region, SP for treatment (long/long scenario), $p=0.3$: Proportion of cases that are resistant over ten years.  We ran the model with initially low levels of resistance and no IPT for 10 years, then begin IPT usage and track next ten years. There are no figures when AL is used for treatment (long/short case) because the resistant strain dies out for this scenario. }
\label{F:ratio1}
\end{figure}

Figure \ref{F:ratio1} shows the impact IPT has on the percent of cases that are resistant for a fixed value of $p$.  In Figure \ref{F:ratio1}(a), we see that for $p=0.3$ (which is right on the border of resistance really dominating), the percent of cases that are resistant will increase steadily over the course of 10 years without IPT. However, the use of IPT (Figure \ref{F:ratio1}(b)) will drive the proportion of resistant cases higher, particularly in children. Thus, IPT usage can work synergistically with treatment to allow resistant strains to increase and take over. However, for $p>0.3$, this effect is muted.

\begin{figure}[h]
\centering
    \subfigure[SP Ratio Resistant]{
  \includegraphics[scale=0.145]{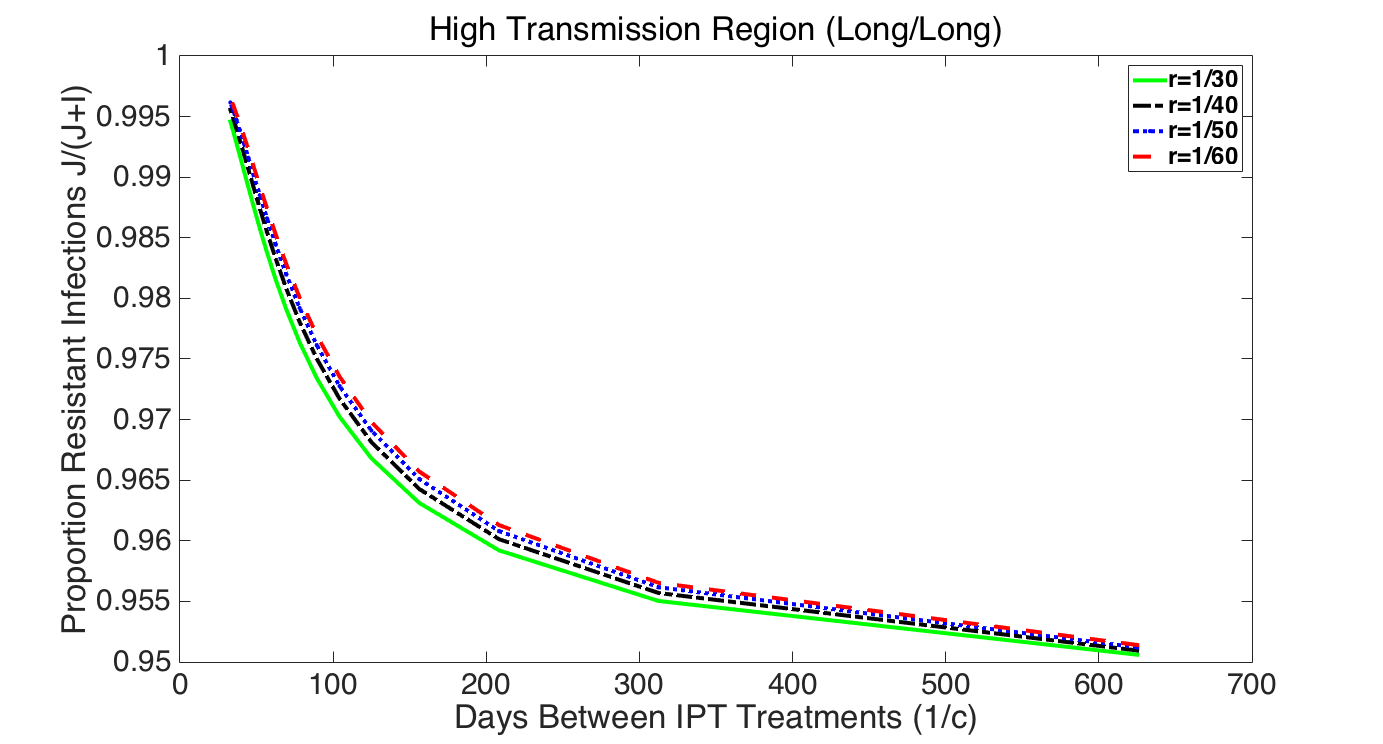}}
    \subfigure[SP Total Child Deaths]{
  \includegraphics[scale=0.145]{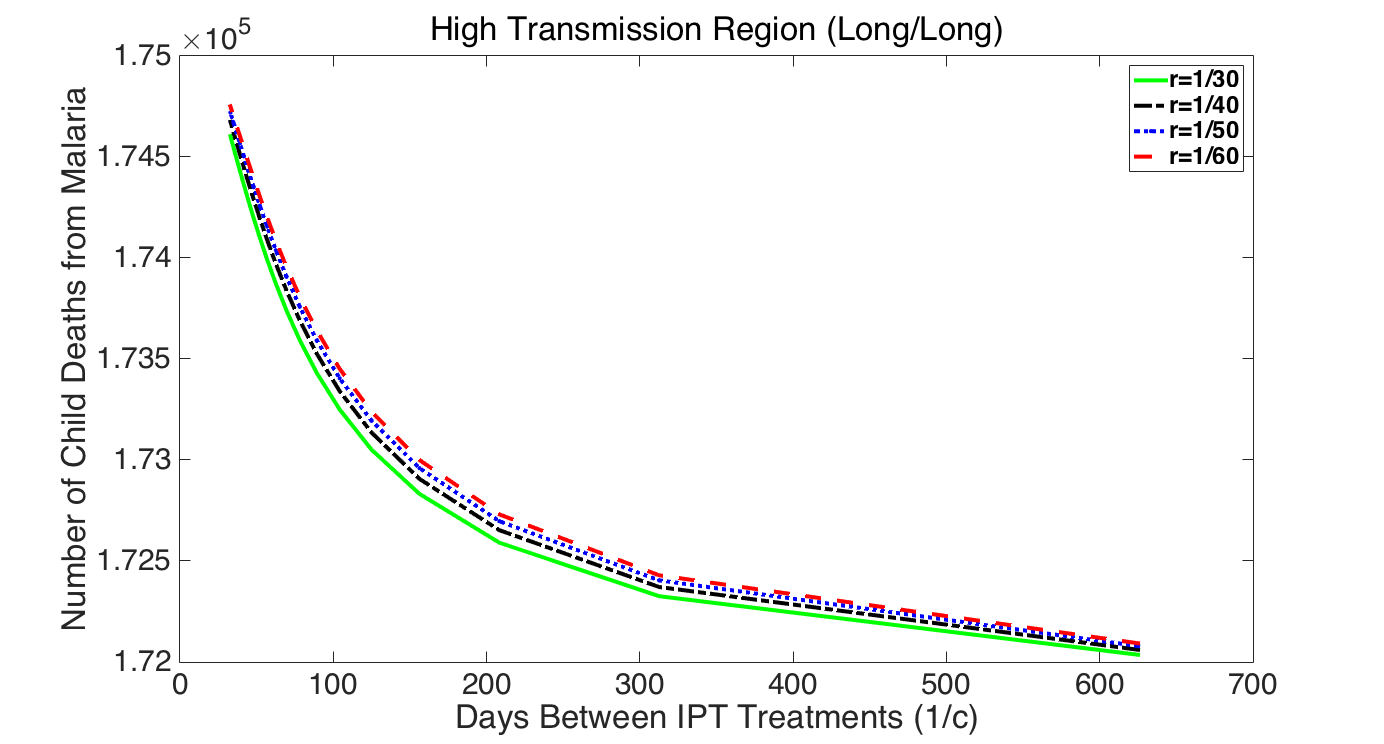}}
      \subfigure[AL Ratio Resistant]{
  \includegraphics[scale=0.14]{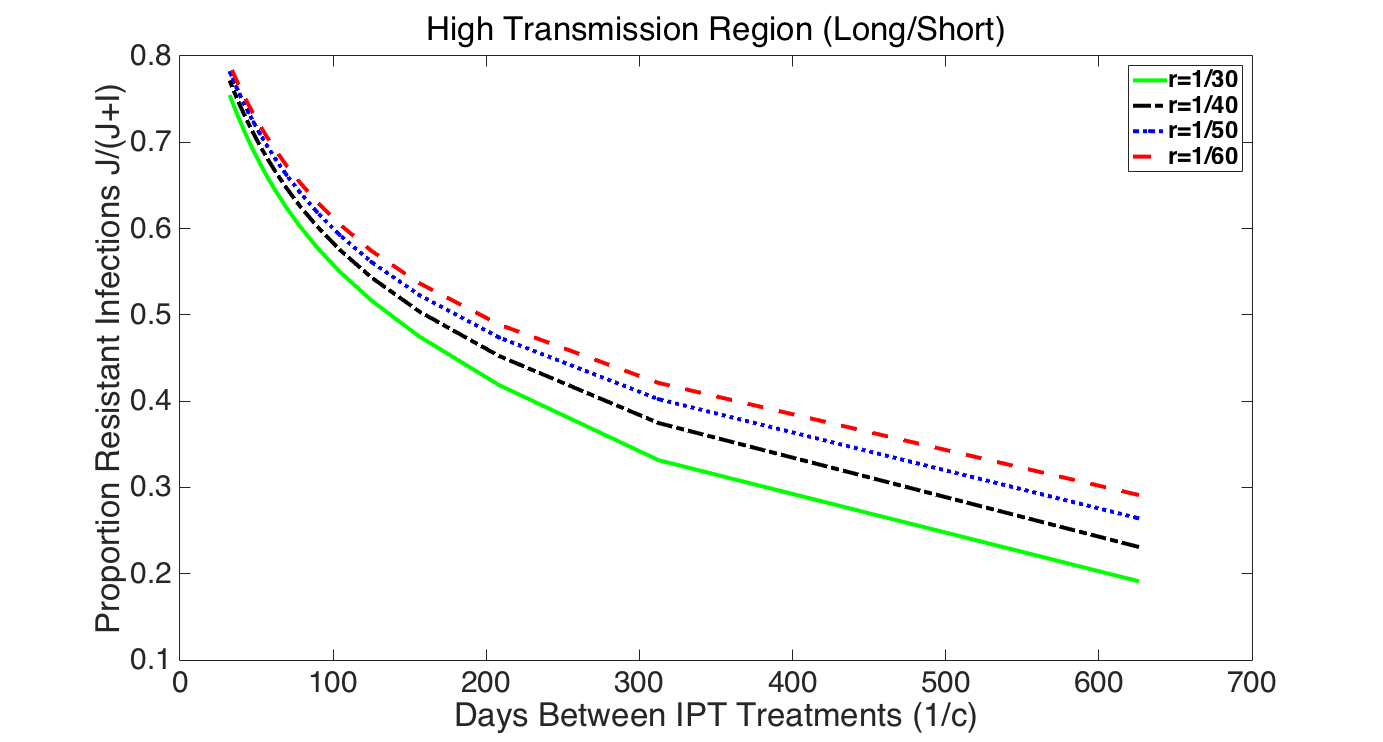}}
    \subfigure[AL Total Child Deaths]{
  \includegraphics[scale=0.145]{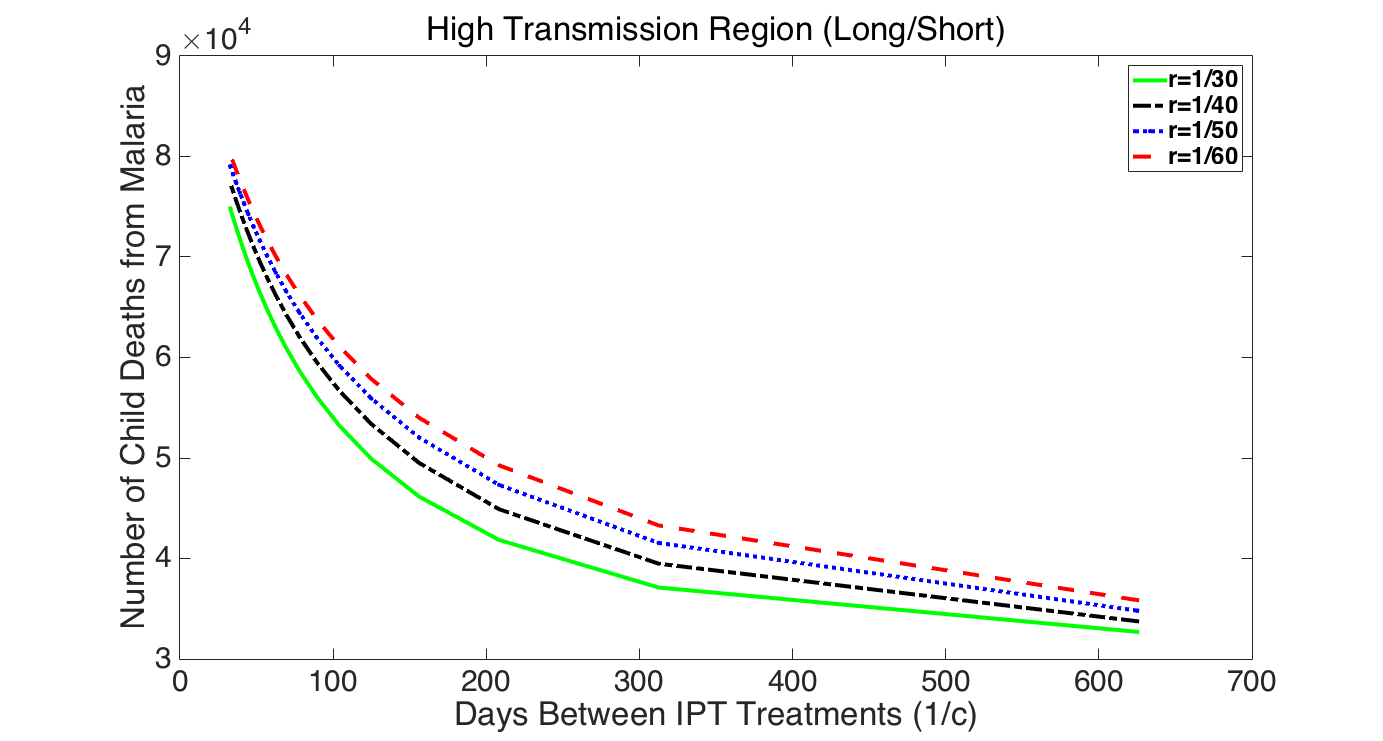}}

  \caption{High transmission region, $p=0.1$: Ratio of resistant infections to total infections (left) and total child deaths (right) after 10 years of IPT for varying time between IPT treatments, $1/c$, and for various values of $r^{-1}$, the time chemoprophylaxis lasts in susceptible IPT treated humans.  \textbf{(Top Row: (long/long) scenario)} Symptomatic treatment is SP. \textbf{(Bottom Row: (long/short scenario))}  Symptomatic treatment is AL. Initial conditions are the same as in Figure \ref{F:ratio1}. For both AL and SP symptomatic treatment, any IPT will result in more resistance and more deaths for $p=0.1$. When the short half-life AL drug, the level of resistance and number of deaths is less than when SP is used for symptomatic treatment (long/long). }
  \label{F:jratiovsrcSP}
\end{figure}

\begin{figure}[h]
\centering
    \subfigure[SP Total Child Deaths]{
  \includegraphics[scale=0.14]{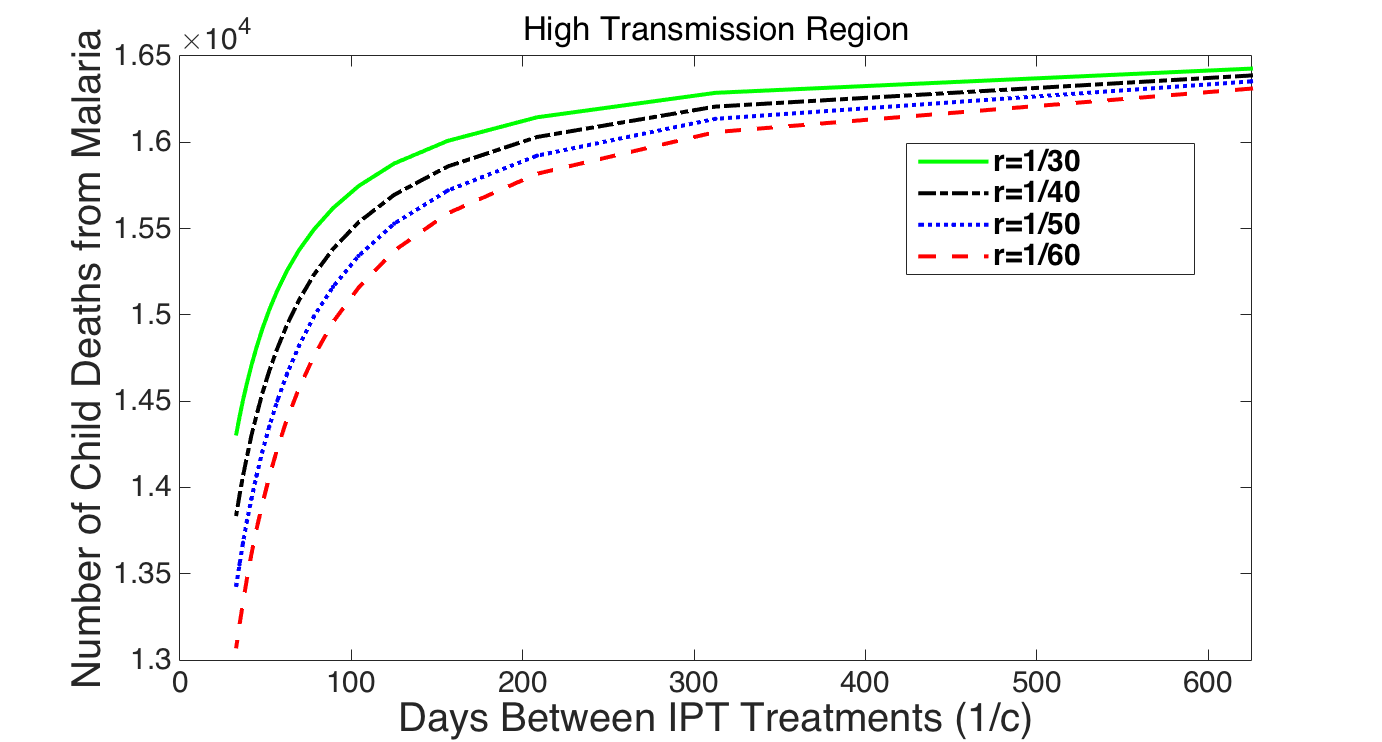}}
  \subfigure[AL Total Child Deaths]{
  \includegraphics[scale=0.14]{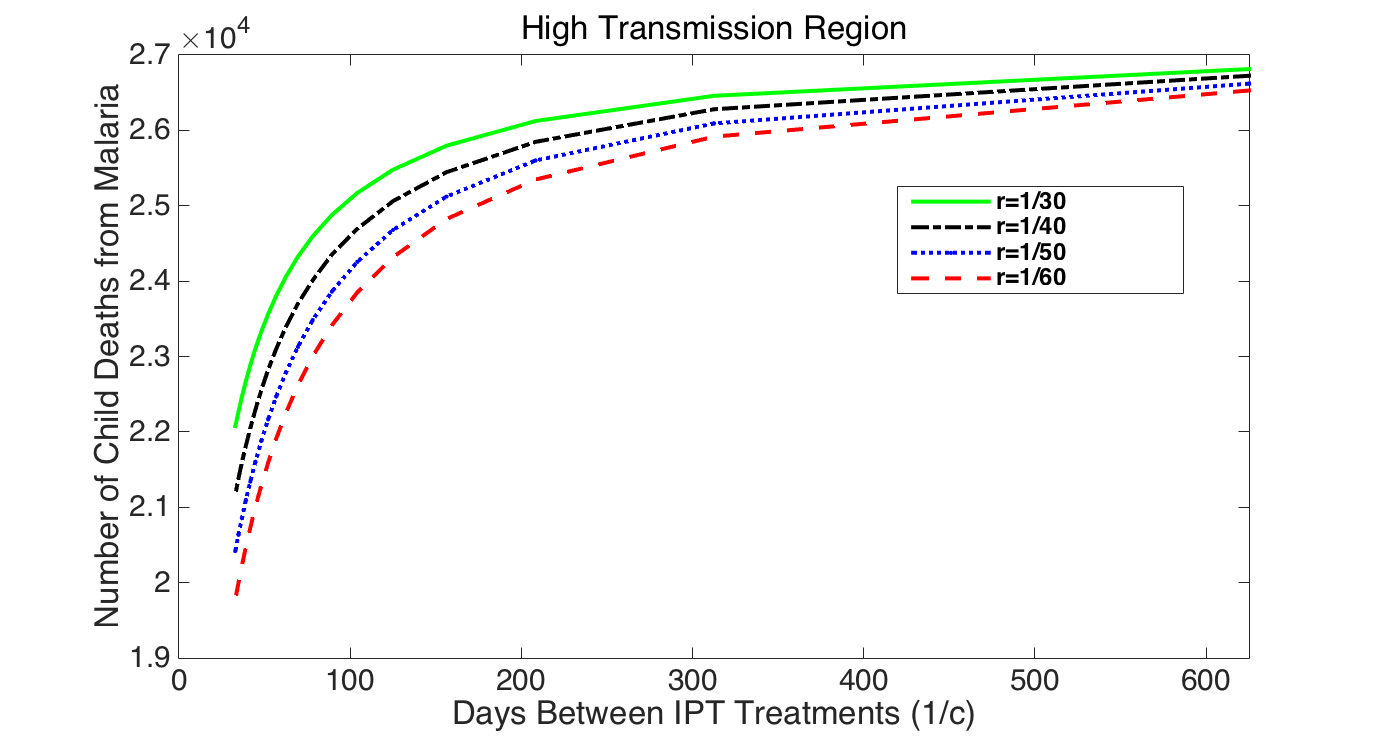}}
  \caption{High transmission region, $p=0.5$: Total child deaths after 10 years of IPT for varying time between IPT treatments, $1/c$, and for various values of $r^{-1}$, the time chemoprophylaxis lasts in susceptible IPT treated humans. \textbf{(Left: long/long scenario)} Symptomatic treatment is SP. \textbf{(Right: long/short scenario)} Symptomatic treatment is AL. Initial conditions are the same as in Figure \ref{F:ratio1}. In this case, both SP and AL scenarios with IPT result in lives saved. However, since resistance is low, using SP for symptomatic treatment is the best choice (saves more total lives). }
  \label{F:jratiovsrcSPp5}
\end{figure}

Figures \ref{F:jratiovsrcSP} and \ref{F:jratiovsrcSPp5} investigate how different rates of IPT treatments and treatment drug half-life influence the dynamics after 10 years.  Figure \ref{F:jratiovsrcSP} shows that in the high transmission region with long/long drug half-lives and $p=0.3$, increases in time between IPT treatments, $1/c$ reduces the effects of malaria. In this scenario the model predicts that the use of IPT has negative consequences, as the number of infections, children death, and percentage of resistant cases is high for low values of $1/c$. Figure \ref{F:jratiovsrcSPp5}(a) shows that in the same scenario but with $p=0.5$, the use of IPT is beneficial. Here increasing time between IPT treatments, $1/c$, increases the number of infections and children deaths. In  high transmission regions using long/long drug half-lives, we see that IPT should only be used for high values of $p$.  Figures \ref{F:jratiovsrcSP}(c)-(d) and \ref{F:jratiovsrcSPp5}(b) show that in the high transmission region with long/short drug half-lives, IPT is beneficial in reducing the number of infections and child deaths. However, when $p=0.3$ although the total number of infections is low for low values of $1/c$, the percentage of resistant infections is high (Figure \ref{F:jratiovsrcSP}(c)-(d)).


We see in Table \ref{T:totaldeaths} that the total number of deaths of children from malaria increases dramatically as the value of $p$ decreases for long/long drug half-lives. So, as strains develop more resistance to the drug used for treatment (low values of $p$), the number of deaths will increase if no new effective drug is available or put into use. For example, in the high transmission region, for $p=0.1$, there are nearly 10 times the number of deaths as for $p=0.5$. For high transmission regions, this effect is much more pronounced and occurs for higher values of $p$. For high transmission, number of deaths start drastically increasing for $p<0.3$, but for low transmission, this occurs for $p<0.11$. We can also see that IPT only results in significant ($>10\%$) reductions in total number of children deaths for $p>0.4$ and over 10 years in the high transmission region. For low transmission, if $p>0.11$, then a $>10\%$ reduction in child deaths occurs over 5 or more years. It is also interesting to note the distinctly non-linear relationship between $p$ and number of lives saved/lost due to IPT.

\begin{figure}[htpb]
\centering
  \includegraphics[scale=0.7]{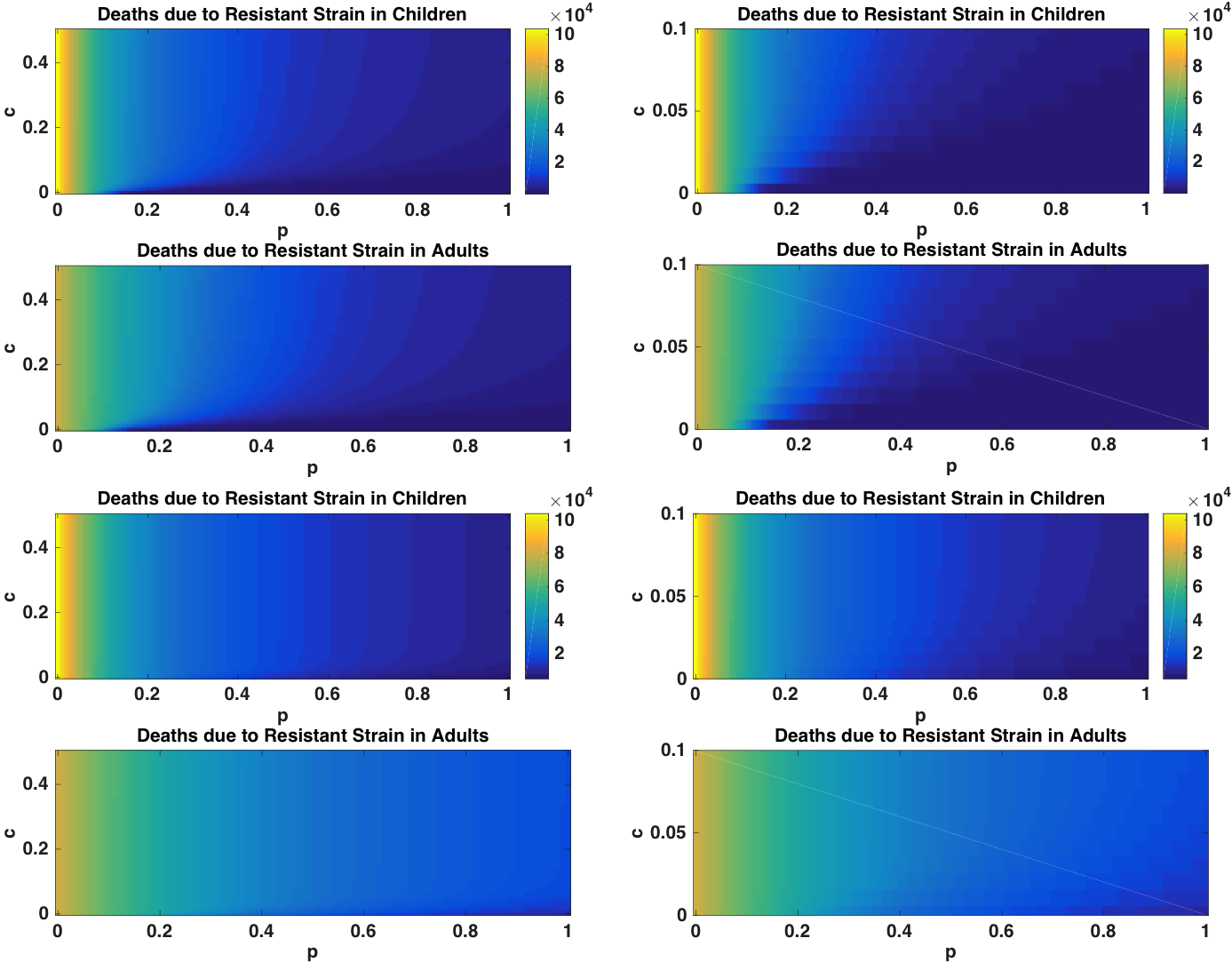}
  \caption{Heatmap of the \textit{number} of deaths from the resistant strain for the high transmission region. \textbf{(Top Row: (long/short) scenario)} AL for symptomatic treatment and \textbf{(Bottom Row:(long/long) scenario)} SP for symptomatic treatment. The right column is a zoom-in of the left column to show more realistic values of $c$, the rate at which IPT is given. The parameter $p$ is the effectiveness of the treatment drug on the resistant strain, so $p=0$ is fully resistant and $p=1$ is fully sensitive. Number of deaths is dependent almost exclusively on $p$.}
  \label{F:heatmapnumhigh}
\end{figure}

\begin{figure}[htpb]
\centering
  \includegraphics[scale=0.7]{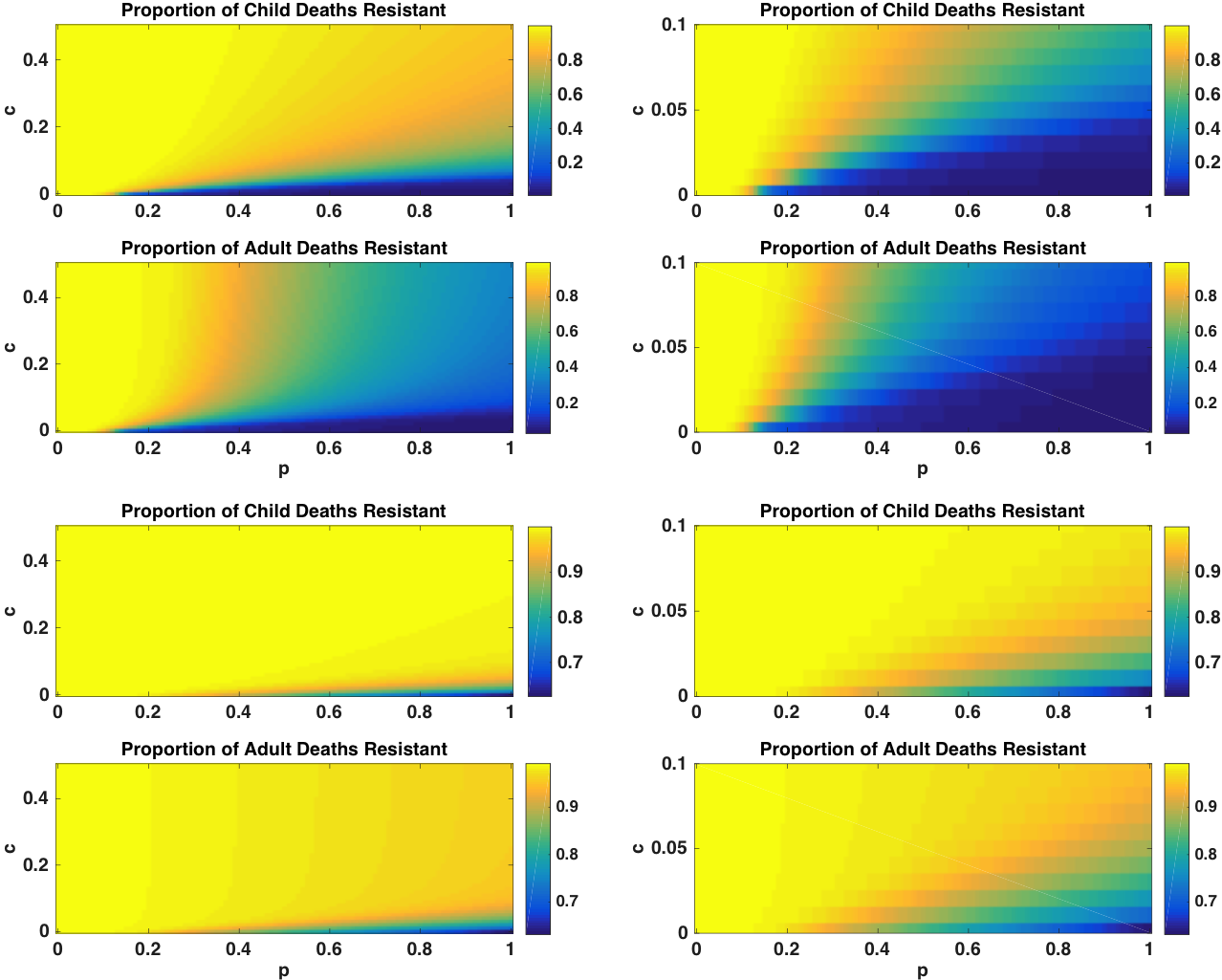}
  \caption{Heatmap of the \textit{proportion} of deaths from the resistant strain for the high transmission region and for \textbf{(Top Row: (long/short) scenario)} AL for symptomatic treatment  and \textbf{(Bottom Row: (long/long) scenario)} SP for symptomatic treatment. The right column is a zoom-in of the left column to show more realistic values of $c$, the rate at which IPT is given.  Note different scales for top and bottom rows. The proportion of deaths from the resistant strain is dependent on both $p$ and $c$, showing that IPT schedule can increase resistance.}
  \label{F:heatmapprophigh}
\end{figure}

Finally, Figures \ref{F:heatmapnumhigh} and \ref{F:heatmapprophigh} illustrate how differences in IPT and treatment half-lives can change results in the high transmission region. In Figure \ref{F:heatmapnumhigh}, the top row shows the relationship between IPT treatment frequency, $c$, and resistance strength, $p$, with number of deaths from the resistant strain of malaria at the endemic equilibrium for long/short half-lives. The left figure is for a wide range (with some unrealistically high application rates) of IPT, while the right figure looks at more realistic values of $c<0.1$ corresponding to IPT treatment schedules of greater than 10 days. The total number of deaths from the resistant strain is almost exclusively dependent on the value of $p$, with some slight change as $c$ increases for the long/short IPT/treatment half-lives. The bottom row is the same scenario except for long/long IPT/treatment half-lives. In this case, $c$ has no discernible effect. This figure also shows the extremely wide range of total number of child and adult deaths as resistance levels increase, or values of $p$ decrease. It also illustrates that the long/short scenarios requires lower values of $p$ to result in higher numbers of deaths than the long/long scenario. Figure \ref{F:heatmapprophigh}, showing dependence of proportion of deaths that are resistant on $c$ and $p$, highlights the difference between long/short (top row) and long/long (bottom row) regimes. We see that if both IPT and treatment have long half-lives (bottom row), then the space where the resistant strain dominates is much larger. When IPT is long half-life but treatment is short half-life (top row) there is a wide range of parameter space for which the proportion resistant is low. It is also important to note that for this high transmission region, use of IPT affects proportion of deaths from the resistant strain in both adults and children. This implies that IPT is directly changing the dynamics of the resistant strain.

\clearpage

\clearpage

\subsection{Numerical Results: Low Transmission Region}

For the low transmission region we changed the parameters  to match the low transmission parameters in Tables \ref{T:parmvalues} and \ref{T:parmvalues2}. For this scenario, the total number of child deaths from malaria are at least an order of magnitude smaller than in the high transmission region (see Table \ref{T:totaldeaths}). In sheer numbers, then, IPT and treatment will have a lower impact in the low transmission region. The basic reproduction numbers for the sensitive and resistant strains are less than one at our low transmission baseline parameters, Table \ref{T:RoValues}.   Figure \ref{F:RoPlots} (top left) shows that for very low values of $p$, indicating very high resistance to the treatment drug, the resistant strain has  $R_0>1$, greater than the sensitive strain reproduction number.   In Figure \ref{F:RoPlots} (top right), the sensitive strain reproduction number is slightly reduced by $c$ at very low values of $c$, corresponding to very infrequent IPT, but remains unchanged after that. The resistant reproduction number is unchanged by $c$. This means that frequency of IPT application has very little impact on either reproduction number for the low transmission region.

\begin{figure}[htbp]
\centering
\subfigure[SP Treatment]{\includegraphics[scale=0.15]{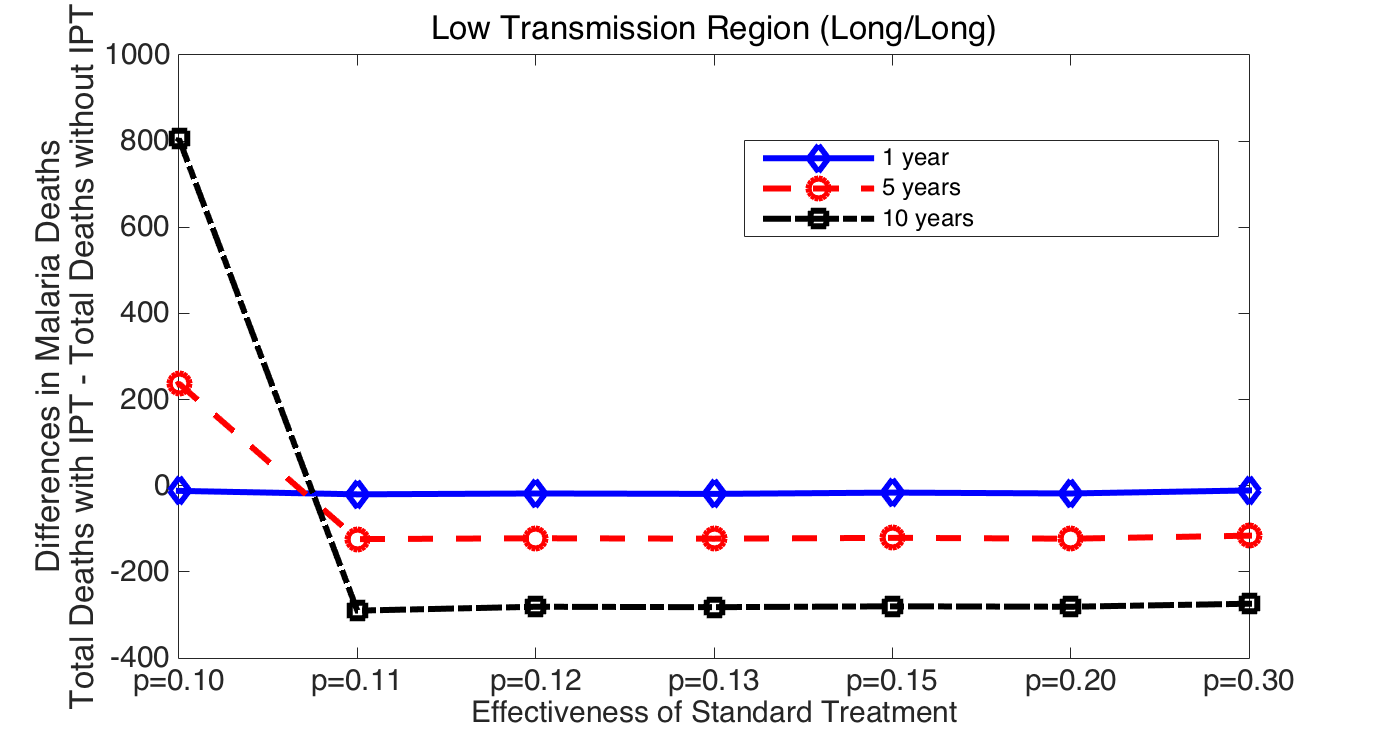}}
\subfigure[AL Treatment]{\includegraphics[scale=0.15]{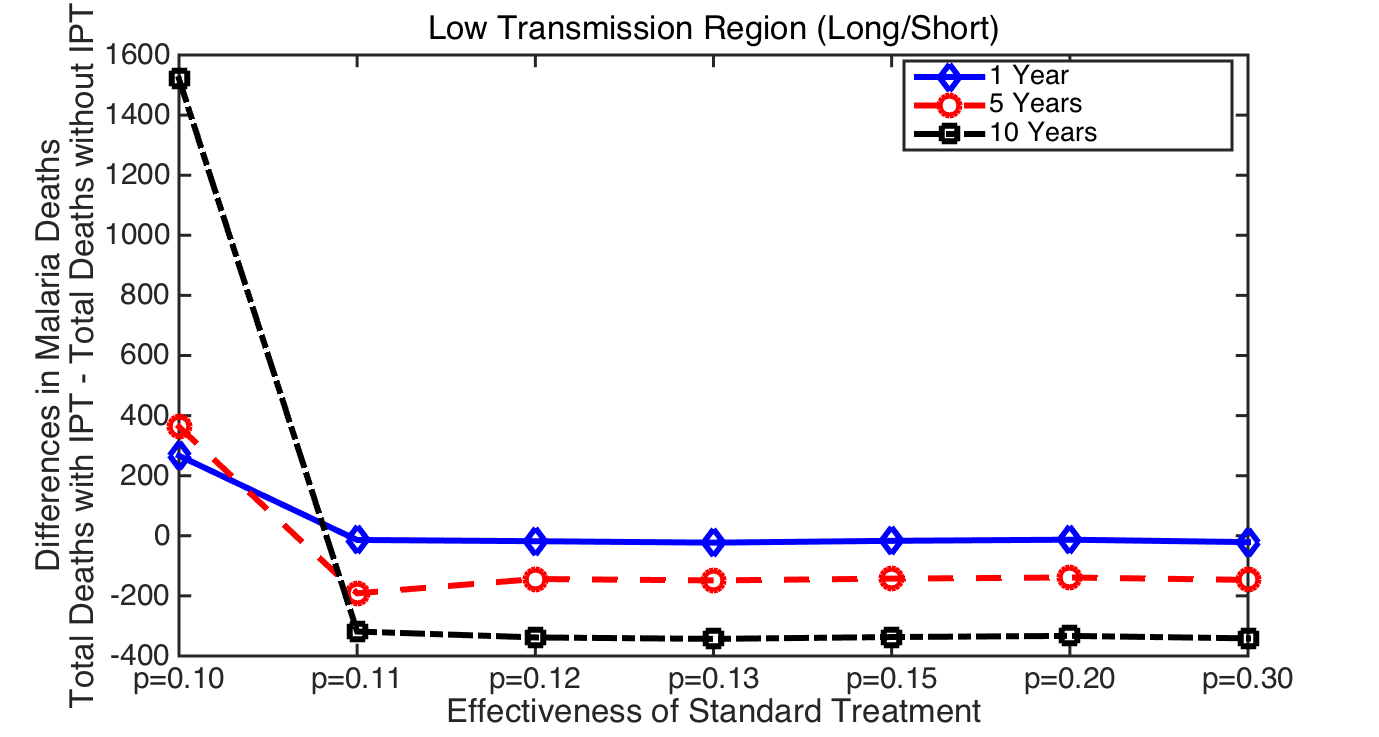}}
\caption{Low transmission region: Net increase in deaths due to IPT usage, or \textit{(Total child deaths due to sensitive and resistant strains of malaria with IPT) - (total child deaths without IPT)} for 1 year, 5 years, and 10 years of IPT use for different levels of standard treatment effectiveness against the resistant strain, $p$. (a) SP used as the symptomatic treatment drug and (b)  AL used for symptomatic treatment.  Negative numbers indicate lives saved due to IPT while positive numbers indicate more deaths from using IPT.}
\label{F:lowtotal}
\end{figure}

In Figure \ref{F:lowtotal}(a), for  $p>0.11$, IPT results in a net gain of lives saved for 1 year, 5 years, and 10 years for the long half-life drug SP used as treatment and as IPT. Past that point, in fact, there is very little difference across all values of $p$, unlike the high transmission scenario. However, as expected, the number of lives saved is an order of magnitude less than for the high transmission region, Figure \ref{F:hightotal}. For $p<0.11$, application of IPT results in an increase in deaths over 5 and 10 years. There is a bifurcation point for $p$ where the dominant strain switches from the sensitive to the resistant strain. Once the resistant strain is dominant, widespread use of the drug that it is resistant to leads to more rather than fewer deaths. When the short half-life drug AL is used for treatment and SP for IPT, Figure \ref{F:lowtotal}(b), we see a very similar bifurcation point at $p=0.11$ below which the resistant strain takes over and spreads, resulting in IPT being not only ineffective, but damaging. It is interesting to note that the increase in number of deaths from using IPT at $p=0.10$ for AL treatment is double the increase in deaths from IPT when SP is used for treatment. This is in contrast to the high transmission region where using SP as treatment  results in a higher increase in deaths resulting from IPT usage (Figures \ref{F:hightotal}(a) and (b)). However, it should be noted that although the increase in deaths from using IPT is larger for AL treatment, the \textit{total} number of deaths is larger when SP is used for both treatment and IPT, Table \ref{T:totaldeaths}.

In Table \ref{T:totaldeaths}, we see that the resistant strain only dominates after introduction in the low transmission region for very low values of $p$, which equates to very high resistance to the drug used for treatment in the resistant strain. For the \textbf{long/long} IPT/treatment half-life scenario, the total number of deaths  jumps by more than a factor of 3 when $p=0.09$. For the \textbf{long/short} scenario, a smaller jump in cases is seen at $p=0.09$. In absolute numbers, IPT saves more lives in the high transmission region, but as a percent reduction of total deaths, IPT does better in the low transmission region. Another interesting pattern is that for higher values of $p$, using short half-life treatment results in more deaths than using long half-life treatment. However, once a highly resistant strain is circulating, the long/short regime has lower total deaths than long/long. For example, in the low transmission region, when $p=0.10$, there are  1,599 deaths without IPT and 1,836 deaths with IPT for long/long after 5 years. By contrast, for long/short there were  698 deaths without IPT and 1,060 deaths with IPT. If a very resistant strain is circulating it is better to use a short half-life treatment drug.

\begin{figure}[h]
\centering
       \subfigure[No IPT]{
    \includegraphics[scale=0.145]{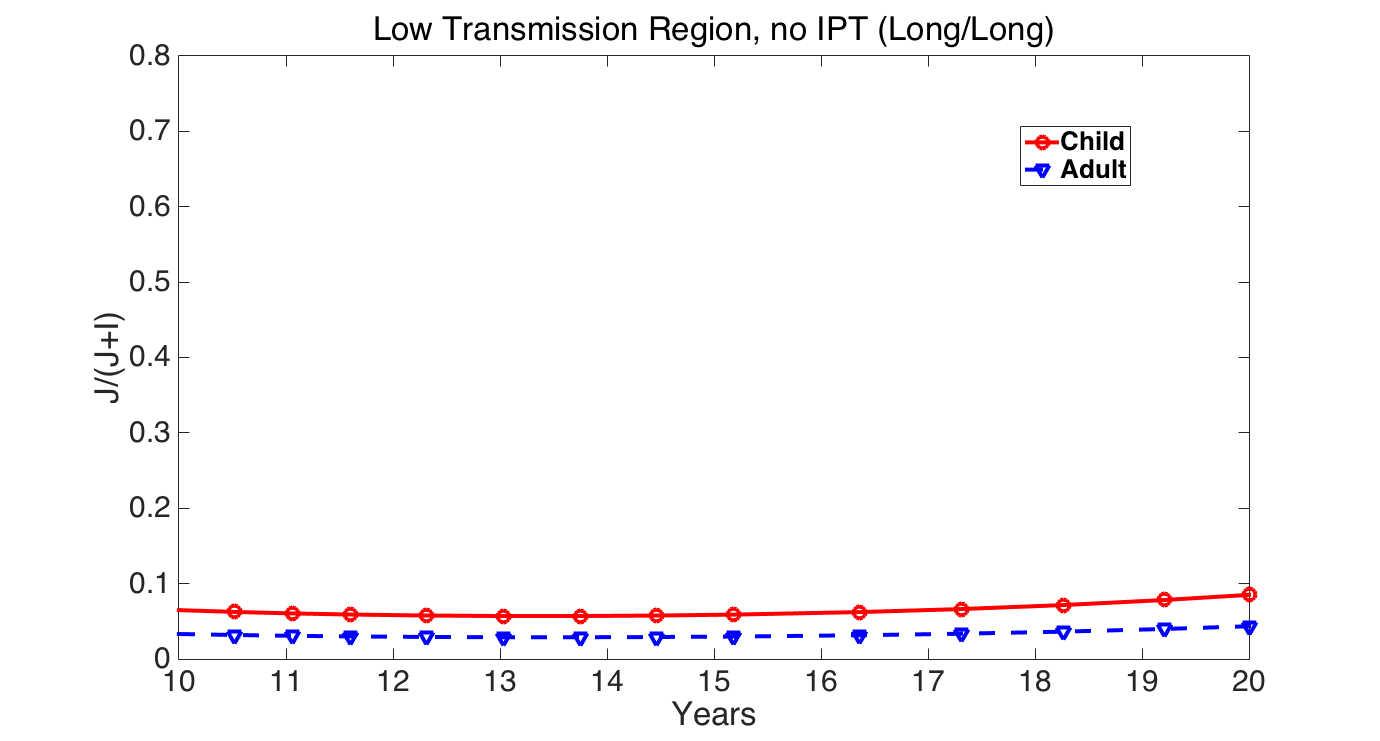} }
   \subfigure[With IPT]{
    \includegraphics[scale=0.145]{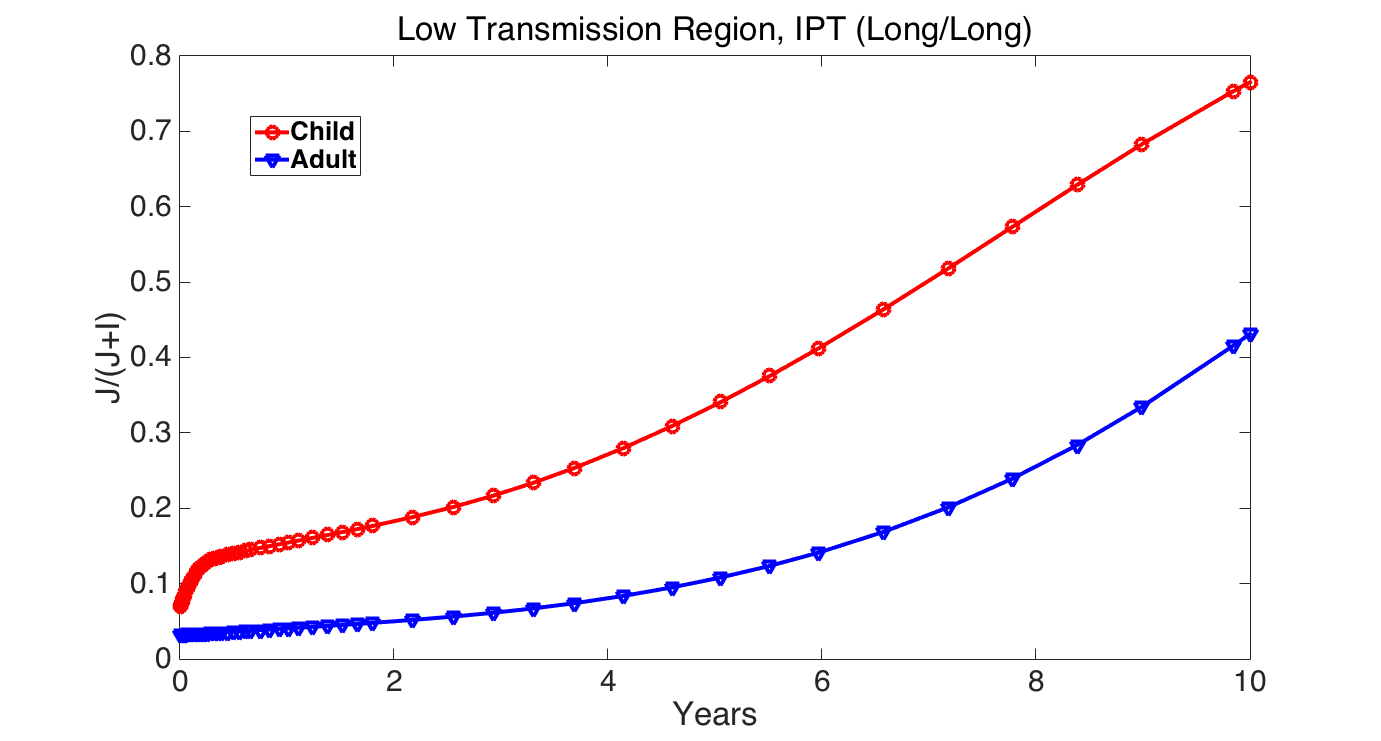} }
\caption{ Low transmission region, SP for symptomatic treatment, $p=0.11$.  Proportion of infections that are resistant (ratio of resistant strains of infection to the sum of resistant and sensitive infections) over ten years.}
\label{F:lowratio1}
\end{figure}
\begin{figure}[h]
\centering
       \subfigure[No IPT]{
    \includegraphics[scale=0.145]{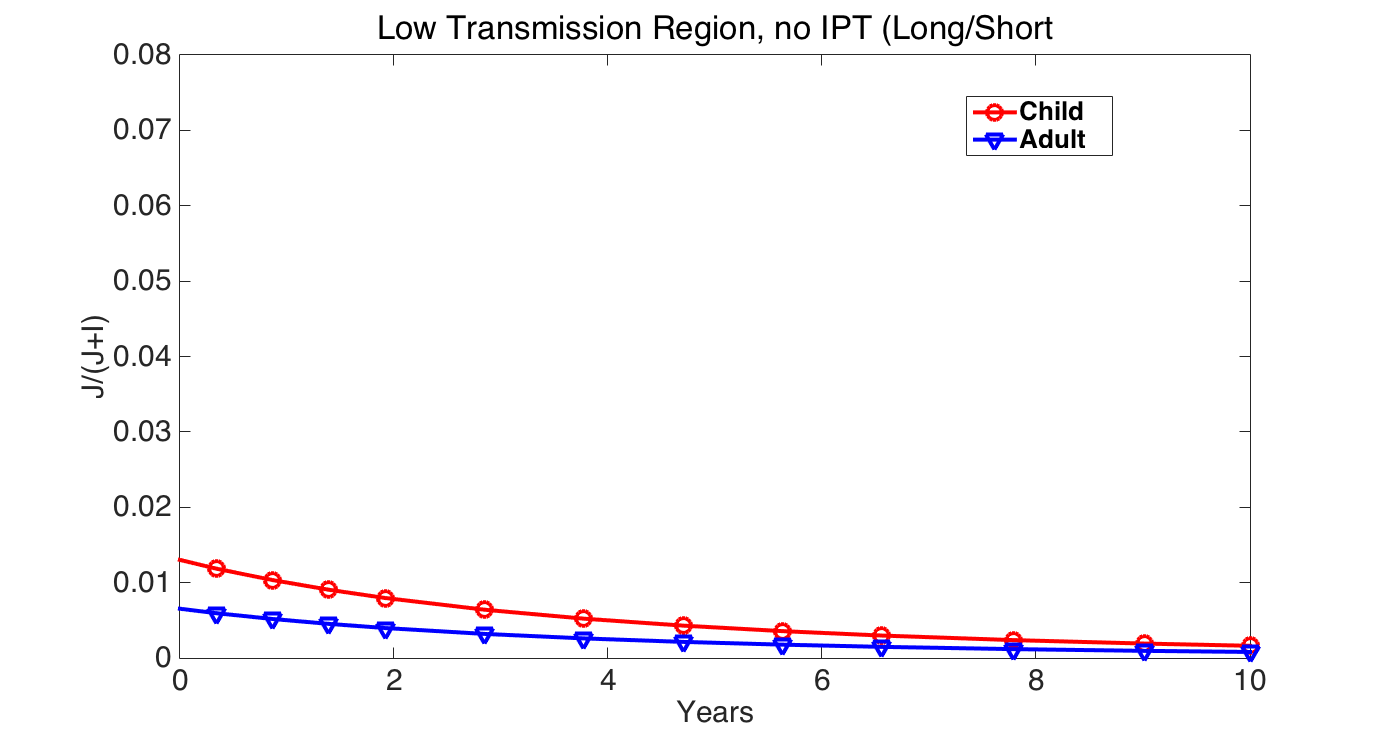} }
   \subfigure[With IPT]{
    \includegraphics[scale=0.145]{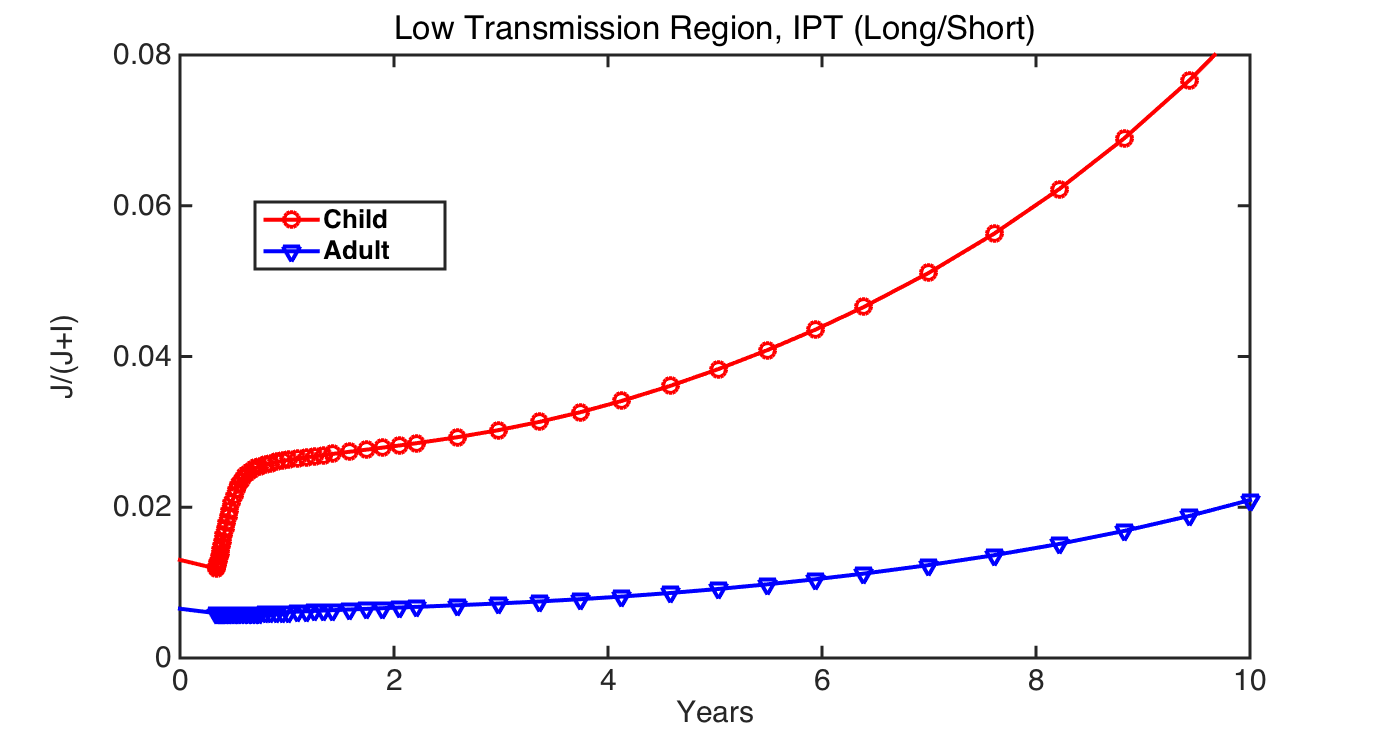} }
\caption{Low transmission region, AL for symptomatic treatment, $p=0.11$.  Proportion of infections that are resistant (ratio of resistant strains of infection to the sum of resistant and sensitive infections) over ten years.   Notice that the scale here is $1/10$ of that in Figure \ref{F:lowratio1}.}
\label{F:lowratioAL}
\end{figure}
\begin{figure}[h]
\centering
 \subfigure[SP treatment]{ \includegraphics[scale=0.147]{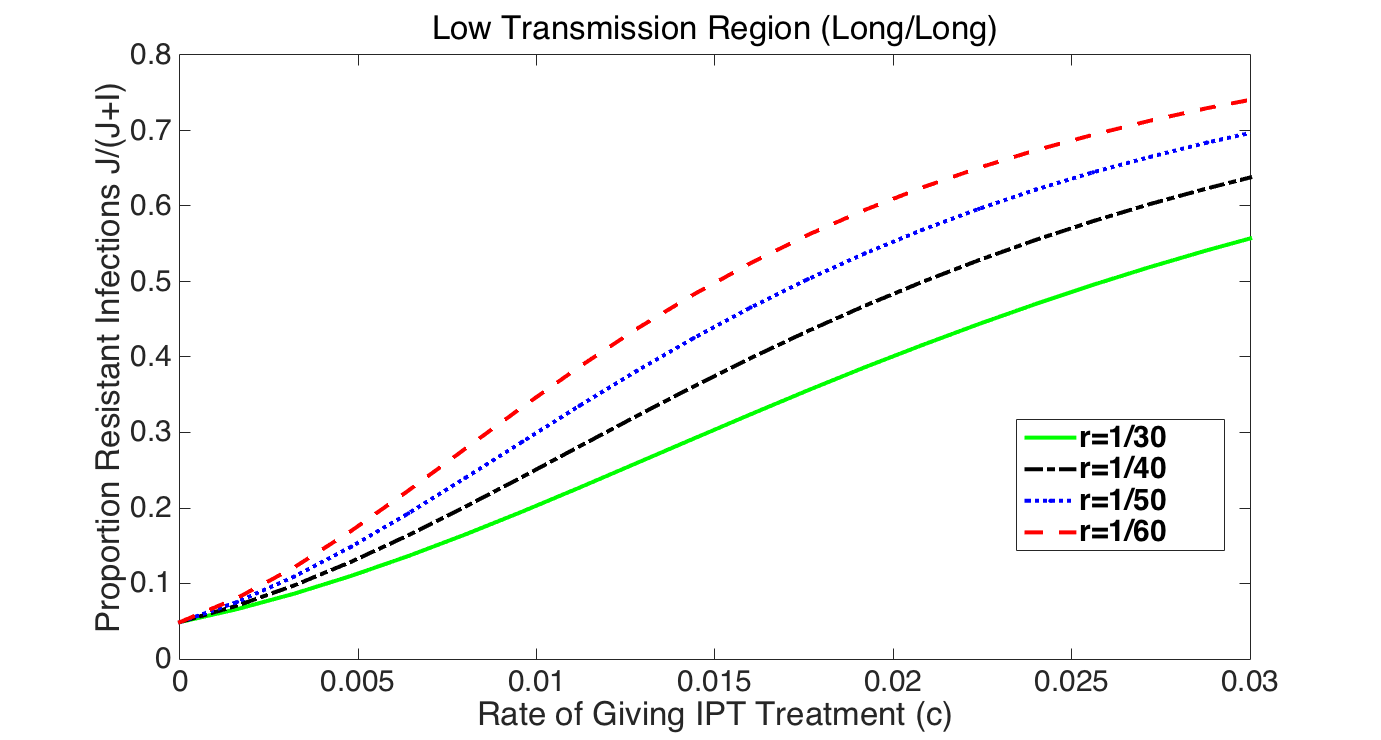}}
  \subfigure[AL treatment]{ \includegraphics[scale=0.147]{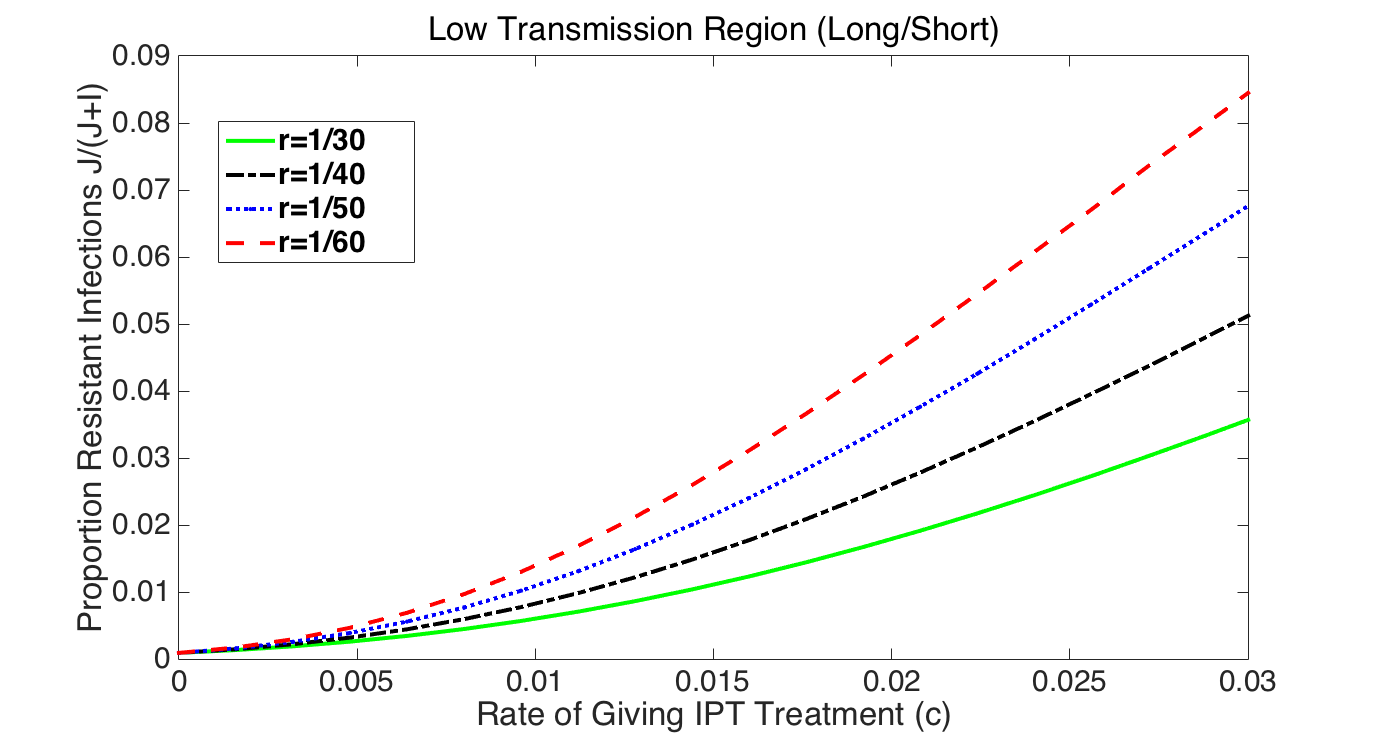}}
  \caption{  Low transmission region, $p=0.11$:  Ratio of resistant infections to total infections (left) and total child deaths (right) after 10 years of IPT for varying time between IPT treatments, $1/c$, and for various values of $r^{-1}$, the time chemoprophylaxis lasts in susceptible IPT treated humans.
  (a) is SP symptomatic treatment and  (b) is AL symptomatic treatment. Note that the y-axis in (a) is 10 times that in (b). }
  \label{F:lowjratiovsrc}
\end{figure}

At the inflection point $p=0.11$, Figure \ref{F:lowratio1}(a) shows that the proportion of cases resistant stays at a low and constant level over 10 years. However, when IPT is applied, it will slowly increase the proportion of resistant cases for children and adults  after 10 years, Figure \ref{F:lowratio1}(b). In this case, since IPT is applied only to children, the proportion of resistant cases is in children is roughly double that in adults after 10 years. In Figure \ref{F:lowratioAL}(a), with short half-life AL used for treatment, the resistant strain decreases to zero with time rather than staying steady. Secondly, in Figure \ref{F:lowratioAL}(b), the proportion resistant does increase with the use of IPT, but at a very slow rate compared to the high transmission region (notice the difference in scales). In general, with a strain that is very resistant to the treatment, it is better to employ the short half-life drug and not to use IPT in the low transmission region to control the spread of resistance. In fact, at $p=0.11$, lower levels of resistance are always obtained by using an IPT drug with the shortest half-life and at very infrequent intervals, Figure \ref{F:lowjratiovsrc}. Again, the short half-life drug used for treatment results in an order of magnitude lower level of resistance than the long half-life drug.

\begin{figure}[h]
\centering
    \subfigure[SP Number Child Deaths]{
  \includegraphics[scale=0.147]{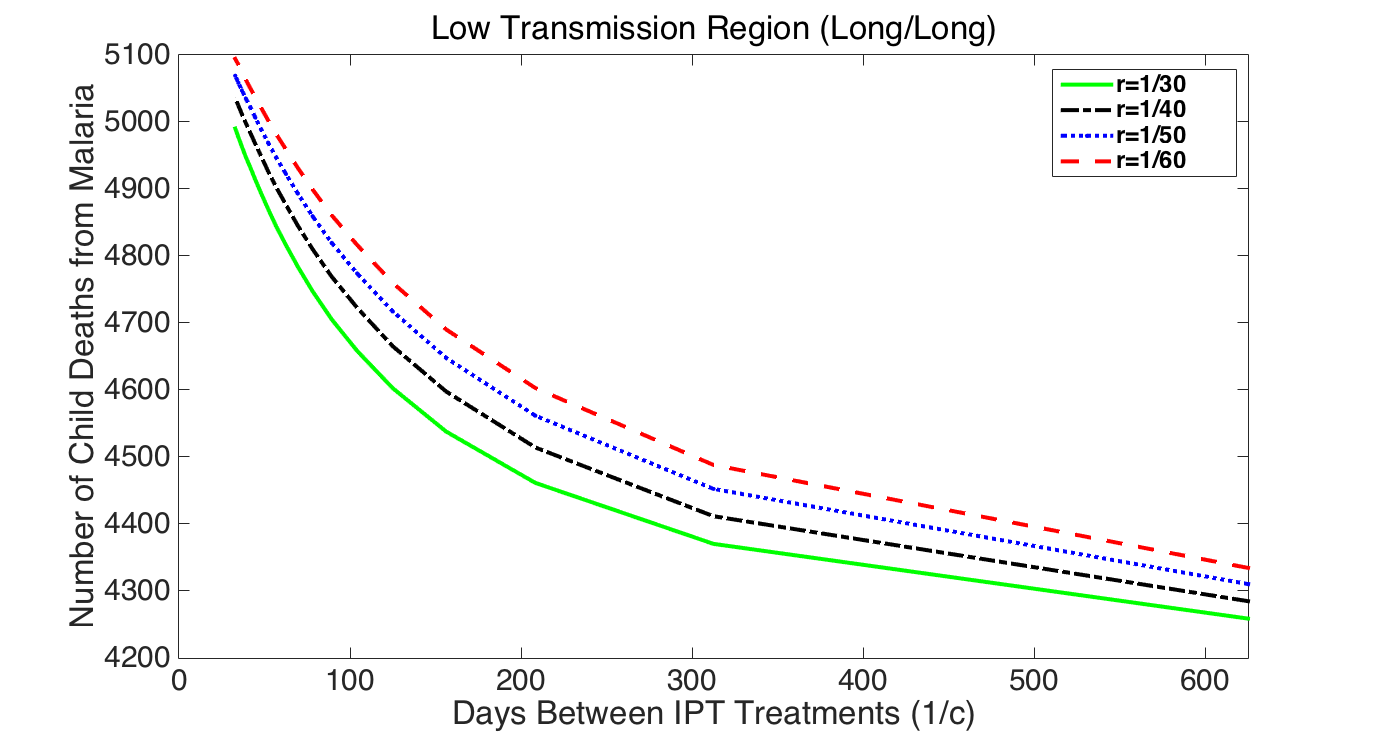}}
   \subfigure[AL Number Child Deaths]{
  \includegraphics[scale=0.147]{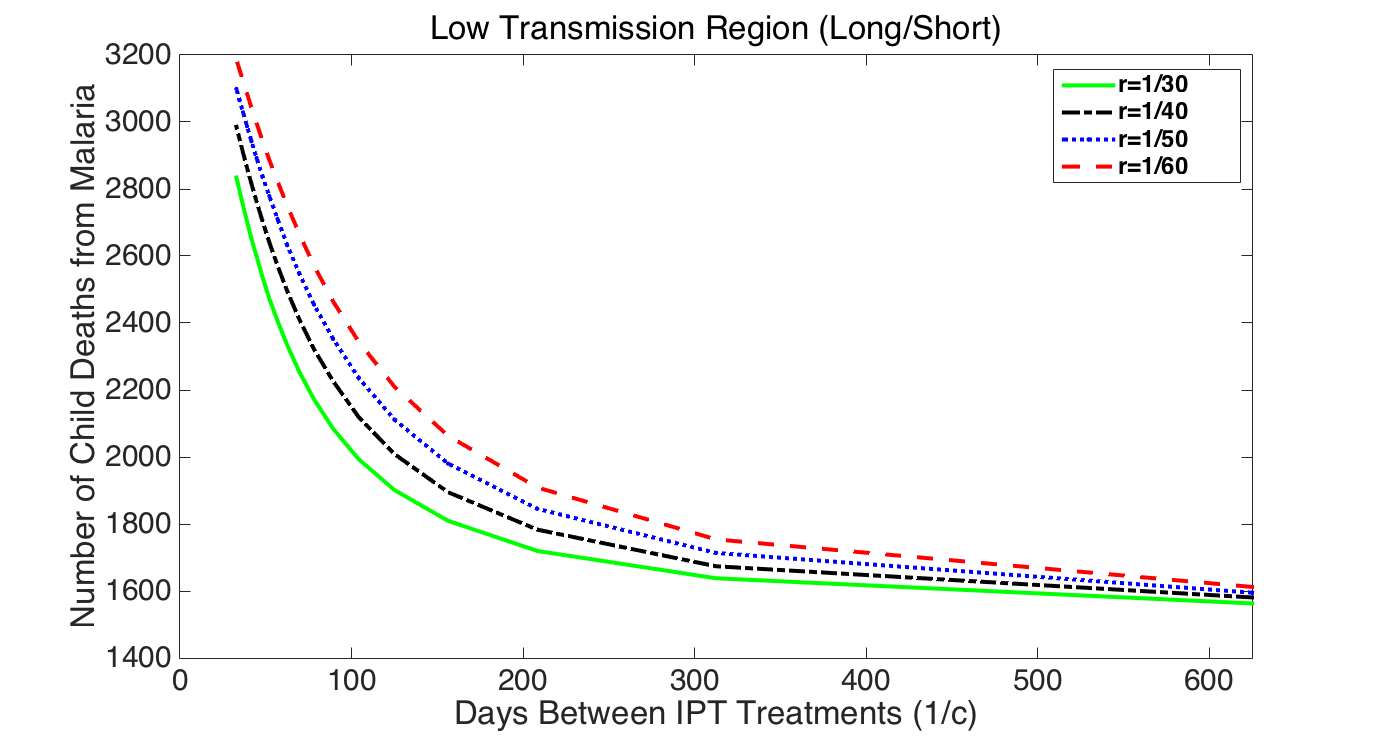}}
  \caption{Low transmission region, $p=0.1$: Total child deaths after 10 years of IPT for varying time between IPT treatments, $1/c$, and for various values of $r^{-1}$, the time chemoprophylaxis lasts in susceptible IPT treated humans.  \textbf{(Left)} is for SP used for symptomatic treatment, \textbf{(Right)} is AL used for symptomatic treatment.  Initial conditions are the same as in Figure \ref{F:ratio1}. In this case, both SP and AL scenarios with IPT result in an increase in deaths due to the circulation of a highly resistant strain.}
  \label{F:jratiovsrclowSPp1}
\end{figure}

\begin{figure}[h]
\centering
    \subfigure[SP Total Child Deaths]{
  \includegraphics[scale=0.147]{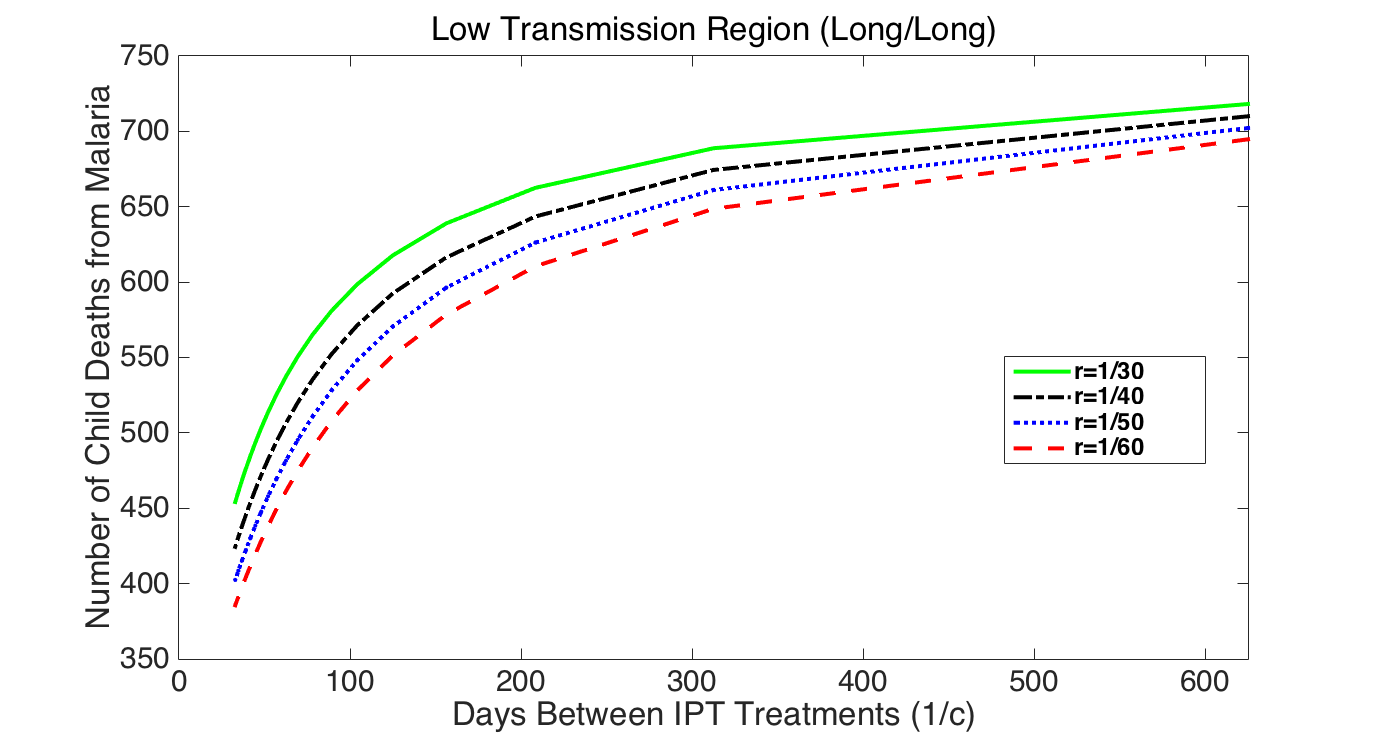}}
      \subfigure[AL Total Child Deaths]{
  \includegraphics[scale=0.147]{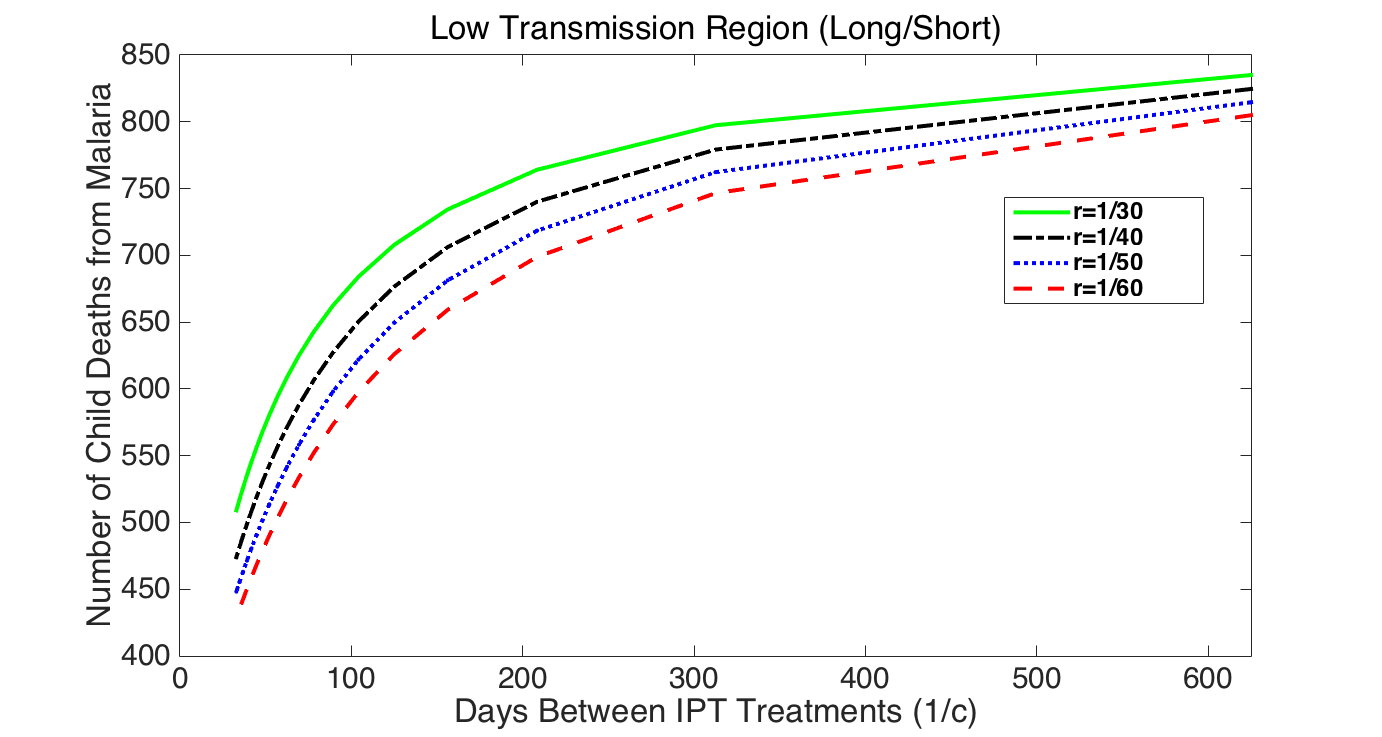}}
  \caption{Low transmission region, $p=0.3$: Total child deaths after 10 years of IPT for varying time between IPT treatments, $1/c$, and for various values of $r^{-1}$, the time chemoprophylaxis lasts in susceptible IPT treated humans.  \textbf{(Left)} is for SP used for symptomatic treatment, \textbf{(Right)} is AL used for symptomatic treatment.  Initial conditions are the same as in Figure \ref{F:ratio1}. In this case, both SP and AL scenarios with IPT result in saved lives.}
  \label{F:jratiovsrclowSP}
\end{figure}

When the treatment used is still partially effective against the resistant strain ($p=0.3$), then applying IPT more frequently and for longer half-live drugs will lead to lower total number of infections and childhood deaths, Figure \ref{F:jratiovsrclowSP}. However, when the treatment drugs are very ineffective against the resistant strain ($p=0.1$) then longer time between IPT application and a shorter IPT drug half-life always leads to a decrease in resistance, infections, and deaths, Figure \ref{F:jratiovsrclowSPp1}. There are  similar patterns for this behavior when a short half-life treatment drug, AL, is used, Figures \ref{F:jratiovsrclowSPp1}(b) and \ref{F:jratiovsrclowSP}(b). In short, there is a sharp regime change above which frequent IPT and long-half life drugs are  useful but below which they can be deleterious.

%

\begin{figure}[h]
\centering
  \includegraphics[scale=0.7]{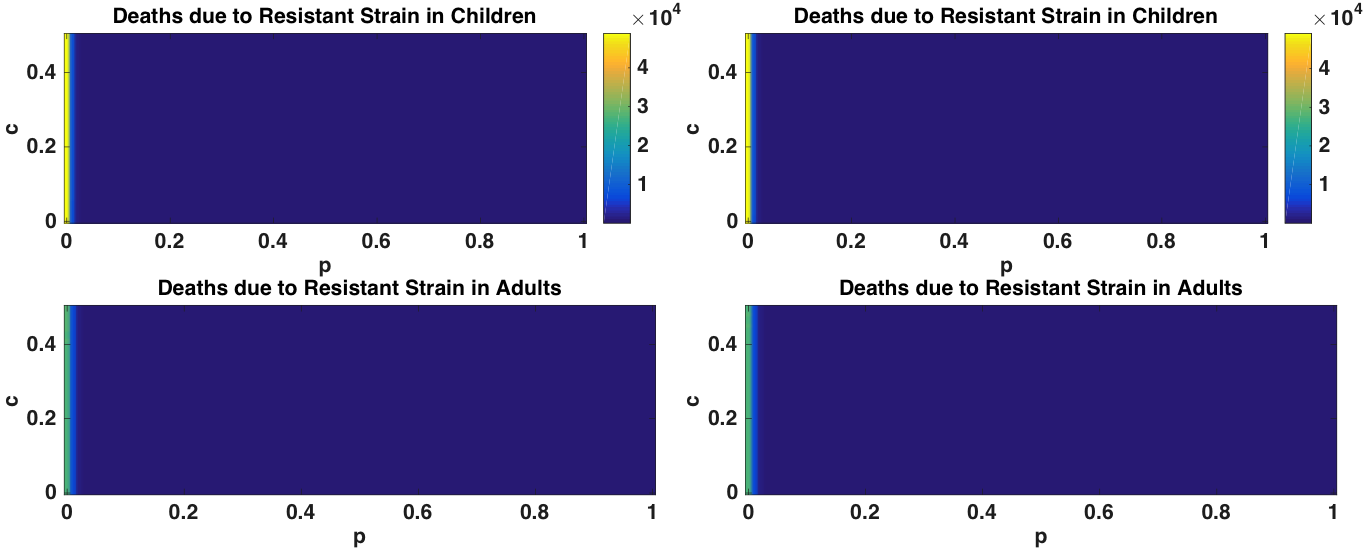}
  \caption{Heatmap of the \textit{number} of deaths from the resistant strain for the low transmission region. \textbf{(Left: (long/short)} AL for treatment  and \textbf{(Right: long/long)} SP for treatment. The parameter $p$ is the effectiveness of the treatment drug on the resistant strain, so $p=0$ is fully resistant and $p=1$ is fully sensitive. Number of deaths is dependent almost exclusively on $p$ and is much lower than the high transmission region. The zoom for realistic values of c looks very similar to the shown figures, so is omitted.}
  \label{F:heatmapnumlow}
\end{figure}

\begin{figure}[h]
\centering
  \includegraphics[scale=0.7]{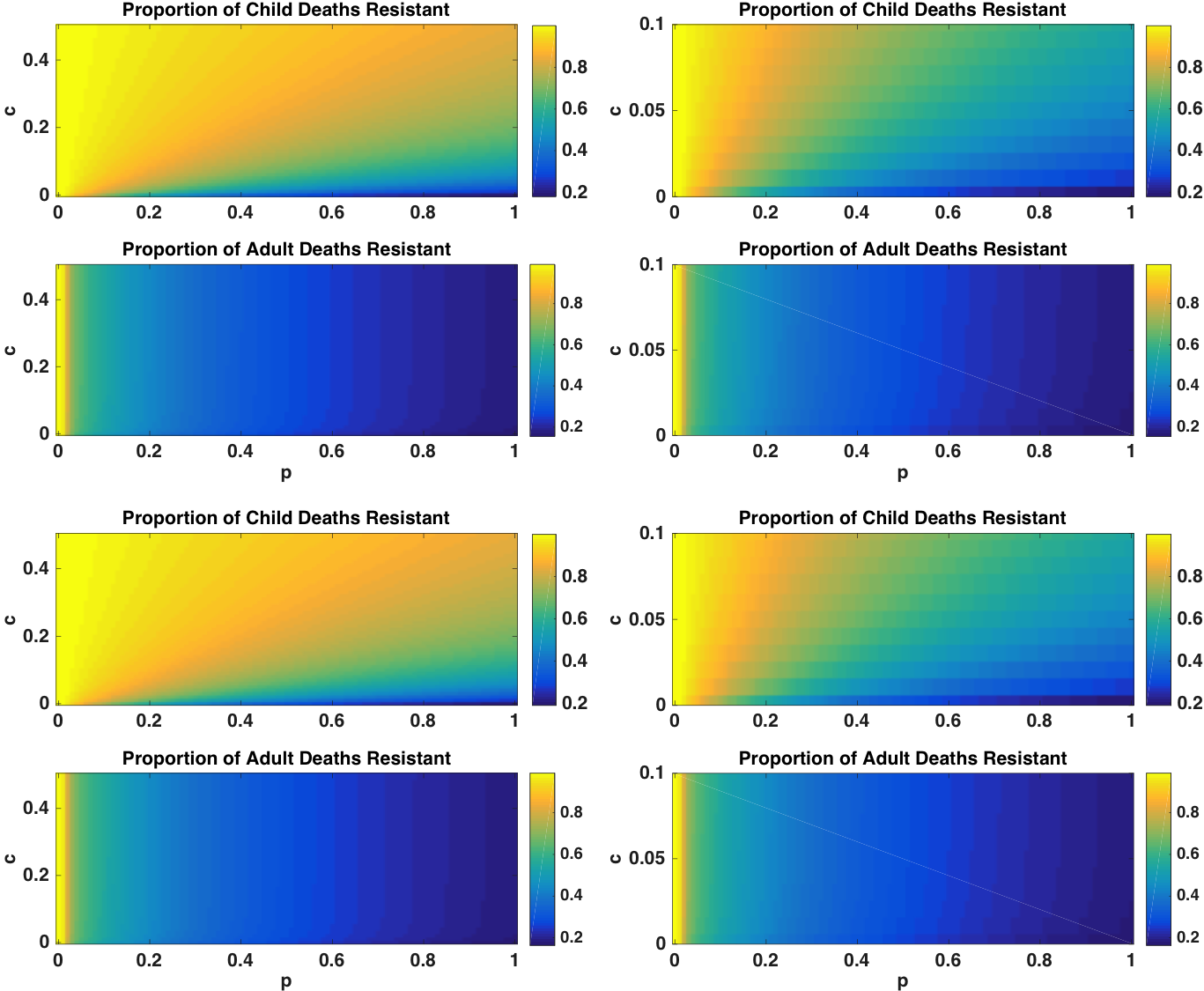}
  \caption{Heatmap of the \textit{proportion} of deaths from the resistant strain for the low transmission region and for \textbf{(Top Row: long/short)} AL for treatment and \textbf{(Bottom Row: long/long)} SP for treatment. The right column is a zoom-in of the left column to show more realistic values of $c$, the rate at which IPT is given.  Note different scales for top and bottom rows. The proportion of deaths from the resistant strain is dependent on both $p$ and $c$, showing that IPT schedule can increase resistance. However, unlike for high transmission, in this case the adult population is not affected by IPT.}
  \label{F:heatmapproplow}
\end{figure}

Next we present heatmaps of number of child deaths from malaria across $p$ and $c$ space for the low transmission region for long/short (Figure \ref{F:heatmapnumlow}, left) and for long/long (Figure \ref{F:heatmapnumlow}, right) IPT/treatment half-lives. For both scenarios, the number of deaths depends almost exclusively on the value of $p$, or resistance to the treatment drug. However, the proportion of deaths from the resistant strain, Figure \ref{F:heatmapproplow}, does depend on $c$, or the frequency of IPT application, particularly as values of $p$ increase. Also, unlike the high transmission region, the number of deaths from malaria in adults is unchanged by IPT usage.

\begin{figure}[h]
\centering
      \subfigure[High Transmission Region (SP)]{
    \includegraphics[scale=0.3]{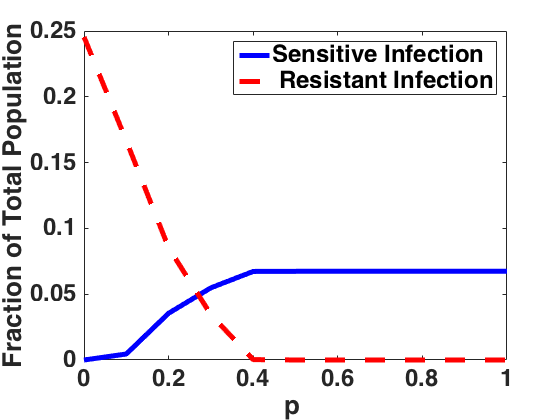} }
      \subfigure[Low Transmission Region (SP)]{
    \includegraphics[scale=0.3]{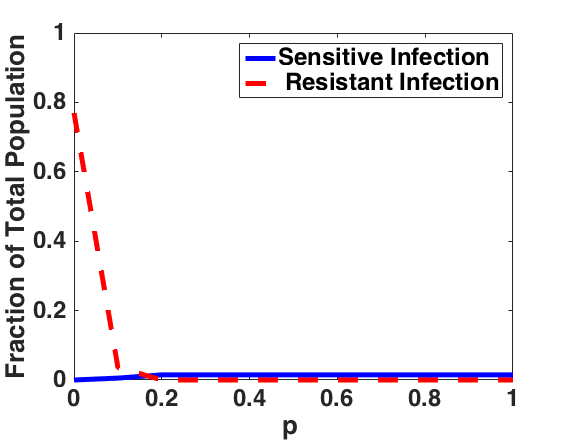} }
         \subfigure[High Transmission Region (AL)]{
    \includegraphics[scale=0.3]{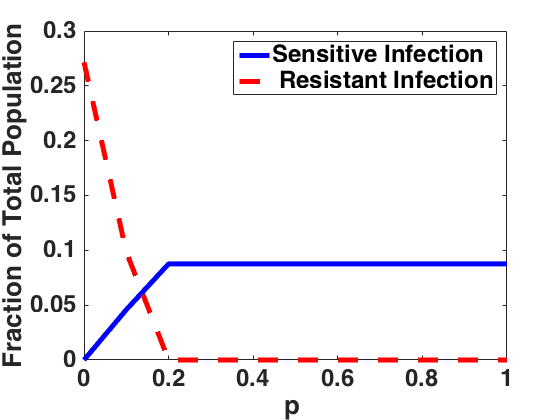} }
      \subfigure[Low Transmission Region (AL)]{
    \includegraphics[scale=0.3]{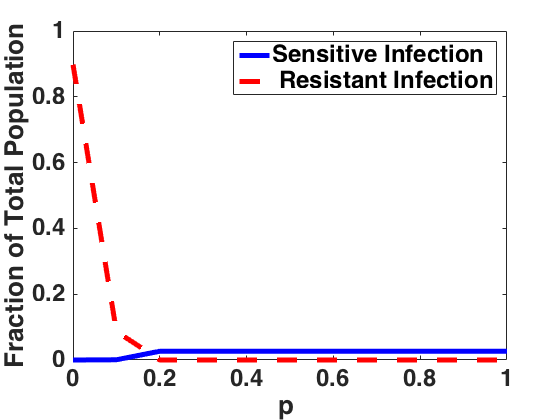} }
\caption{ Fraction of the total population infected with sensitive and resistant strains at $t$=10 years when both treatment and IPT are applied the whole time. Note that the region for coexistence of the sensitive and resistant strains has a small range.  As $p$ increases, more people with the symptomatic resistant strain get effective treatment, thereby shortening the infectious period.  The scale for each $y$-axis is different.   }
\label{F:IJversusp}
\end{figure}

Figure \ref{F:IJversusp} gives more information about why we see some distinct impacts on levels of resistance and number of deaths as $p$ varies for the high and low transmission regions. For the high transmission region (Figure \ref{F:IJversusp}(a)), resistance dominates to the exclusion of the sensitive strain for $p<0.1$ (long/long). For approximately $0.1<p<0.4$, the fraction of sensitive increases with $p$ while fraction resistant decreases but both strains coexist. Finally, for $p>0.4$, the sensitive strain is dominant to the exclusion  of the resistant strain and the sensitive strain persists at endemic but relatively low levels due to treatment. For the low transmission region (Figure \ref{F:IJversusp}(b)), the resistant strain dominates until about $p=0.1$, at which point it drops precipitously while the sensitive strain increases for $0.1<p<0.2$ after which the resistant strain is extinct and the sensitive strain persists at low and steady levels due to treatment. The scales are again different for high and low transmission regions, which reflects the much higher prevalence of malaria in the high transmission regions.

\begin{table}[htbp]
\caption{Total number of child deaths from malaria for various values of $p$ and either no IPT or IPT used. The (IPT/treatment) half-lives are also noted where the long half-life drug is SP and the short half-life drug is (AL). For high resistance to treatment (low values of $p$), the total number of deaths is much higher than for lower resistance to treatment. The cut-off for this dramatic increase in number of deaths is at about $p=0.3$ for the high transmission region and about $p=0.1$ for the low transmission region for long/long scenario. }
\label{T:totaldeaths}
\rowcolors{2}{gray!25}{white}
\begin{tabular}{ l|  c c|| c c|| c c}
\hline
&\multicolumn{2}{l||}{   \textbf{Year 1}} &\multicolumn{2}{l||}{   \textbf{         Year 5}} &\multicolumn{2}{l}{   \textbf{Year 10}} \\
\hline\hline
$p$ &{    No IPT    } & {    IPT    } & {    No IPT    } &  {    IPT    }&{    No IPT    } & {    IPT    }\\
\hline\hline
\multicolumn{7}{|l|}{\textbf{     High transmission region, long/long}}\\
\hline
0.1 &	 	77,743&	77,823&	118,505&	119,455&	171,716&	174,047\\
0.2&		26,806&	27,103&	43,929&	46,795&	67,258&	73,304\\
0.25&	14,860&	14,973&	25,622&	28,221&	40,901&	46,687\\
0.3&		9,158&	9,052&	14,772&	16,771&	24,720&	29,759\\
0.35&	8,533&	8,407&	12,167&	11,375&	16,959&	18,864\\
0.4&		8,394&	8,141&	12,021&	10,990&	16,717&	14,742\\
0.5&		8,254&	8,052&	11,878&	10,888&	16,575&	14,605\\
\hline\hline
\multicolumn{7}{|l|}{\textbf{     Low transmission region, long/long}}\\
\hline
0.09&	2,309&	2,308&	4,950&	4,961&	9,179&	9,192\\
0.1&  	696	&	684	&	1,599&	1,836&	4,125&	4,930\\
0.11&	323	&	303	&	503	&	379	&	787	&	497\\
0.12&	313	&	295	&	495	&	373	&	784	&	503\\
0.13&	300	&	281	&	482	&	359	&	772	&	490\\
0.15&	301	&	285	&	484	&	363	&	774	&	494\\
0.2  &	288	&	270	&	471	&	348	&	760	&	479\\
0.3  &	279	&	268	&	462	&	346	&	751	&	477\\
\hline\hline\hline
\multicolumn{7}{|l|}{\textbf{     High transmission region, long/short}}\\
\hline
0.1 &	 13,500 & 13,147 & 19,596 & 17,080&   27,522 & 22,776\\
0.2 &	 13,341 & 12,955 & 19,440 & 16,714 & 27,367 & 22,292\\
0.25 & 13,275 & 12,842 & 19,371 & 17,134 & 27,318 & 22,822\\
0.3 &	 13,235 & 12,572 & 19,331 & 16,847 & 27,258 & 22,425 \\
0.35 & 13,208 & 12,649 & 19,304 & 16,762 & 27,230 & 22,508\\
0.4 &	 13,187 & 12,856 & 19,284 & 16,784 & 27,212& 22,424\\
0.5 &	 13,164 & 12,793 & 19,260 & 17,068 & 27,187& 22,678 \\
\hline\hline\hline
\multicolumn{7}{|l|}{\textbf{     Low transmission region, long/short}}\\
\hline
0.09& 1,539 & 1,536 & 3,741 & 3,866 & 7,688 & 7,877 \\
0.10& 499 & 765 & 698 & 1,060 & 1,137 & 2,658 \\
0.11& 380 & 366 & 650 & 459 & 940 & 622 \\
0.12& 358 & 340 & 572 & 428 & 915 & 577 \\
0.13& 349 & 326 & 563 & 415 & 906 & 563 \\
0.15& 340 & 323 & 554 & 411 & 896 & 559 \\
0.2& 329 & 316 & 543  & 404 & 886 & 552 \\
0.3& 321& 300 & 535 & 388 & 878 & 537 \\
\hline
\end{tabular}
\end{table}

\clearpage

\section{Parameter Sensitivity}\label{S:sensitivity}
\label{LHS}

Latin hypercube sampling (LHS) \cite{McKay1979}, is a technique that uses stratified sampling without replacement. The LHS technique takes $n_p$ parameter distributions, divides them into $N$ predetermined equally probable intervals, and then draws a sample from each interval. For the system described by equations \eqref{juvenilehuman_odesfirst}-\eqref{juvenilehuman_odeslast}, \eqref{Maturehuman_odesfirst}-\eqref{Maturehuman_odeslast} and \eqref{susmosquito_odes}-\eqref{resinfmosquito_odes}, with $n_p=18$  parameters, the technique generates a hypercube of size $N$, chosen to be 5000 row by 18 column matrix of parameter values.  Each set of 18 parameter values is then used  to generate a solution for the system given in equations \eqref{juvenilehuman_odesfirst}-\eqref{juvenilehuman_odeslast}, \eqref{Maturehuman_odesfirst}-\eqref{Maturehuman_odeslast} and \eqref{susmosquito_odes}-\eqref{resinfmosquito_odes} for a total of 5000 simulations.  The LHS method performs an unbiased estimate of the average model output, sampling each parameter interval shown as ranges in Tables \ref{T:parmvalues}  and \ref{T:parmvalues2} exactly once.

\begin{figure}[h]
\centering
\subfigure[Child, 1 Year]{\includegraphics[scale=.28]{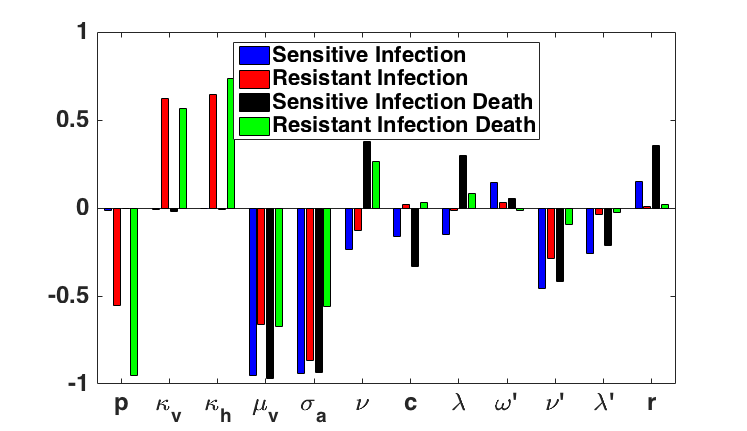}}
\subfigure[Child, 5 Years]{\includegraphics[scale=.28]{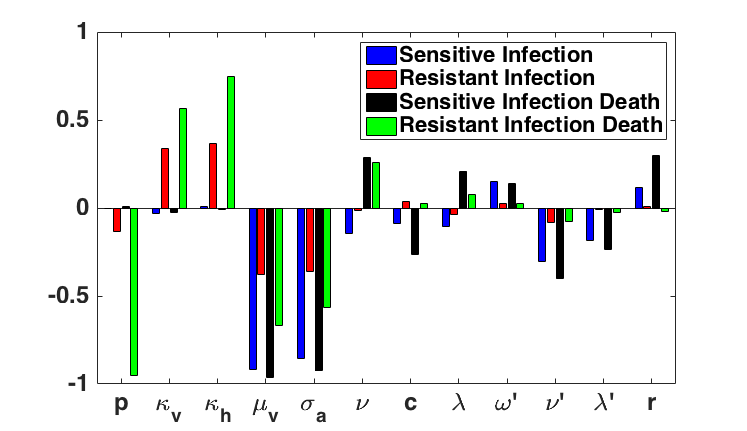}}
\caption{Note that as time increases, the sensitivity to $p$ decreases for infections, but not for deaths. Each parameter has a quartet of bars representing the PRCC values for sensitive child infections, resistant child infections, sensitive child deaths, and resistant child deaths.}
\label{F:prcc_child}
\end{figure}

\begin{figure}[h]
\centering
\subfigure[Adult, 1 Year]{\includegraphics[scale=.28]{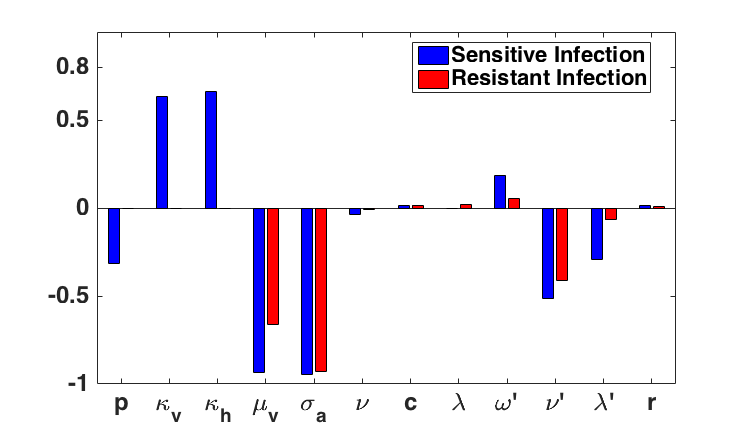}}
\subfigure[Adult, 5 Years]{\includegraphics[scale=.28]{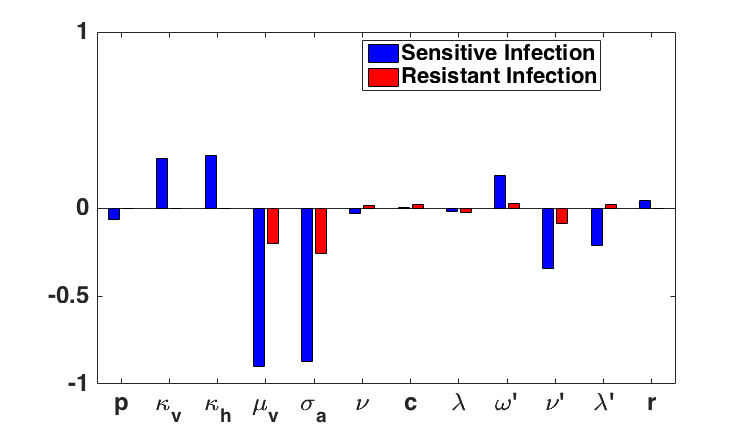}}
\caption{Note that as time increases, the sensitivity to $p$,   $\kappa_v$, and $\kappa_h$ decrease. Each parameter has a doublet of bars representing the PRCC values for sensitive  and  resistant  infections.}
\label{F:prcc_adult}
\end{figure}

Figures \ref{F:prcc_child} and \ref{F:prcc_adult} show only the statistically significant parameters ($p$-test value $< 0.01$).  Note that as time increases from 1 year to 5 years to 10 years since the start of IPT, the significance of $p$ decreases for the sensitive and resistant infections.  This is expected as the reproduction numbers $\mathcal{R}_S$ and $\mathcal{R}_R$ do not depend on $p$.  However, the PRCC plot illustrates that the number of child deaths due to the resistant strain greatly decreases as $p$ increases.  This is a result we have seen repeatedly in our numerical simulations, illustrating that numerical simulations add to our understanding of the dynamical progression of IPT and its influence on death prevention and disease resistance.  The PRCC plots for the high and low transmission regions show the same sensitivities as we have the same model for both regions with only changes in parameter values.

We can see in Figures \ref{F:prcc_child} and \ref{F:prcc_adult} that, for all QOI, $\mu_v$ and $\sigma_a$, the death rate of mosquitoes and rate at which asymptomatic juveniles clear infection naturally, are extremely important. As the lifespan of the mosquito decreases (or $\mu_v$ increases), the QOI all decrease. As the time spent asymptomatic but still infectious for juveniles decreases (so $\sigma_a$ increases), the QOI all decrease. Additional important parameters are $p$, $\kappa_v$, and $\kappa_h$. The number of child deaths from resistant infection is particularly sensitive to $p$ and as $p$ increases, that number decreases. $\kappa_v$, and $\kappa_h$ are measures of the competitive disadvantage of the resistant strain. As they increase towards 1 (so the competitive disadvantage decreases), the resistant infections and resistant deaths increase significantly.


\section{Discussion and Conclusion}\label{S:discussion}

There are a few general patterns in our simulations. First, using a short half-life treatment drug, assumed here to be effective against both sensitive and resistant symptomatic infections, decreases the advantage of the resistant strain, so also reduces the dependence of resistant emergence on IPT. Second, all the results are highly sensitive to $p$ and the value of $p$ at which the resistant strain dominates depends on whether it's a low or high transmission region.
There are strong non-linear relationships between $p$, $c$, and the IPT and treatment drug half-lives. There are bifurcations in realistic parameter regimes that suggest IPT should be applied with caution and with a good knowledge of the background levels of resistance in the region. Finally, we specifically considered both short- and long-range results (1 - 10 years) to inform the sustainability of current IPT and treatment programs. Particularly as new drugs are not quickly developed, it will be important to know if our current protocols will result in high levels of resistance in the future.

In the high transmission region, successful invasion of resistant strains is mostly driven by the drug(s) used for symptomatic treatment. Over the first year, IPT has a 0.1\%-5\% effect (both increases and reductions) on the total number of deaths from malaria for all scenarios. When a short half-life drug such as AL or ACT is used for treatment, IPT usage always results in lives saved with a 16.5\%-18.5\% reduction in total child deaths over 5 years (around 4,500-5,000 lives saved). However, when a long half-life drug such as SP is used for symptomatic treatment, use of IPT results range from a 13\% increase in deaths to an 8.5\% decrease in deaths over 5 years (from 2,900 additional deaths to 1,000 lives saved). When resistance to the treatment drug is high (p is low) then IPT use results in faster takeover of the resistant strain, thus causing in more deaths.  Initially, one would then recommend using a short half-life treatment drug whenever possible while applying IPT with a long half-life drug such as SP.

However, it is important to note the effect that the half-life of the symptomatic treatment drug has on total number of deaths. In particular, a short half-life treatment drug gives very similar total number of deaths across the resistance level spectrum, from partially to nearly fully resistant. The long half-life drug used as treatment gives order of magnitude differences in total deaths depending on the level of resistance. When $p=0.10$ (resistance is high), there are 119,000 total deaths over 5 years, whereas when $p = 0.50$ (low resistance) there are about 11,000 deaths over 5 years. For the short half-life treatment drug scenario, the total number of deaths over 5 years is about 17,000 for all levels of resistance considered and thus gives much lower number of deaths than the long/long scenario for highly resistant strains, but higher total deaths if resistance is weak.

The take-home message is that (1) treatment drugs are generally driving resistance in high transmission areas and the role of IPT in driving resistance tends to be minor comparatively, (2) however, when a highly resistant strain is circulating, IPT can indeed result in increased levels of resistance and loss of lives, particularly over longer time periods, and (3) in general, when short half-life drugs such as AL or ACT are used for treatment and SP is used for IPT, as is currently the case, regular use of IPT in children will result in potentially thousands of lives saved over the course of 5 to 10 years. We point out that the dynamics can be complex, so there are levels of resistance for which IPT saves lives over a short time period, but results in a cumulative loss of lives over 5-10 year periods as resistance levels ramp up. Therefore, our model suggests caution in using IPT without a corresponding heightened surveillance and awareness of changes in the circulating resistant strains over time. If resistance were to be significantly increasing over time, then evaluation of both the treatment drug and IPT usage would be warranted. Finally, we measured the effectiveness of IPT in lives saved. There may also be other benefits, such as a shortened length of asymptomatic malaria infections, that are not measured here.

In low transmission regions, we see different patterns in the costs and benefits of IPT. Here, IPT can have a much larger role in driving resistance when highly resistant strains are circulating. For example, in the long/long scenario with a highly resistant strain circulating, the proportion of resistant cases stays low when IPT is not used, but rises to over 70\% in children over the course of 10 years when IPT is used (Figure 13). For the long/short scenario, IPT also results in an increase in proportion resistant that would not otherwise occur, but at a greatly reduced rate of increase (Figure 14). However, for all but the most highly resistant strains, IPT usage in low transmission regions results in lives saved and does not drive take over of resistant strains. IPT generally results in a 24-26\% reduction in deaths in the long/long scenario over 5 years (about 120 lives saved) and in 26-29\% decrease in deaths for the long/short scenario over 5 years (about 140 lives saved). Thus, in general, it is better to use the short half-life treatment drug with a long half-life IPT in the low transmission regions. Although it is not as critical as in the high transmission regions, our model does suggest some caution and an increased awareness of circulating resistant strains is warranted when IPT is used in a low transmission region.

	A more complete cost/benefit analysis that includes cost of IPT and treatment drugs per dose, total number of doses needed, and a broader definition of benefits including not only deaths averted but severe and asymptomatic cases averted and reductions in total time infected would be interesting. We have not considered how IPT might directly change the age at which children gain the ``mature'' status based on a combination of many previous exposures to malaria and general improvement in the immune system due to age. Effective use of IPT could in fact increase that age, resulting in more serious cases of malaria in older than usual children. This could result again in increases of deaths or serious disease in what we are now calling the mature age group. We have focused solely on the use of SP as the IPT drug while varying the drugs used for treatment. While this is generally true currently, considering additional drugs for potential use as IPT could be useful. We are looking at holoendemic regions with no seasonality (year-round transmission) and it would be interesting to extend to regions with seasonal malaria transmission.

\clearpage

\newpage
\section{Acknowledgements}
The authors would like to acknowledge the support of the American Institute of Mathematics through a AIM SquAREs grant.  Additional grant acknowledgements: Katharine Gurski was supported by NSF grant 1361209 and Simons Foundation grant 245237, and Carrie Manore was supported by NSF SEES grant CHE-1314029 and by a Los Alamos National Laboratory Director's Postdoctoral Fellowship.


\bibliographystyle{plain}
\bibliography{IPTMalariaReferences}
\newpage

\appendix

\section{}
\begin{table}[htbp]
\begin{tabular}{|c|c|p{2cm}|p{2cm}|p{2cm}|p{2cm}|}
\hline
&&{\bf Valley bottom} & {\bf Middle Hill} & {\bf Hilltop} & {\bf Asymp. Prevalence}\\
\hline\hline
Altitude in meters & & 1430 & 1500 & 1580 & \\
(Village) & & (Iguhu) & (Makhokho) & (Sigalagala) & \\
\hline
\multirow{3}{*}{\parbox{3cm}{\centering Duration (in months) of parasitemia by age}} &{\bf Age 5-9} & 6  & 4  & 3 & 34.4 \%\\
&{\bf Age 10-14} & 6 & 4 & 3 & 34.1 \%\\
&{\bf Age $>$14} & 1 & 1 & 1 & 9.1\%\\
\hline
\multirow{3}{*}{\parbox{3cm}{\centering \% asymptomatic by region}} & & \multirow{3}{*}{52.4\%} &  & \multirow{3}{*}{23.3\%} &\\
&&&&&\\
&&&&&\\
\hline
\multirow{3}{*}{\parbox{3cm}{\centering \% of vectors found in region}} & & \multirow{3}{*}{\parbox{3cm}{\centering 98\%}} & \multirow{3}{*}{\parbox{3cm}{\centering 1\%} }& \multirow{3}{*}{\parbox{3cm}{\centering 1\% }}\\
&&&&&\\
&&&&&\\
\hline
\multirow{3}{*}{\parbox{3cm}{\centering \% of 334 asymptomatic episodes in region}} & & \multirow{3}{*}{\parbox{3cm}{\centering 44\%}} & \multirow{3}{*}{\parbox{3cm}{\centering 24.9\%} }& \multirow{3}{*}{\parbox{3cm}{\centering 31.1\% }}\\
&&&&&\\
&&&&&\\
\hline
\end{tabular}
\caption{Duration (in months) of asymptomatic parasitemia by age and microgeographic locale; prevalence of asymptomatic malaria by age and region; and percent of vector population found in each locale.  This region is considered hypoendemic. 15\% of asymptomatic episodes lasted 1 month.  38.1\% of episodes lasted 2-5 months and 14.2\% of episodes lasted 6-12 months.  32.5\% experienced no infection episode. Note: Iguhu is near the Yala River, a major breeding site for {\it An. gambiae} mosquitoes \cite{baliraine2009high}.}
\end{table}

\section{Basic reproductive numbers }\label{A:R0}

The basic reproductive numbers for the sensitive parasite strain $\mathcal{R}_s$ and the resistant parasite strain $\mathcal{R}_r$ were computed using the next generation matrix.
The next generation matrix (NGM) is
$$K=\begin{pmatrix}
0 & K_{1,2}\\
K_{2,1} & 0
\end{pmatrix},$$
where
$$K_{1,2}=\begin{pmatrix}
\frac{\beta_h \lambda S0}{\mu_m N_0} & 0 & 0 & 0 & 0 & 0 \\
  \frac{\beta_h (1-\lambda) S_0}{\mu_m N_0} & 0 & 0 & 0 & 0 & 0 \\
  0 & \frac{\beta_h k_h \lambda (S_0+T_0)}{\mu_m N_0} & 0 & 0 & 0 & 0 \\
 0 & \frac{\beta_h k_h (1-\lambda) (S_0+T_0)}{\mu_m N_0} & 0 & 0 & 0 & 0 \\
 \frac{\beta_h \lambda' S_{m0}}{\mu_m N_0} & 0 & 0 & 0 & 0 & 0 \\
 \frac{\beta_h (1-\lambda') S_{m0}}{\mu_m N_0} & 0 & 0 & 0 & 0 & 0 \\
 0 & \frac{\beta_h k_h \lambda_p (S_{m0}+T_{m0})}{\mu_m N_0} & 0 & 0 & 0 & 0 \\
  0 & \frac{\beta_h k_h (1-\lambda') (S_{m0}+T_{m0})}{\mu_m N_0} & 0 & 0 & 0 & 0
\end{pmatrix}$$
and
$$K_{2,1}=
\begin{pmatrix}
k_{9,1} &k_{9,2} & 0 & 0 & k_{9,5} &
  k_{9,6} & 0 & 0 \\
 0 & 0 & k_{10,3}& k_{10,4}& 0 & 0 &
   k_{10,7} &k_{10,8}\\
 0 & 0 & 0 & 0 & 0 & 0 & 0 & 0\\
 0 & 0 & 0 & 0 & 0 & 0 & 0 & 0\\
 0 & 0 & 0 & 0 & 0 & 0 & 0 & 0\\
 0 & 0 & 0 & 0 & 0 & 0 & 0 & 0
\end{pmatrix}
$$
\begin{align*}
k_{9,1}&= \frac{\beta_m S_{v0}}{N_0}\left(1+\frac{\eta}{A_{ms}}\right) \\
k_{9,2}&= \frac{\beta_m  S_{v0}}{A_a
   N_0}\left(1+\frac{\nu}{A_s} + \frac{\eta  (A_{ma} \nu +A_s \nu')}{A_s A_{ms} A_{ma}}+\frac{ \eta}{A_{ma}}\right)\\
 k_{9,5}&=\frac{\beta_m S_{v0}}{A_{ms} N_0} \\
 k_{9,6}&= \frac{\beta_m S_{v0}}{A_{ma} N_0}\left(1+\frac{\nu'}{A_{ms}}\right)\\
 k_{10,3}&=\frac{\beta_m k_m S_{v0}}{B_s N_0}\left(1+\frac{\eta}{B_{ms}} \right)\\
 k_{10,4} &=  \frac{\beta_m k_m
   S_{v0}}{B_a N_0} \left(1+\frac{ \eta (A_{ma} \nu +B_s \nu')}{B_s A_{ma}
   B_{ms}} + \frac{\eta  k_m}{A_{ma}} + \frac{\nu}{B_s} \right) \\
 k_{10,7}&= \frac{\beta_m k_m S_{v0}}{B_{ms} N_0}\\
 k_{10,8} &=  \frac{b_m k_m S_{v0}}{A_{ma} N_0}\left(1+\frac{\nu'}{B_{ms}}\right)
 \end{align*}

In addition to the next generation matrix approach, the reproductive numbers were derived based on the biological interpretation of the model.

\subsubsection*{Sensitive reproduction number  $\mathcal{R}_s$}
Let $\mathcal{R}_{s-asym}^{naive}$ and $\mathcal{R}_{s-sym}^{naive}$ denote the reproduction numbers for the sensitive strain of infection associated with asymptomatic and symptomatic cases in naive humans, respectively.  Let $\mathcal{R}_{s-asym}^{mature}$ and $\mathcal{R}_{s-sym}^{mature}$ denote the reproduction numbers for the sensitive strain of infection associated with asymptomatic and symptomatic cases, in mature humans respectively.

At the beginning of an outbreak the proportion of the population susceptible to the sensitive parasite is $S_0+S_{m0}$. A portion of this sensitive population will become asymptomatically infected and either remain asymptomatic or transition to a symptomatic case (there is no transition from symptomatic to asymptomatic in this model). A portion of these infected individuals will age into the mature population. The sensitive reproductive number for the asymptomatic cases in the naive population over the full course of infection, i.e., the number of naive human asymptomatic cases resulting from one initially sensitive case, is then given by
\begin{align*}
{\mathcal R}_{s-asym}^{naive}&= \underbrace{\left(1-\lambda \right)}_\text{\parbox{1.7cm}{\centering fraction that are asym.}}
\underbrace{\left(\beta_m\right)}_\text{\parbox{1.5cm}{\centering trans. rate to vectors}}
\underbrace{ \Bigg[\frac{1}{A_a}}_\text{\parbox{1.75cm}{\centering duration of naive asym.}} +
\underbrace{\frac{\nu}{A_a}}_\text{\parbox{2.cm}{\centering fraction that become sym.}}
\underbrace{\Bigg(\frac{1}{A_s}}_\text{\parbox{1.75cm}{\centering duration of naive sym.}}+
\underbrace{\Big(\frac{\eta}{A_s}\Big)}_\text{\parbox{1.75cm}{\centering fraction of sym. that age}}
\underbrace{\Big(\frac{1}{A_{ms}}\Big)\Bigg)}_\text{\parbox{1.75cm}{\centering duration of mature sym.}}
\\+&  \underbrace{\Big(\frac{\eta}{A_a}\Big)}_\text{\parbox{2cm}{\centering fraction of asym. that age}}
\underbrace{\Bigg(\frac{1}{A_{ma}}}_\text{\parbox{1.75cm}{\centering duration of mature asym.}}+
\underbrace{\frac{\nu'}{A_{ma}}}_\text{\parbox{2.cm}{\centering fraction that become sym.}}
\underbrace{\Big(\frac{1}{A_{ms}}\Big)\Bigg)\Bigg]}_\text{\parbox{2.cm}{\centering duration of mature sym.}}
\underbrace{\left( \beta_h\right)}_\text{\parbox{1.5cm}{\centering trans. rate  to hosts}}
\underbrace{\left( \frac{S_{v0}}{N_0}\right)}_\text{\parbox{1.6cm}{\centering vector to host ratio}}
\underbrace{\left( \frac{1}{\mu_m}\right)}_\text{\parbox{2cm}{\centering duration  of vector infection}}
\underbrace{\left(\frac{S_0}{N_0}\right)}_\text{\parbox{1.5cm}{\centering susceptible proportion}}
\end{align*}
The sensitive symptomatic reproductive number for the naive population, or the number of naive human cases resulting from one initial symptomatic individual, is given by
\begin{equation*}
\mathcal{R}_{s-sym}^{naive}= \underbrace{\lambda}_\text{\parbox{1.7cm}{\centering fraction that are sym.}}
\underbrace{\left(\beta_m\right)}_\text{\parbox{1.5cm}{\centering trans. rate to vectors}}
 \underbrace{\Bigg(\Big(\frac{1}{A_s}\Big)}_\text{\parbox{2cm}{\centering duration of naive sym.}}+
\underbrace{\Big(\frac{\eta}{A_s}\Big)}_\text{\parbox{1.75cm}{\centering fraction of sym. that age}}
\underbrace{\Big(\frac{1}{A_{ms}}\Big)\Bigg)}_\text{\parbox{1.75cm}{\centering duration of mature sym.}}
 \underbrace{\left( \beta_h\right)}_\text{\parbox{1.5cm}{\centering trans. rate to hosts}}
\underbrace{\left( \frac{S_{v0}}{N_0}\right)}_\text{\parbox{1.6cm}{\centering vector to host ratio}}
\underbrace{\left( \frac{1}{\mu_m}\right)}_\text{\parbox{2cm}{\centering duration  of vector infection}}
\underbrace{\left(\frac{S_0}{N_0}\right).}_\text{\parbox{1.4cm}{\centering susceptible proportion}}
\end{equation*}
The sensitive reproductive number for the asymptomatic cases in the mature population over the full course of infection, i.e., the number of mature human asymptomatic cases resulting from one initially sensitive case, is then given by
\begin{align*}
{\mathcal R}_{s-asym}^{mature}&= \underbrace{\left(1-\lambda' \right)}_\text{\parbox{1.7cm}{\centering fraction that are asym.}}
\underbrace{\left(\beta_m\right)}_\text{\parbox{1.5cm}{\centering trans. rate to vectors}}
 \Bigg(\underbrace{\Big(\frac{1}{A_{ma}}\Big)}_\text{\parbox{2cm}{\centering duration of mature asym. }} +
\underbrace{\Big(\frac{\nu'}{A_{ma}}\Big)}_\text{\parbox{2.5cm}{\centering fraction that become sym.}}
\underbrace{\Big(\frac{1}{A_{ms}}\Big)}_\text{\parbox{2cm}{\centering duration of mature sym.}} \Bigg)
\\&\underbrace{\left( \beta_h\right)}_\text{\parbox{1.6cm}{\centering trans. rate  to hosts}}
\underbrace{\left( \frac{S_{v0}}{N_0}\right)}_\text{\parbox{1.6cm}{\centering vector to host ratio}}
\underbrace{\left( \frac{1}{\mu_m}\right)}_\text{\parbox{2cm}{\centering duration  of vector infection}}
\underbrace{\left(\frac{S_{m0}}{N_0}\right)}_\text{\parbox{1.5cm}{\centering susceptible proportion}}
\end{align*}
The sensitive symptomatic reproductive number for the mature population, or the number of mature human cases resulting from one initial symptomatic individual, is given by
\begin{equation*}
\mathcal{R}_{s-sym}^{mature}= \underbrace{\lambda'}_\text{\parbox{1.7cm}{\centering fraction that are sym.}}
\underbrace{\left(\beta_m\right)}_\text{\parbox{1.8cm}{\centering trans. rate to vectors}}
 \underbrace{\left(\frac{1}{A_{ms}}\right)}_\text{\parbox{2cm}{\centering duration of mature sym.}}
 \underbrace{\left( \beta_h\right)}_\text{\parbox{1.8cm}{\centering trans. rate to hosts}}
\underbrace{\left( \frac{S_{v0}}{N_0}\right)}_\text{\parbox{1.6cm}{\centering vector to host ratio}}
\underbrace{\left( \frac{1}{\mu_m}\right)}_\text{\parbox{2cm}{\centering duration  of vector infection}}
\underbrace{\left(\frac{S_{m0}}{N_0}\right).}_\text{\parbox{1.4cm}{\centering susceptible proportion}}
\end{equation*}
Then, the reproduction number for the sensitive strain of infection takes the following form:
\begin{equation}
\begin{aligned}
\mathcal{R}^2_s &= \mathcal{R}_{s-asym}^{naive}+ \mathcal{R}_{s-sym}^{naive}+ \mathcal{R}_{s-asym}^{mature}+ \mathcal{R}_{s-sym}^{mature} \\
&=\frac{\beta_m\beta_hS_0S_{v0}}{\mu_mN_0^2}\left[\frac{1-\lambda}{A_a}+ \frac{\nu(1-\lambda)}{A_aA_s} + \frac{\eta\nu(1-\lambda)}{A_aA_{ms}A_s} + \frac{\eta(1-\lambda)}{A_aA_{ma}}
+\frac{\eta\nu'(1-\lambda)}{A_aA_{ma}A_{ms}} +\frac{\lambda}{A_s}+\frac{\eta\lambda}{A_sA_{ms}}   \right]
\\&+\frac{\beta_m\beta_hS_{m0}S_{v0}}{\mu_mN_0^2}\left[\frac{1-\lambda'}{A_{ma}}+ \frac{\nu'(1-\lambda')}{A_{ma}A_{ms}} +\frac{\nu'}{A_{ms}}  \right].
\end{aligned}
\end{equation}
The above reproduction number $\mathcal{R}_s$ was also computed using the next generation matrix approach.

\subsubsection*{Resistant reproduction number  $\mathcal{R}_r$}
Let $\mathcal{R}_{r-asym}^{naive}$ and $\mathcal{R}_{r-sym}^{naive}$ denote the reproduction numbers for the resistant strain of infection associated with asymptomatic and symptomatic cases in naive humans, respectively.  Let $\mathcal{R}_{r-asym}^{mature}$ and $\mathcal{R}_{r-sym}^{mature}$ denote the reproduction numbers for the resistant strain of infection associated with asymptomatic and symptomatic cases, in mature humans respectively.

At the beginning of an outbreak the proportion of the population susceptible to the resistant parasite is $S_0+S_{m0}+T_0+T_{m0}$. The resistant reproductive number for the asymptomatic cases in the naive population over the full course of infection, i.e., the number of naive human asymptomatic cases resulting from one initially resistant case, is then given by

\begin{align*}
{\mathcal R}_{r-asym}^{naive}&= \underbrace{\left(1-\lambda \right)}_\text{\parbox{1.7cm}{\centering fraction that are asym.}}
\underbrace{\left(\beta_m\kappa_m\right)}_\text{\parbox{1.5cm}{\centering trans. rate to vectors}}
\underbrace{ \Bigg[\frac{1}{B_a}}_\text{\parbox{1.75cm}{\centering duration of naive asym.}} +
\underbrace{\frac{\nu}{B_a}}_\text{\parbox{2.cm}{\centering fraction that become sym.}}
\underbrace{\Bigg(\frac{1}{B_s}}_\text{\parbox{1.75cm}{\centering duration of naive sym.}}+
\underbrace{\Big(\frac{\eta}{B_s}\Big)}_\text{\parbox{1.75cm}{\centering fraction of sym. that age}}
\underbrace{\Big(\frac{1}{B_{ms}}\Big)\Bigg)}_\text{\parbox{1.75cm}{\centering duration of mature sym.}}
\\+&  \underbrace{\Big(\frac{\eta}{B_a}\Big)}_\text{\parbox{2cm}{\centering fraction of asym. that age}}
\underbrace{\Bigg(\frac{1}{A_{ma}}}_\text{\parbox{1.75cm}{\centering duration of mature asym.}}+
\underbrace{\frac{\nu'}{A_{ma}}}_\text{\parbox{2.cm}{\centering fraction that become sym.}}
\underbrace{\Big(\frac{1}{B_{ms}}\Big)\Bigg)\Bigg]}_\text{\parbox{2.cm}{\centering duration of mature sym.}}
\underbrace{\left( \beta_h\kappa_h\right)}_\text{\parbox{1.5cm}{\centering trans. rate  to hosts}}
\underbrace{\left( \frac{S_{v0}}{N_0}\right)}_\text{\parbox{1.6cm}{\centering vector to host ratio}}
\underbrace{\left( \frac{1}{\mu_m}\right)}_\text{\parbox{2cm}{\centering duration  of vector infection}}
\underbrace{\left(\frac{S_0+T_0}{N_0}\right)}_\text{\parbox{1.5cm}{\centering susceptible proportion}}
\end{align*}
The resistant symptomatic reproductive number for the naive population, or the number of naive human cases resulting from one initial symptomatic individual, is given by
\begin{equation*}
\mathcal{R}_{r-sym}^{naive}= \underbrace{\lambda}_\text{\parbox{1.7cm}{\centering fraction that are sym.}}
\underbrace{\left(\beta_m\kappa_m\right)}_\text{\parbox{1.5cm}{\centering trans. rate to vectors}}
 \underbrace{\Bigg(\Big(\frac{1}{B_s}\Big)}_\text{\parbox{2cm}{\centering duration of naive sym.}}+
\underbrace{\Big(\frac{\eta}{B_s}\Big)}_\text{\parbox{1.75cm}{\centering fraction of sym. that age}}
\underbrace{\Big(\frac{1}{B_{ms}}\Big)\Bigg)}_\text{\parbox{1.75cm}{\centering duration of mature sym.}}
 \underbrace{\left( \beta_h\kappa_h\right)}_\text{\parbox{1.5cm}{\centering trans. rate to hosts}}
\underbrace{\left( \frac{S_{v0}}{N_0}\right)}_\text{\parbox{1.6cm}{\centering vector to host ratio}}
\underbrace{\left( \frac{1}{\mu_m}\right)}_\text{\parbox{2cm}{\centering duration  of vector infection}}
\underbrace{\left(\frac{S_0+T_0}{N_0}\right).}_\text{\parbox{1.4cm}{\centering susceptible proportion}}
\end{equation*}
The resistant reproductive number for the asymptomatic cases in the mature population over the full course of infection, i.e., the number of mature human asymptomatic cases resulting from one initially resistant case, is then given by
\begin{align*}
{\mathcal R}_{r-asym}^{mature}&= \underbrace{\left(1-\lambda' \right)}_\text{\parbox{1.7cm}{\centering fraction that are asym.}}
\underbrace{\left(\beta_m\kappa_m\right)}_\text{\parbox{1.5cm}{\centering trans. rate to vectors}}
 \Bigg(\underbrace{\Big(\frac{1}{A_{ma}}\Big)}_\text{\parbox{2cm}{\centering duration of mature asym. }} +
\underbrace{\Big(\frac{\nu'}{A_{ma}}\Big)}_\text{\parbox{2.5cm}{\centering fraction that become sym.}}
\underbrace{\Big(\frac{1}{B_{ms}}\Big)}_\text{\parbox{2cm}{\centering duration of mature sym.}} \Bigg)
\\&\underbrace{\left( \beta_h\kappa_h\right)}_\text{\parbox{1.6cm}{\centering trans. rate  to hosts}}
\underbrace{\left( \frac{S_{v0}}{N_0}\right)}_\text{\parbox{1.6cm}{\centering vector to host ratio}}
\underbrace{\left( \frac{1}{\mu_m}\right)}_\text{\parbox{2cm}{\centering duration  of vector infection}}
\underbrace{\left(\frac{S_{m0}+T_{m0}}{N_0}\right)}_\text{\parbox{1.5cm}{\centering susceptible proportion}}
\end{align*}
The resistant symptomatic reproductive number for the mature population, or the number of mature human cases resulting from one initial symptomatic individual, is given by
\begin{equation*}
\mathcal{R}_{r-sym}^{mature}= \underbrace{\lambda'}_\text{\parbox{1.7cm}{\centering fraction that are sym.}}
\underbrace{\left(\beta_m\kappa_m\right)}_\text{\parbox{1.8cm}{\centering trans. rate to vectors}}
 \underbrace{\left(\frac{1}{B_{ms}}\right)}_\text{\parbox{2cm}{\centering duration of mature sym.}}
 \underbrace{\left( \beta_h\kappa_h\right)}_\text{\parbox{1.8cm}{\centering trans. rate to hosts}}
\underbrace{\left( \frac{S_{v0}}{N_0}\right)}_\text{\parbox{1.6cm}{\centering vector to host ratio}}
\underbrace{\left( \frac{1}{\mu_m}\right)}_\text{\parbox{2cm}{\centering duration  of vector infection}}
\underbrace{\left(\frac{S_{m0}+T_{m0}}{N_0}\right).}_\text{\parbox{1.4cm}{\centering susceptible proportion}}
\end{equation*}
Then, the reproduction number for the resistant strain of infection takes the following form:
\begin{equation}
\begin{aligned}
\mathcal{R}^2_r &= \mathcal{R}_{r-asym}^{naive}+ \mathcal{R}_{r-sym}^{naive}+ \mathcal{R}_{r-asym}^{mature}+ \mathcal{R}_{r-sym}^{mature} \\
&=\frac{\kappa_m\beta_m\kappa_h\beta_h(S_0+T_0)S_{v0}}{\mu_mN_0^2}\left[\frac{1-\lambda}{B_a}+ \frac{\nu(1-\lambda)}{B_aB_s} + \frac{\eta\nu(1-\lambda)}{B_aB_{ms}B_s} + \frac{\eta(1-\lambda)}{B_aA_{ma}}
+\frac{\eta\nu'(1-\lambda)}{B_aA_{ma}B_{ms}} +\frac{\lambda}{B_s}+\frac{\eta\lambda}{B_sB_{ms}}   \right]
\\&+\frac{\kappa_m\beta_m\kappa_h\beta_h(S_{m0}+T_{m0})S_{v0}}{\mu_mN_0^2}\left[\frac{1-\lambda'}{A_{ma}}+ \frac{\nu'(1-\lambda')}{A_{ma}B_{ms}} +\frac{\nu'}{B_{ms}}  \right].
\end{aligned}
\end{equation}
The above reproduction number $\mathcal{R}_r$ was also computed using the next generation matrix approach.


\end{document}